\documentclass[fleqn,usenatbib]{mnras}

\usepackage{newtxtext,newtxmath}
\usepackage{lscape}

\usepackage[T1]{fontenc}

\DeclareRobustCommand{\VAN}[3]{#2}
\let\VANthebibliography\thebibliography
\def\thebibliography{\DeclareRobustCommand{\VAN}[3]{##3}\VANthebibliography}

\usepackage{graphicx}	% Including figure files
\usepackage{amsmath}	% Advanced maths commands
\def\cm2{cm$^2$ }
\def\se1{s$^{-1}$ }

\def\arcsec{\hbox{$^{\prime\prime}$} }

\usepackage{xfrac}

\usepackage{threeparttable}
\usepackage{tablefootnote}
\usepackage{multirow}
\usepackage{booktabs}
\usepackage{float}

\newcommand{\I}{~\textsc{i}}
\newcommand{\ii}{~\textsc{ii}}
\newcommand{\iii}{~\textsc{iii}}
\newcommand{\iv}{~\textsc{iv}}
\newcommand{\V}{~\textsc{v}}

\title[{\rm O/H} and $L_{\rm X}$--$Z_{\rm NLR}$ relation in  Seyferts]{Oxygen abundances in the narrow line regions of Seyfert galaxies and the metallicity--luminosity relation}

\author[M. Armah et al.]{
Mark Armah,$^{1}$\thanks{E-mail: armah@ufrgs.br} Rog\'erio Riffel,$^{1,2}$\thanks{E-mail: riffel@ufrgs.br
} 
O. L. Dors,$^3$
Kyuseok Oh$^{4,5}$,
Michael J. Koss$^{6,7}$,
Claudio Ricci$^{8,9}$,
\newauthor
~Benny Trakhtenbrot$^{10}$,
Mabel Valerdi$^{11}$,
Rogemar A. Riffel$^{2,12}$ and
Angela C. Krabbe$^3$
\\
$^1$Departamento de Astronomia, Instituto de Física, Universidade Federal do Rio Grande do Sul, CP 15051, 91501-970, Porto Alegre, RS, Brazil \\
$^{2}$Laborat\'orio Interinstitucional de e-Astronomia - LIneA, Rua General Jos\'e Cristino 77, Rio de Janeiro, RJ - 20921-400, Brazil\\
$^3$ Universidade do Vale do Paraíba. Av. Shishima Hifumi, 2911, CEP: 12244-000, São José dos Campos, SP, Brazil\\
$^{4}$Korea Astronomy and Space Science Institute, Daedeokdae-ro 776, Yuseong-gu, Daejeon 34055, Republic of Korea\\
$^{5}$Department of Astronomy, Kyoto University, Kitashirakawa-Oiwake-cho, Sakyo-ku, Kyoto 606-8502, Japan\\
$^{6}$Eureka Scientific, 2452 Delmer Street Suite 100, Oakland, CA 94602-3017, USA\\
$^{7}$Space Science Institute, 4750 Walnut Street, Suite 205, Boulder, Colorado 80301, USA\\
$^{8}$N\'ucleo de Astronom\'ia de la Facultad de Ingenier\'ia, Universidad Diego Portales, Av. Ej\'ercito Libertador 441, Santiago 22, Chile\\
$^{9}$Kavli Institute for Astronomy and Astrophysics, Peking University, Beijing 100871, People's Republic of China\\
$^{10}$School of Physics and Astronomy, Tel Aviv University, Tel Aviv 69978, Israel\\
$^{11}$Instituto Nacional de Astrof{\'i}sica, \'Optica y Electr\'onica (INAOE), Luis E. Erro No. 1, Sta. Ma. Tonantzintla, Puebla, C.P. 72840, M\'exico \\
$^{12}$Departamento de F\'isica, CCNE, Universidade Federal de Santa Maria, 97105-900, Santa Maria, RS, Brazil\\
}

\date{Accepted XXX. Received YYY; in original form ZZZ}

\pubyear{2022}

\begin{document}

\label{firstpage}
\pagerange{\pageref{firstpage}--\pageref{lastpage}}
\maketitle

\begin{abstract}
We present  oxygen abundances relative to hydrogen (O/H) in the narrow line regions (NLRs) gas phases of Seyferts 1 (Sy 1s) and Seyferts 2 (Sy 2s) Active Galactic Nuclei (AGNs).  We used  fluxes of the optical narrow  emission line intensities  [$3\,500<\lambda($\AA$)<7\,000$]   of 561 Seyfert nuclei in the local universe ($z\lesssim0.31$) from the second catalog and data release  (DR2) of the BAT AGN Spectroscopic Survey, which focuses on the \textit{Swift}-BAT  hard X-ray ($\gtrsim10$ keV) detected AGNs.  We derived  O/H from  relative intensities of the emission lines via the strong-line methods.  We find that the AGN O/H abundances are related to their hosts stellar masses and that they follow a downward redshift evolution.
The derived O/H together with the hard X-ray  luminosity ($L_{\rm X}$)  were used to study the X-ray luminosity-metallicity ($L_{\rm X}$-$Z_{\rm NLR}$)  relation  for the first time in Seyfert galaxies.  In contrast to the broad-line focused ($L_{\rm X}$-$Z_{\rm BLR}$) studies, we find that  the $L_{\rm X}$-$Z_{\rm NLR}$ exhibit significant anti-correlations with the Eddington ratio  ($\lambda_{\rm Edd}$)  and these correlations vary with redshifts. This result indicates that the low-luminous AGNs are more actively undergoing Interstellar Medium (ISM) enrichment through star formation in comparison with the more luminous X-ray sources. Our results suggest that the AGN is somehow driving the galaxy chemical enrichment, as a result  of the inflow of pristine gas that is diluting the metal rich gas, together with a recent cessation on the circumnuclear star-formation.

\end{abstract}


\date{Released 2021 Apr 20}

%%%%%%%%%%%%%%%%%%%%%%%%%%%%%%%%%%%%%%%%%%%%%%%%%%

%%%%%%%%%%%%%%%%% BODY OF PAPER %%%%%%%%%%%%%%%%%%

\section{Introduction}
There are significant observational data which point to a direct link between the host galaxies and the accretion onto supermassive black holes (SMBHs). For instance, the masses of star bulges and the masses of their central SMBHs are  tightly correlated \citep[see][for review]{2013ARA&A..51..511K,2016ASSL..418..263G}.  Active Galactic Nuclei (AGNs) activity can impact the host galaxy and the tenuous environment by ionizing or photo-dissociating the gas, heating the halo gas and reducing the rate of cold accretion onto the galaxy, and/or driving fast outflows that eject gas to large galactocentric distances and, thus in powerful AGNs, temporarily or permanently suppressing  star formation (SF) by removing the gas supply from massive galaxies \citep[see][for a review]{2015ARA&A..53...51S}. 

Global changes in the metallicity of galaxies are driven by stellar nucleosynthesis, and thus, it is fundamental to connect it with the  star formation history (SFH) in galaxies. AGNs are  crucial components of theoretical models of galaxy formation and evolution used in regulating the SF in galaxies.  The constant gas ejection or heating and, as a result, SF suppression is expected in massive galaxies ($\sim 10^{10}$ $\rm M_{\odot}$) because of AGN nuclear emission. Although this mechanism is an important component of numerical simulations of galaxy formation, observational studies have not yet conclusively supported it. The negative AGN feedback from observational study is required for many findings to be reproduced by theoretical models, such as the substantial suppression of SF in the most massive galaxies \citep[e.g.][]{springel05b,hopkins10,2020MNRAS.498L..66B,2020ApJ...904...83S}.
AGN-driven winds, shock compression or gas accretion may also cause SF in the host galaxy, a phenomenon known as positive  feedback \citep[e.g.][]{elbaz09,2013ApJ...772..112S, 2017Natur.544..202M, 2017A&A...608A..98S,2019MNRAS.485.3409G,koss2021,2021ApJS..257...64K}.  However, some studies, particularly for moderate-luminosity AGNs, claim that star formation is independent of AGN activity \citep[e.g.][]{2015MNRAS.453..591S,2019ApJ...872..168S}. From the foregoing, it is safe to posit that galaxy-scale warm-ionized  and cold molecular outflows, which are powered by actively accreting  SMBHs and traced by rest-frame optical \citep[][]{1981ApJ...247..403H} and carbon
monoxide \citep[CO;][]{2014A&A...562A..21C} emission
lines  respectively, can be associated with both the suppression and triggering of star formation.  However, a thorough understanding of the interactions between AGN accretion and  star formation processes of the host galaxy  is required to fully understand galaxy formation and evolution.

The elemental abundance and metallicity ($Z$)\footnote{The gas-phase metallicity is usually calculated as the oxygen abundance relative to hydrogen, and defined in units of [12 + log(O/H)]. Oxygen is used to define the overall
gas-phase metallicity because it presents prominent temperature-sensitive collisionally excited lines (e.g. $\rm O^{+}$: [O\ii]$\lambda3727$ and $\rm O^{2+}$: [O\iii]$\lambda4363$, [O\iii]$\lambda5007$) in the optical spectrum of gaseous nebulae. For brevity, oxygen abundance and metallicity are used interchangeably in this work.}  estimations  in the gas phase of AGNs are essential
in the study of the chemical evolution of  galaxies. The gas phase  metallicity can be 
derived in both star-forming regions (SFs, i.e. H{\ii} regions and star-forming galaxies) and AGNs through direct
estimation of the electron temperature, usually known as $T_{\rm e}$-method\footnote{For a review of the $T_{\rm e}$-method for SFs see \citet{2017PASP..129h2001P}, \citet{2017PASP..129d3001P} and AGNs see \citet{2020MNRAS.496.3209D}.} and strong-line methods\footnote{For a review of strong-line methods for SFs and AGNs see \citet{2010arXiv1004.5251L}
and \citet{2020MNRAS.492..468D}, respectively.}.  It is worthwhile to note that SFs are  ionized by massive stars (e.g. O- or early B-type stars), while AGNs have harder radiation fields as their ionization sources. Therefore, applying SFs metallicity formalism to AGNs will yield systematically biased metallicities \citep[e.g.][]{2020MNRAS.496.3209D}.   There is consensus that the  $T_{\rm e}$-method 
is the most reliable approach to estimate metallicity
and elemental abundances in  SFs \citep{2003A&A...399.1003P,2017MNRAS.467.3759T}. Recently,
\citet{2020MNRAS.496.3209D} proposed a new methodology of
 the $T_{\rm e}$-method for AGNs which produces reliable O/H abundances
 lower ($\lesssim \: 0.2$ dex) than those derived from detailed photoionization models. For the first time, \citet{2021MNRAS.508..371A}, \citet{2021MNRAS.508.3023M} and
 \citet{2022MNRAS.514.5506D}, motivated by this new methodology
 derived the neon, argon and helium abundances, respectively,  in 164 local Seyfert 2 nuclei ($z\lesssim0.25$). 
The  $T_{\rm e}$-method requires measurements of auroral lines (e.g.  [O\iii]$\lambda$4363, [N\ii]$\lambda$5755, [S\iii]$\lambda$6312) which are weak ($\sim 100$ times weaker than H$\beta$) and
detected only in objects with high ionization and/or low metallicity \citep[e.g.][]{van1998spectroscopy, 2007MNRAS.382..251D, 2008A&A...482...59D}.  Thus, it is impossible to derive a direct measurement of abundance from the emission lines if the critical emission-line diagnostics for electron temperature and, to some extent,  electron density are unavailable. To circumvent this
problem, \citet{pagel1979composition}, following the original
idea by \citet{1976ApJ...209..748J}, proposed a calibration between
strong-emission lines, in the case of [O\iii]($\lambda4959+\lambda$5007)/H$\beta$, and the O/H abundance. Thereafter this pioneering work, several authors
have proposed calibrations for SFs \citep[e.g.][and references therein]{ mar13,pil16, cur17, 2019ApJ...872..145J} and for AGNs \citep[e.g.][among others]{sb98, dors2014, 2019MNRAS.486.5853D, castro2017, 2020MNRAS.492.5675C, dors2021}. 

Therefore, by comparing AGN luminosity ($L_{\rm AGN}$) and the metallicity, it is possible to study the origin and evolution of SMBHs and  their host galaxies \citep[e.g.][]{2010ApJ...719L.148W}. Previous studies have shown a distinct relation between the metallicity from the narrow-line region ($Z_{\rm NLR}$) and the $L_{\rm AGN}$ \citep[e.g.][]{2004ApJ...614..558N,2006A&A...447..863N,2009A&A...503..721M}. The ionized gas mass of the NLR is $M_{\rm NLR} \sim 10^{5-7}\, {\rm M_\odot}$, which is far greater than that of the broad-line region (BLR) i.e. $M_{\rm BLR} \sim 10^{2-4}\, {\rm M_\odot}$ \citep[e.g.][]{ost1989,2003ApJ...582..590B}, suggesting that the $L_{\rm AGN}$-$Z_{\rm NLR}$ relation is an indication that AGN gas metallicity traces the gas enrichment in its host galaxy. Since the BLR has a very small radius \citep[e.g. $R_{\rm BLR} < 1\,{\rm pc}$;][]{2000ApJ...533..631K,2006A&A...459...55B,2006ApJ...639...46S} because of  its  proximity  to the central source as compared to the NLR   \citep[e.g. $R_{\rm NLR} \sim 10^{1-4}\,{\rm pc}$;][]{2006A&A...456..953B} in galactic nuclei, the $L_{\rm AGN}$-$Z_{\rm NLR}$ relation will be a better tracer of the evolution of the host galaxy than the  metallicity from the BLR ($L_{\rm AGN}$-$Z_{\rm BLR}$ relation).  Also, $Z_{\rm NLR}$ may be uniquely promising and better suited as a proxy for the properties of the host galaxy since the spatial extent of the NLR region is larger than the BLR in low- and high-$z$ as well as high $L_{\rm AGN}$, as observed by the strong correlation between the size of the NLR and the $L_{\rm AGN}$ of the optical [O\iii]$\lambda5007$ emission line \citep[e.g.][]{2018MNRAS.477.4615D,2018MNRAS.480.2302S,2019MNRAS.489..855C}.  Moreover, $Z_{\rm N LR}$ values have been found to be either solar  or near the solar metallicity ($Z_\odot$) i.e.  $0.2 \lesssim \: (Z_{\rm NLR}/Z_{\odot})  \: \lesssim \: 1$ \citep[e.g.][]{2006A&A...447..863N,2019MNRAS.486.5853D}, hence
$Z_{\rm BLR}$ estimates  at all redshifts are higher by a factor of 2-15 times $Z_{\rm N LR}$ \citep[e.g.][]{1993ApJ...418...11H,2003ApJ...583..649B, 2004AJ....128..561B, 2019MNRAS.484.2575T, 2022A&A...667A.105G}, which indicates that the $Z_{\rm BLR}$ is not representative of the metallicity of the host galaxy, which is better traced by the $Z_{\rm NLR}$.  Although the BLRs are located in objects with high mass, where high metallicity values are expected \citep[e.g.][]{2018MNRAS.480..345X}, the  higher $Z_{\rm BLR}$ values which are indirectly inferred from metallicity-sensitive broad emission-line flux ratios seem to be unreal \citep[e.g.][ consideration for changes in the physical conditions of the emitting gas before metallicity estimation]{2021MNRAS.505.3247T} as compared to direct estimates of the $Z_{\rm NLR}$ values. However, most of the $Z_{\rm NLR}$ studies have been based on small samples of objects and/or  with no  $L_{\rm AGN}$-$Z_{\rm NLR}$ relation studies \citep[see][for example]{2020MNRAS.492..468D}.

The stellar mass-metallicity relation \citep[MZR; e.g.][]{2004ApJ...613..898T,2022MNRAS.514.2298B} indicates that the metallicities of galaxies increase with increasing stellar masses, while the fundamental metallicity relation  \citep[FMR; e.g.][]{2013ApJ...772..119L,2022A&A...663A.162P} suggests that, for a given stellar mass, galaxies with higher star formation rates (SFRs) tend to have lower metallicities. 
 For the first time,  \citet{2018A&A...616L...4M}  found  a  correlation between $Z_{\rm NLR}$ and the host galaxy mass ($M_\star$-$Z_{\rm NLR}$ relation) for objects  at $z\sim3$.   A comparison between  observed ultraviolet emission-line flux ratios and photoionization model predictions by these authors showed that $Z_{\rm NLR}$ increases by  $\sim$ 0.7 dex  as $\log(M_\star/ \rm M_{\odot})$ 
increases from $\sim 10$ to $\sim 12$. This relation was also confirmed by \citet{2019MNRAS.486.5853D} for type 2 AGNs in a wider redshift range ($1.6 \: < \: z \: < 3.8$) following a similar methodology by \citet{2018A&A...616L...4M}. Both studies indicate that AGNs and their host galaxies have  similar metal enrichment and the $M_\star$-$Z_{\rm NLR}$ relation seems to complement similar estimates from star-forming  galaxies \citep[see][]{maiolino08} for $z\sim 3.5$ towards higher masses.
 
On the other hand, for the local Universe, a different scenario is found. For instance, \citet{2019ApJ...874..100T}  used the Bayesian code ({\sc NebulaBayes}) which  is based on  photoionization model fitting of several optical emission lines and found that the $Z_{\rm NLR}$ increases by $\rm \Delta (O/H) \sim 0.1$ dex as a function of the $M_\star$ over the range $10.1\lesssim \log(M_\star/ \rm M_{\odot})\lesssim11.3$. However, this $Z_{\rm NLR}$ increase
is  lower than the uncertainty produced by the strong-line methods \citep[e.g.][]{sb98, 2002MNRAS.330...69D})
or by the $T_{\rm e}$-method  \citep[e.g.][]{2020ApJ...893...96B, 2022MNRAS.514.5506D}. Moreover, \citet{2020MNRAS.492..468D} applied  all the  methods available in the literature for deriving  $Z_{\rm NLR}$ using  spectroscopic data from SDSS-DR7 \citep{york00,2009ApJS..182..543A} but  could not confirm the  $M_\star$-$Z_{\rm NLR}$   relation for local objects ($z \: \lesssim \: 0.4)$.

The  aforementioned MZR studies suggest some discrepancies between the results from various studies and raise the question of whether there are differences in the $L_{\rm AGN}$-$Z_{\rm NLR}$ and MZR relations between the global parameters of low or/and high redshift(s) in AGNs, or if they are driven by small samples or due to the effects of the various selection criteria and methods have on the $Z$ of galaxies. In view of this, we apply  the strong-line methods to analyse the $L_{\rm X}$-$Z_{\rm NLR}$ relation using optical and X-ray data from a large sample of  X-ray selected AGNs in this current study. 
In principle, the X-ray luminosity  can yield more reliable results than the stellar mass because X-ray surveys are practically efficient for selecting AGNs since X-ray emissions are generated from the nuclear components with relatively clean signals which are less affected by obscuration and contamination from non-nuclear emissions.   
 Hitherto, almost all the $L_{\rm AGN}$-$Z_{\rm NLR}$ and $L_{\rm AGN}$-$Z_{\rm BLR}$ relations in the literature have been obtained based on either optical or UV selection criteria. However, the hard  ($\gtrsim10$ keV) X-ray luminosity ($L_{\rm X}$) of the \textit{Swift}-BAT AGN Spectroscopic Survey \citep[BASS; see][for details]{Ricci2017b} which have been relied on in this work, measures direct emission from the AGN  which is unaffected by dust or contamination from SF, and is much less sensitive to obscuration in the line-of-sight as compared to  soft X-ray or optical wavelengths, allowing a selection based  on only  the central engine properties.  Compact, nuclear, and luminous X-ray emission is certainly a sign of an AGN  due to the compact and dense plasma required to produce X-ray emission since  at low-z  an X-ray luminosity e.g. $L_X > 10^{42}$ $\rm\,erg\,s^{-1}$ cannot be produced by anything else other than an accreting SMBH. Therefore, we  compare
the metallicity derived from optical emission lines  of these hard X-ray selected AGNs with their  X-ray properties to better understand the AGN $L_{\rm X}$-$Z$  relation.
This paper is organized as follows. We
describe the observational data in \S~\ref{data}. 
Details to the calculations of the total oxygen abundances via  the strong-line methods have been outlined in \S~\ref{meth}. The results and discussions are presented in \S~\ref{resdisc}.  Finally, we summarize our findings in \S~\ref{conc}.  Throughout this paper, we adopt a spatially flat $\Lambda$CDM cosmology with the parameters: $\Omega_{\rm M}$ = 0.3, $\Omega_\Lambda$ = 0.7 and $H_0 = {\rm 70\, km\,s^{-1}\,Mpc^{-1}}$.

\section{Observational Data and Sample Description}
\label{data}

\begin{figure}
\includegraphics[width=1.\columnwidth]{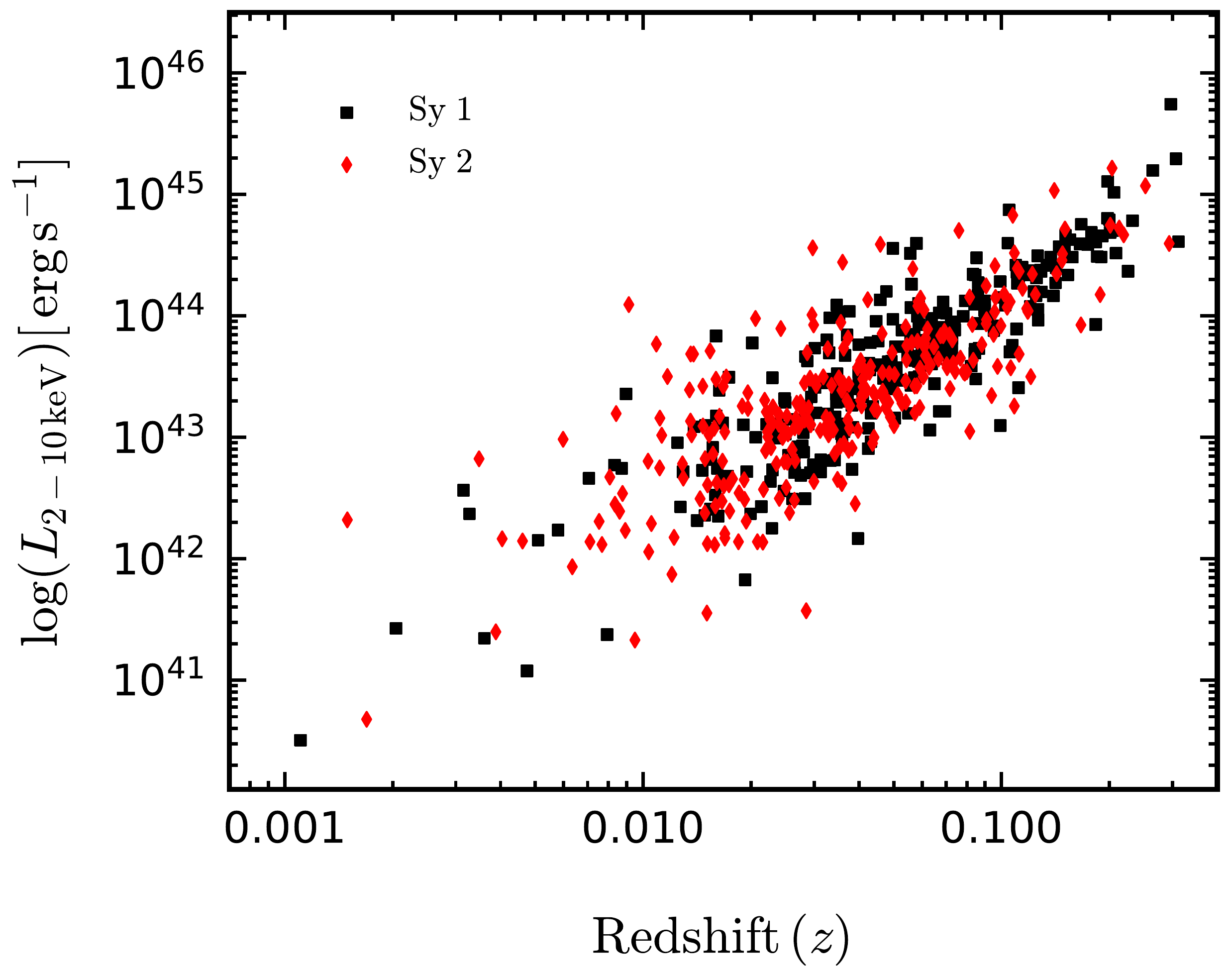}
\caption{Redshift distribution of the rest-frame hard X-ray luminosity based on the 2-10 keV emissions of the BAT AGNs. Points with different colours represent the observed 2-10 keV  luminosities  for Sy 1 and Sy 2 of the sources in our sample with spectroscopic redshifts, as indicated.}
\label{Lz}
\end{figure}

To calculate the oxygen abundances for the narrow line region of a sample of Seyfert  nuclei, optical spectroscopic data taken from the literature are considered.  Therefore, we selected AGN emission line intensities from the BASS DR2 presented by \citep{2022ApJS..261....4O} for the O/H estimations.  It is worth mentioning that these authors have taken special care to remove the broad components from the allowed emission lines as well as removing possible outflow asymmetries effects from the emission line fluxes \citep[for details see][]{2022ApJS..261....4O}. In order to compare these abundances with AGN properties  (e.g. $L_{\rm X}$, Eddington ratios) we  cross-matched this sample with the BASS DR2, which provides  X-ray observed and intrinsic luminosities \citep{Ricci2017b}, resulting in  743 common sources, more details are given below (see \S~\ref{swift}).  The luminosity values for the AGN sample are those measured in the two  X-ray bands, 2-10 keV and 14-195 keV, and the metallicity estimated are calculated through optical emission lines. Figure~\ref{Lz} shows a plot of the selected X-ray luminosities in the Swift-BAT AGN sample together with their corresponding redshifts.
Points with different colours represent the observed 2-10 keV luminosities for all the sources in our sample with spectroscopic redshifts.

\subsection{The 70-month \textit{Swift}-BAT Catalog and the emission line fluxes}
\label{swift}

The Burst Alert Telescope (BAT) instrument onboard the \textit{Swift} satellite \citep{Gehrels2004}, which is undertaking an all-sky survey in the ultra-hard X-ray band ($14-195$ keV),  has identified 1210 objects \citep{Baumgartner2013}, of which 858 are classified as AGNs \citep{2022ApJS..261....2K} based on their cross-correlations with objects in the medium and soft energy X-ray bands.

The optical emission line fluxes (3\,200-10\,000 {\AA}) of 743 sources presented by \cite{2022ApJS..261....4O} in the BASS DR2 were considered in this work.  A complete summary of the instrument setups and observing parameters have been provided by  \citet{2022ApJS..261....2K}.
The BASS DR2 used targeted observations with the Palomar Double Spectrograph (DBSP), which is attached to the Hale 200-inch telescope (36.4\,\%, 271/743), the X-Shooter spectrograph  mounted on the European Southern Observatory’s Very Large Telescope (ESO-VLT) \citep[22.7\,\%, 169/743;][]{vernet11},  the Boller \& Chivens (B \&\ C) spectrograph mounted on the 2.5 m Ir\'{e}n\'{e}e du Pont telescope at the Las Campanas Observatory (5.5\,\%, 41/743),  observations from the Goodman spectrograph \citep[4.3\,\%, 32/743;][]{cemens04} on the Southern Astrophysical Research (SOAR) telescope, and  the low-resolution imaging spectrometer \citep[LRIS: 0.7\,\%, 5/743;][]{oke95} on the Keck telescope. Spectra from publicly available surveys, such as SDSS Data Release 15 \citep[15.9\,\%, 118/743;][]{aguado19} and, 71 AGN spectra \citep[see][]{koss17} that were not in the DR2  were also used.

Due to its selection from the hard X$-$ray band (14--195 keV), the BASS sample is almost insensitive to obscuration up to Compton-thick levels  \citep[$N_{\rm H}$ > 10$^{24}$ cm$^{-2}$,][]{2015ApJ...815L..13R,2016ApJ...825...85K}.
  \citet{koss17,2022ApJS..261....2K} and  \citet{2022ApJS..261....4O} have given a detailed overview of the optical spectroscopic data. Additionally, extended multi-wavelength campaigns from near-infrared (NIR) to soft X-ray wavelengths have made it possible to further characterize the BASS sample, for the scaling between global galaxy properties such as the correlations between X-ray and optical obscuration, NIR lines, X-ray photon index, absorption and coronal properties, AGN mass outflow rates and bolometric luminosity \citep[e.g.][among others]{Berney2015,Lamperti2017,Oh2017,Trakhtenbrot2017, Ricci2017a, 2018MNRAS.480.1819R,rojas2020, liu2020, koss2021,Ananna_2022,kakkad2022, 2022ApJ...938...67R}.

 \subsection{Our final sample}
 \label{sfinal}

We selected Seyfert galaxies from the second catalog and data release of the BASS made available  by \cite{2022ApJS..261....4O}. For a better comparison with previous works with regards to the oxygen abundance, we consider only sources classified as Seyferts based on  diagnostic diagrams (see \S~\ref{diag} below) as suggested by \citet{bpt81}.
We consider the diagrams [O\,{\sc iii}]$\lambda$5007/$\text{H}\beta$ {\it versus} [N\,{\sc ii}]$\lambda$6583/$\text{H}\alpha$, [S\,{\sc ii}]$\lambda\lambda$6717,6731/$\text{H}\alpha$, and [O\,{\sc i}]$\lambda$6300/$\text{H}\alpha$ that require the theoretical demarcation lines used by  \citet{kewley2001, kewley2006} proposed to separate AGNs, transition and star-forming regions and the empirical line proposed by  \citet{schawinski07} to separate Seyferts from LI(N)ERs  \citep[also see][]{2010MNRAS.403.1036C}.   
 
Considering  the selected Seyferts sample, we applied criteria for the  optical data which consists of emission lines observed in the wavelength range of $3500 \: < \: \lambda $({\AA})$ \: < \: 7000$ as well as the $L_{\rm X}$. These criteria are underscored below:
\begin{enumerate}
\item  The objects must have the narrow optical  H$\beta~\lambda4861$, [O\iii]$\lambda$5007, H$\alpha~\lambda6563$, [N\ii]$\lambda6584$    and [S\ii]$\lambda\lambda$6716,6731   emission-line fluxes measured, which gives a total sample consisting of 561 objects. We started with 743 sources, excluded 138 because they were not within Seyfert region (including 56 LI(N)ERs),  44 AGNs with high obscuration and non-detection of H$\beta$ were also discarded.   We define Sy~1 as the sources with Sy~1-1.5 classification as defined  by \cite{2022ApJS..261....4O} based on the strength of [O\iii] to H$\beta$ and  Sy~2 as those with Sy~1.8-2 classification\footnote{This choice was made because objects classified as Sy~1.8  shows  very weak broad H$\beta$ and H$\alpha$  in its optical spectrum while in Sy~1.9 only the weak broad component is detected at H$\alpha$ \citep[e.g][]{1981ApJ...249..462O}, thus their overall spectrum is more close to a type~2 source than to a true type~1.}.
Therefore, the 561 objects comprise of 287 Sy 1s and 274 Sy~2s where the emission  lines were  detected at $\lesssim3\sigma$ significance level for each  target. This selection criterion permits
to estimate the NLRs metallicities based on several strong-line methods \citep[see][for a review of the methods]{2020MNRAS.492..468D}. 

\item  Finally, we selected objects with estimated Eddington ratios, $\mathrm{\lambda_{Edd}} = L_{\rm bol}^{\rm int}$/$L_{\rm Edd}$  to analyse the $L_{\rm X}$-$\lambda_{\rm Edd}$ relation. We obtained a sub-sample for 282 Sy~1s and 264 Sy~2s from \citet{2022ApJS..261....2K} with   Eddington ratio estimates based on the intrinsic X-ray emission and black hole masses using velocity dispersions \citep{2022ApJS..261....6K}  and broad Balmer lines \citep{2016MNRAS.460..187M}. Thus, we note that the analysis involving these quantities are with smaller samples than those  with only $L_{X}$.

We chose only the strong-line methods to determine the abundances because when the presence of the [O\iii]$\lambda$4363   line
is considered in the selection criteria for the sources, the sample will considerably be reduced to 218/561 sources. Moreover, considering  the $T_{\rm e}$-method will lead to selection effect, specifically the high degree of uncertainties in the measurements of the strengths of the weak temperature-sensitive auroral oxygen line [O\iii]$\lambda$4363 given that [O\iii]$\lambda$4363   emission line should
be at rest with respect to the $T_{\rm e}$-method  associated rest-optical nebular emission lines, which could potentially introduce similar margin of high degree of uncertainties in the oxygen abundances derived using the  $T_{\rm e}$-method.

\end{enumerate}

 \subsection{Diagnostic diagrams}
 \label{diag}

\begin{figure*}
\includegraphics[width=2.1\columnwidth]{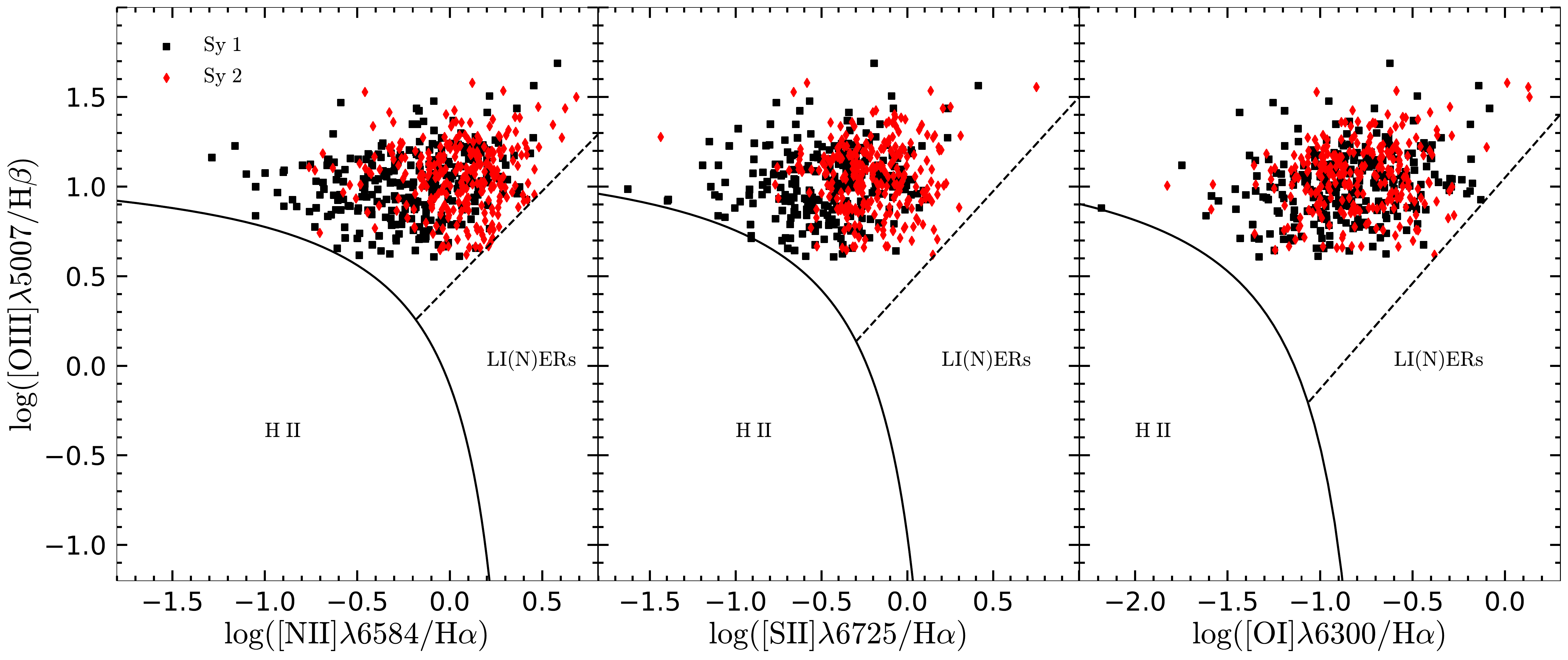}
\caption{Diagnostic diagrams for emission-line ratios of log([O\iii]$\lambda5007$/H$\beta$) versus log([N\ii]$\lambda6584$/H$\alpha$),   log([S\ii]$\lambda$6725/H$\alpha$) 
and log([O\I]$\lambda$6300/H$\alpha$).
[S\ii]$\lambda$6725 represents the
sum of the lines [S\ii]$\lambda$6717 and [S\ii]$\lambda$6731.
Points represent objects of our sample (see \S~\ref{data}).
Solid lines, by  \citet{kewley2001} and represented by Eqs.~\ref{eq1}, \ref{eq2} and \ref{eq3}, separate objects ionized by massive stars   from those ionized by AGN-like objects, as indicated. The dashed lines represent the
criterion proposed by \citet{schawinski07} to  separate AGN-like and low ionisation (nuclear) emission line regions [LI(N)ERs] objects, given by Eq.~\ref{eq4}. Red and black points represent the emission lines from the NLRs of Seyferts~1 and 2 respectively, as indicted.}
\label{fig1}
\end{figure*}

Although the objects considered in the present work have been classified as AGNs by the original authors, for consistency with our previous works we produced an additional
test for a homogeneous sample selection based on the standard Baldwin-Phillips-Terlevich  \citep[BPT;][]{bpt81, 1987ApJS...63..295V} diagrams. Here we adopted  the theoretical criterion which relied on photoionization model results, proposed by \citet{kewley2001,kewley2006}, where emission line objects with 
 \begin {equation}
\label{eq1}
\rm log([O\iii]\lambda5007/H\beta)  >  \frac{0.61}{[log([N\ii]\lambda6584/H\alpha)-0.47]} + 1.19, 
\end {equation}
\begin {equation}
\label{eq2}
\rm log([O\iii]\lambda5007/H\beta)  >  \frac{0.72}{[log([S\ii]\lambda6725/H\alpha)  - 0.32]} + 1.30
\end {equation} 
and
\begin {equation}
\label{eq3}
\rm log([O\iii]\lambda5007/H\beta)  >  \frac{0.73}{[log([O\I]\lambda6300/H\alpha)  + 0.59]} + 1.33
\end {equation}
are classified as AGNs. The  [S\,\ii]$\lambda$6725 line  represents the sum 
of the   [S\,\ii]$\lambda$6717, $\lambda$6731   doublet.  Fig.~\ref{fig1} shows the diagnostic diagrams using the  emission-line ratios of  log([O\iii]$\lambda5007$/H$\beta$) versus log([N\ii]$\lambda6584$/H$\alpha$),   log([S\ii]$\lambda$6725/H$\alpha$) 
and log([O\I]$\lambda$6300/H$\alpha$) for our sample of galaxies. 
The dashed line shown in this figure  represents   the criterion proposed by \citet{schawinski07} to  separate AGN-like and low ionisation (nuclear) emission line regions [LI(N)ERs] objects, given by
\begin{equation} 
\label{eq4}
\rm log([O\iii]\lambda5007/H\beta) \: > \: 0.45+log([N\ii]\lambda6584/H\alpha)\times1.05.
\end{equation}
We notice 
that the objects cover a large range of ionization degree and
metallicity since a wide range of [O\iii]/H$\beta$ 
and [N\ii]/H$\alpha$ are observed   \citep[e.g.][]{feltre2016nuclear, castro2017, 2020MNRAS.496.1262J, 2021ApJ...922..156A}.

\subsection {Reddening correction} 
\label{corr}

Seyfert galaxies can be strongly affected by internal reddening \citep[see][for example]{2016MNRAS.462.3570S,2019MNRAS.483.1722L}. However, the internal extinction can be derived from the ratio of the strongest recombination lines of hydrogen in the optical spectrum i.e.  H$\alpha$/H$\beta$.  For a pure recombination and a temperature of $10^4$ K, this ratio is expected to have the value H$\alpha$/H$\beta$ = 2.86 \citep[Case B recombination;][]{ost1989}. An observed ratio higher than this value can thus be attributed to reddening from dust. The corresponding colour
excess from this ratio is expressed as, 
\begin{equation}
\begin{split}
     E (B-V) ~=~ & \frac{E(\mathrm{H}{\beta} - \mathrm{H}{\alpha}) }{ f_{\lambda} ( \mathrm{H}{\beta}) - f_{\lambda} ( \mathrm{H}{\alpha})}
    \\
    ~=~  &  \frac{2.5}{R_{\lambda} \left[f_{\lambda} ( \mathrm{H}\beta) - f_{\lambda} ( \mathrm{H}\alpha)\right]} \left[\frac{(F_{\mathrm{H}\alpha}/F_{\mathrm{H}\beta})^{\rm obs}}{(F_{\mathrm{H}\alpha}/F_{\mathrm{H}\beta})^{\rm int}}\right],
    \end{split}
\end{equation}
where $R_{\lambda} \left[f_{\lambda} ( \mathrm{H}\beta) - f_{\lambda} ( \mathrm{H}\alpha)\right]$ are the Galactic extinction coefficients,  $f_{\lambda} ( \mathrm{H}{\beta})$ and $f_{\lambda} (\mathrm{H}{\alpha})$ are the reddening curve values at the  $\mathrm{H}{\beta}$ and $\mathrm{H}{\alpha}$ wavelengths, respectively.  The  logarithmic extinction coefficient, c(H$\beta$), is the reddening constant  simply defined as
\begin{equation}
c(\mathrm{H\beta}) = \frac{R_{\lambda} \left[f_{\lambda} ( \mathrm{H}\beta) \right]}{2.5} E (B-V) 
\end{equation}
The final logarithmic extinction coefficient can be thus calculated from the following relation as:
\begin{equation}\label{eq:7}
c(\mathrm{H\beta})  = -\frac{1}{f(\lambda)-f({\rm H}\beta)}\cdot \left[ \log \left(\frac{F(\lambda)}{F({{\rm H}\beta})}\right)- \log \left(\frac{I(\lambda)}{I({{\rm H}\beta})}\right) \right].
\end{equation}
An extinction correction can then be applied to all observed emission line fluxes, normalized to the H$\beta$ flux using
\begin {equation}
\frac {I (\lambda)} {I\mathrm{(H\beta)}} = \frac{F(\lambda)} {F\mathrm{(H\beta)}} \times 10^{-c\mathrm{(H\beta)} [f(\lambda) -f\mathrm{(H\beta)}]},
\end {equation}
where $I(\lambda)$ is the intensity (reddening corrected) of the emission line at a given wavelength $\lambda$, $F(\lambda)$ is the observed flux of the emission line, $ f (\lambda)$ is the adopted reddening curve normalized to H$\beta$ and $c$(H$\beta$) is the interstellar extinction coefficient. The $[f( \lambda)  -f\mathrm{(H\beta)}]$ values were compiled from Table 7.1  by \citet{ost1989}. In order to provide  supplementary interpolation data for
further important nebular emission lines
between different  speciﬁc wavelength-dependent extinction factor  $f(\lambda)$ values since not all lines were listed, we have derived the following reddening function\footnote{The wavelength dependence in the optical domain, $f(\lambda)$ is the reddening value for the line derived from the curve given by $[f(\lambda) -f\mathrm{(H\beta)}]$, such that $f(\infty) = -1 $ and $f(\text{H}\beta) = 0 $ yields to the curve, considering the rest-frame wavelengths for permitted and forbidden lines  from \citet{1945CoPri..20....1M} and  \citet{1960ApJ...132....1B}, respectively.}:
\begin {equation}
f(\lambda) = 2.967 \lambda^2 - 5.454 \lambda + 1.953,
\end {equation}
with $\lambda$ in units of micrometers within the range $\mathrm{0.35 \: \la \: \lambda (\mu m)  \: \la \: 0.70}$.

 \citet{halpern1982xray}  and  \citet{halpern1983ionization} used photoionization models and found that $I$(H$\alpha$/H$\beta$) is  $\sim3.10$ in AGNs with high and low ionization degree  in comparison to the canonical Case B recombination value of 2.86.
Therefore, $I$(H$\alpha$/H$\beta$) = 2.86  and   3.10 intrinsic ratios are usually considered to be estimations for galaxies dominated by star formation and for galaxies dominated by AGNs, respectively \citep{ferland1983shock,  1982PASP...94..891G, gaskell1984red, gaskell1984theo, 1987ApJS...63..295V, wysota88}. Particularly, in AGNs, there is a large transition zone, or partly ionized region, in which H$^0$ coexists with H$^+$ and free electrons \citep[see][and references therein]{2022MNRAS.514.5506D}. In this zone, collisional excitation is also important in addition to recombination  \citep{ferland1983shock, halpern1983ionization}. The main effect of collisional excitation is to enhance the intensity of H$\alpha$. The higher Balmer lines are less affected because of their large re-excitation energies and smaller excitation cross-sections.   With this in mind, we have corrected the emission-line fluxes for extinction using the Balmer decrement and the \cite{cardelli89} reddening curve for the internal and  Galactic reddening, respectively. We assumed an $R_V = A_V/E(B-V) =3.1$  and an intrinsic H$\alpha$/H$\beta$=3.1. 

Since some measurements for the  emission lines  do not have their uncertainties listed  (H$\beta$: 29/561, H$\alpha$: 1/561, [O\iii]$\lambda$4959: 25/561, [O\iii]$\lambda$5007: 5/561, [N\ii] $\lambda$6548: 63/561, [N\ii] $\lambda$6584: 6/561 and [S\ii]$\lambda\lambda$6716, 6731:  31/561 each), thus, we adopted a typical  error of  $10\,\%$  (see, for instance, \citealt{kraemer1994spectra, 2008MNRAS.383..209H}). These errors were propagated through the estimations of the uncertainties in the derived values of the  metallicity (in order of 0.1 dex). It worth to note that, for a lower redshift limit of $z < 0.04$,  extraction aperture of 1.5-2$\arcsec$ \citep[][]{2022ApJS..261....2K} in radius and the aperture covering fraction of $\sim20\,\%$  are sufficient for avoiding aperture effects on our metallicity estimates \citep[see][]{2005PASP..117..227K,kew08}. Additionally, these authors posited that the derived metallicity can vary by  $\sim$0.14 dex from the value determined when the total galaxy emission is taken into account  for apertures that capture less than $20\,\%$ of the total galaxy emission. However, only the nuclear region abundances are considered in our study, so the aperture effect on our metallicity estimates is insignificant.

%%%%%%%%%%%%%%%%%%%%%%%%%%%%%%%%%%%%%%%%%%%%%%%%%%%%%%%%%%%%%%%%%%%%

\section{Methodology}
\label{meth}

 The main goal of this work is to derive the oxygen abundance relative to hydrogen (O/H) from the NLRs of a sample of Seyfert  nuclei in order to  analyse the  $L_{\rm X}$-$Z_{\rm NLR}$ relation. In view of that,  we adopted  the strong-line methods for optical emission lines to be applied in the studies of the NLR of Seyfert nuclei.
In the following sections we describe to somewhat details of the aforementioned methods  and   the oxygen abundance computed from the data under consideration in this work.

\subsection{AGN calibrations}
\label{abund}

In recent optical surveys such as the Sloan Digital Sky Survey \citep[SDSS;][]{york00}, the [O\ii]$\lambda3727$ line is measured in very few objects.  Moreover, [O\ii]$\lambda3727$ line is more in the bluer part of the spectrum and it is more effectively scattered or affected by interstellar reddening. Furthermore, more than one-half of the current strong-line calibrations derived for AGNs consider the [O\ii]$\lambda3727$ line as data selection criterion \citep[e.g.][]{1994ApJ...429..572S, castro2017,dors2021}. Therefore, we note that  when the presence of the [O\ii]$\lambda3727$ line
is considered in the selection criteria of objects, the sample is considerably reduced.
Hitherto, the only AGN calibrations which solely rely on [N\ii]$\lambda$$\lambda$6548,6584/H$\alpha$ and/or [O\iii]$\lambda$$\lambda$4959,5007/H$\beta$ are the two calibrations proposed  by \citet{sb98} and  \citet{2020MNRAS.492.5675C}. We provide descriptions of these calibrations in the following sections.

\subsubsection{\citet{sb98} calibration}
\label{sb}
\citet{sb98} used grid of photoionization models  by assuming a typical AGN continuum, which were built with the {\sc Cloudy} code \citep{2017RMxAA..53..385F} and, for the first time proposed two  AGN theoretical calibrations between  the NLRs emission line ratios
[N\ii]$\lambda\lambda$6548,6584/H$\alpha$, [O\ii]$\lambda3727$/[O\iii]$\lambda\lambda4959,5007$
as well as  [O\iii]$\lambda\lambda4959,5007$/H$\beta$ and the metallicity (traced by the oxygen abundance).   The models were constructed assuming a gas density value of $300\, \rm cm^{-3}$, and the calibrations were fitted within $\sim0.05$ dex of the models. These calibrations are valid for the range of $\rm 8.4 \: \lesssim \: [12+\log(O/H)] \: \lesssim \:  9.4$ and the oxygen abundances obtained from these calibrations  differ systematically by only $\sim0.1$ dex \citep{sb98,2020MNRAS.492..468D}.

In this work we used only one calibration proposed by \citet[][hereafter  \textcolor{blue}{SB98f1}]{sb98}
 because  using the [O\ii]$\lambda3727$ line as a selection criterion 
will  reduce the number of sources to 396/561.  The  \textcolor{blue}{SB98f1}
 calibration is defined by:
 \begin{eqnarray}
       \begin{array}{l@{}l@{}l}
\rm 12+(O/H)\, & =\, &  8.34  + (0.212 \, x) - (0.012 \,  x^{2}) - (0.002 \,  y)  \\  
         & + & (0.007 \, xy) - (0.002  \, x^{2}y) +(6.52 \times 10^{-4} \, y^{2}) \\  
         & + & (2.27 \times 10^{-4} \, xy^{2}) + (8.87 \times 10^{-5} \, x^{2}y^{2}),   \\
     \end{array}
\label{cal_SB}
\end{eqnarray}
where $x$ = [N\ii]$\lambda$$\lambda$6548,6584/H$\alpha$ and 
$y$ = [O\iii]$\lambda$$\lambda$4959,5007/H$\beta$.

It is important to apply the correction proposed by these authors to  Eq.~\ref{cal_SB}  in order to account for the deviations from the assumed  gas density, therefore, the final calibration is given by 
\begin{equation}
\label{cal_SB2}
{\rm \log(O/H)_{SB98f1}=[\log(O/H)}]-\left[0.1 \: \times \: \log \frac{N_{\rm e}({\rm cm^{-3}})}{300 \: ({\rm cm^{-3}})}\right],
\end{equation}
 where $N_{\rm e}$ is the electron density and Eq.~\ref{cal_SB2} is valid for $10^2\lesssim N_{\rm e} \rm (cm^{-3})\lesssim 10^4$.

\subsubsection{Electron density }
\label{eden}

The electron density ($N_{\rm e}$), for each object, was calculated from the 
[S\ii]$\lambda\lambda$6717,6731 doublet flux ratio, using the  1.1.16 version of {\sc PyNeb} code \citep{2015A&A...573A..42L} and assuming a constant electron temperature $T_{\rm e} = 10^{4}$ K, typical
for photoionized gas in the NLRs.  
The {\sc PyNeb} code  allows an interactive procedure where  $N_{\rm e}$ is computed  from the required forbidden lines  depending on specific value of $T_{\rm e}$.  Moreover, it is worth mentioning that   [S\ii]$\lambda\lambda$6717,6731  can only be applied to $0.45 \lesssim R_{\rm [S\ii]} \lesssim 1.45$  (where $R_{\rm [S\ii]} = F_{\lambda6717}/F_{\lambda6731}$, see \citealt{ost06,san16a}),  which  translates into electron density values in the range of  $5\,000\: \gtrsim N_{\rm e}(\rm cm^{-3}) \: \gtrsim \: 50$, respectively. 
 Therefore, for the objects with emission line ratios outside these theoretical constraints, we assumed the electron density to be $N_{\rm e} =2\,000 \: \rm cm^{-3}$ if $R_{\rm [S\ii]}\lesssim0.45$, in order to avoid collisional de-excitation\footnote{ The critical density values for [S\ii]$\lambda6716$ and [S\ii]$\lambda6731$ are $10^{3.2}$ and
$10^{3.6}$ cm$^{-3}$, respectively, see \citet{2012MNRAS.427.1266V}.} effect and $N_{\rm e} =100 \: \rm cm^{-3}$ if $R_{\rm [S\ii]}\gtrsim1.45$, which is the minimum electron density value for the calibration by \textcolor{blue}{SB98f1}. From the 561 objects considered for the strong-line method calibrations\footnote{It is worthwhile to note that only the calibration by  \textcolor{blue}{SB98f1}
  considers electron density in the formalism given by Eq.~\ref{cal_SB2}.}, there are  5  with $R_{\rm [S\ii]}\lesssim0.45$ and 62 with $R_{\rm [S\ii]}\gtrsim1.45$ i.e. $\sim12\,\%$ of the total sample. While some of the $R_{\rm [S\ii]}$ values are extreme  and  beyond the low-density regime other emission lines from the sources  are not aberrant.
  
\begin{figure}
\includegraphics[width=1.\columnwidth]{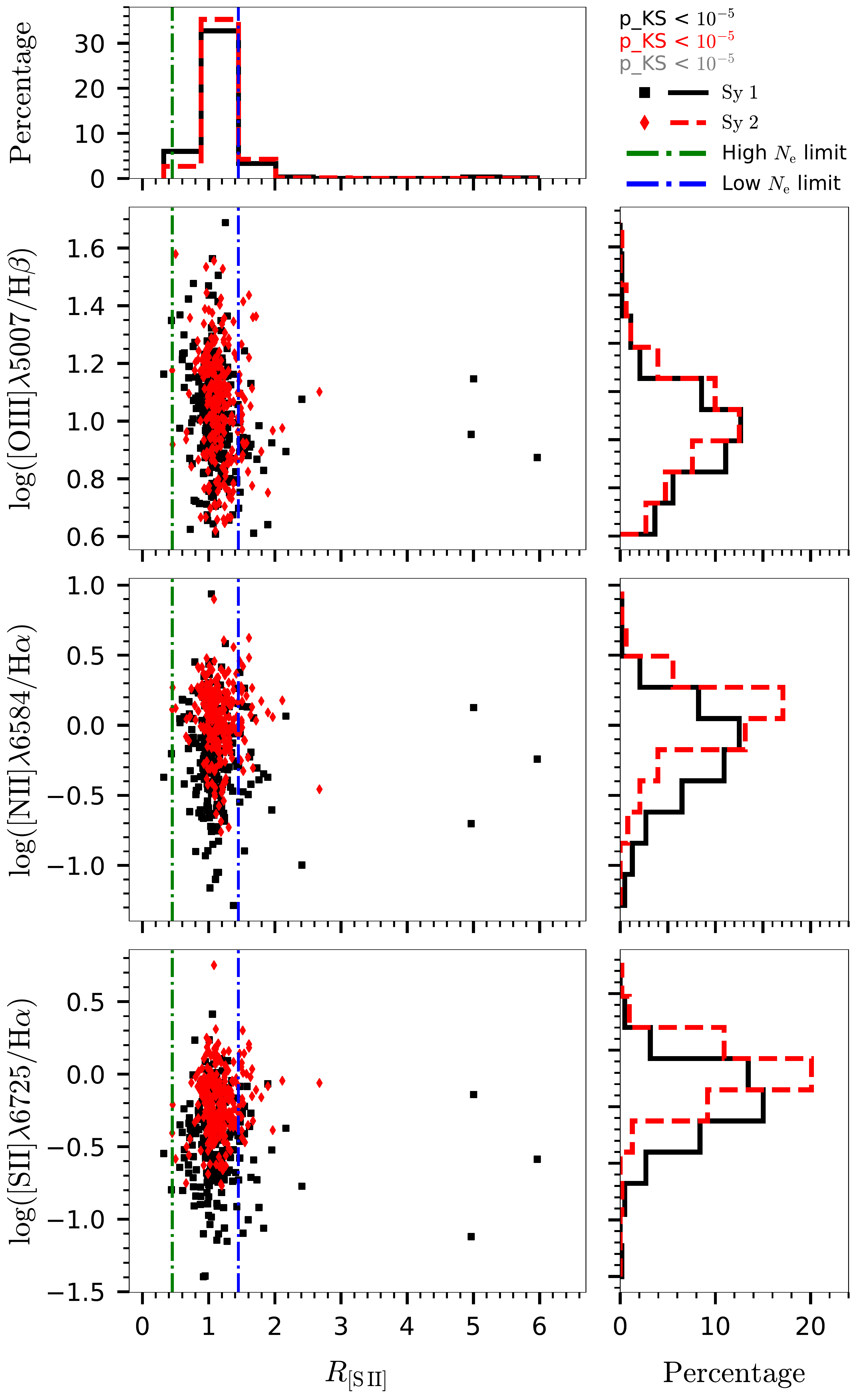}
\caption{ Bottom panel: the distribution of logarithmic of the diagnostic
[S\ii]($\lambda\lambda$6717 + 31)/H$\alpha$ versus the  [S\ii]$\lambda\lambda$6717, 6731 doublet flux ratio ($R_{\rm [S\ii]}$) estimated from   287 Sy 1s and 274 Sy~2s. The green and blue dashdot lines represent the minimum and maximum  flux ratios corresponding to the theoretical high- and low-density limits (see \S~\ref{eden}),
respectively.  Histograms of
log([S\ii]$\lambda6725$/H$\alpha$) and $R_{\rm [S\ii]}$  are plotted in
the right panel. Middle panel: same as bottom panel but for  log([N\ii]$\lambda6584$/H$\alpha$) versus  $R_{\rm [S\ii]}$. Top panel:  same as bottom panel but for log([O\iii]$\lambda5007$/H$\beta$) versus  $R_{\rm [S\ii]}$. Black and red points with corresponding coloured {\it p-values} denote Sy~1 and Sy~2 sources respectively, while the  grey coloured  {\it p-value}  is obtained from  the combined data-set.}
\label{elec}
\end{figure}

  In order to verify if there is a correlation between the electron density and the intensity of the strong emission line ratios  
  involved in the metallicity calculations, in  Fig.~\ref{elec}, we plotted the logarithm of [O\iii]$\lambda5007$/H$\beta$ (top panel), [N\ii]$\lambda6584$/H$\alpha$ (middle panel) and [S\ii]$\lambda6725$/H$\alpha$ (bottom panel) versus  $R_{\rm [S\ii]}$. Also in Fig.~\ref{elec}, we have presented histograms showing the distributions of values from the emission lines considered, where Sy~1 and Sy~2  are indicated by black and red colours, respectively. 
  We applied the two-sample Kolmogorov--Smirnov (KS) statistical test to the frequency distributions of the line ratios in
  Fig.~\ref{elec} for the Sy~1 and Sy~2 nuclei of our sample. 
The KS tests show that the probability of any two distributions being taken from the same parent distribution is lower than $10^{-5}$, which suggests that the difference in the 
 distributions of the density sensitive $R_{\rm [S\ii]}$ and the diagnostic emission line ratios  between Sy~1 and Sy~2 is statistically significant. However, this difference does not necessarily translate into a similar  significant difference in the metallicity distributions of Sy~1 and Sy~2 due to the electron density, in fact, it signifies that the electron density from $R_{\rm [S\ii]}$ has no effect on any discrepancy which may arise from the metallicities between Sy~1 and Sy~2. The KS test values (black, red and grey coloured {\it p-values} are for  Sy~1, Sy~2 and the combined data-set, respectively) from all the diagnostics, shown in Fig.~\ref{elec}, are lower  than  $10^{-5}$, therefore, the electron density has no significant effect on the metallicity values  derived.
  
We  derived electron density values from Sy~1s and Sy~2s  in the range $ 100\lesssim N_{\rm e} \rm (cm^{-3})\lesssim 2\,000$, with  median values of $\rm \sim 480 \: cm^{-3}$  and  $\rm \sim 390 \: cm^{-3}$, respectively  (see  Fig.~\ref{SLDenSy12}).

\subsubsection{\citet{2020MNRAS.492.5675C} calibration}

\citet[][hereafter \textcolor{blue}{C20}]{2020MNRAS.492.5675C} presented a comparison using  photoionization model predictions
 built with the {\sc Cloudy} code \citep{2017RMxAA..53..385F},  considering a wide range of nebular parameters, and a  [O\iii]$\lambda5007$/[O\ii]$\lambda3727$ versus [N\ii]$\lambda$6584/H$\alpha$ diagram obtained from observational data of 
 463 Seyfert~2 nuclei ($z\lesssim0.4$).
 From this comparison, these authors derived a semi-empirical calibration 
 between the $N2$=log([N\ii]$\lambda$6584/H$\alpha$) line ratio and the metallicity $Z_{\rm NLR}$, given 
 by  
\begin{equation}
(Z_{\rm NLR}/{\rm Z_{\odot}}) = (4.01\pm0.08)^{N2} - 0.07\pm0.01, 
\label{c20}
\end{equation}
which is valid for $0.3 \: \lesssim \: (Z_{\rm NLR}/Z_{\odot}) \:\lesssim  \: 2.0$. 
The metallicity results obtained from Eq.~\ref{c20} can be converted to oxygen abundance via the relation:
\begin{equation}
12+\log({\rm O/H})_{\rm C20}=12+\log[(Z_{\rm NLR}/Z_{\odot}) \: \times \: 10^{\log(\rm O/H)_{\odot}}],    
\end{equation}
where $\log(\rm O/H)_{\odot}=-3.31$ is the solar oxygen abundance value taken from \citet{2001ApJ...556L..63A}.
The $N2$ index has an advantage  over other metallicity indicators such as [N\ii]$\lambda6584$/[O\ii]$\lambda3727$ 
because  it involves emission lines with very close wavelengths:  thus, $N2$
is not strongly affected by dust extinction and uncertainties produced
by flux calibration \citep{mar13,castro2017}.

\section{Results and Discussion}
\label{resdisc}

\subsection{Electron  density effect}

Since the \textcolor{blue}{SB98f1} calibration depends on the electron density, as shown above (see \S~\ref{sb}), it is important to  analyse  the potential impact of the electron density on our metallicity results. 

\begin{figure}
\includegraphics[width=1.\columnwidth]{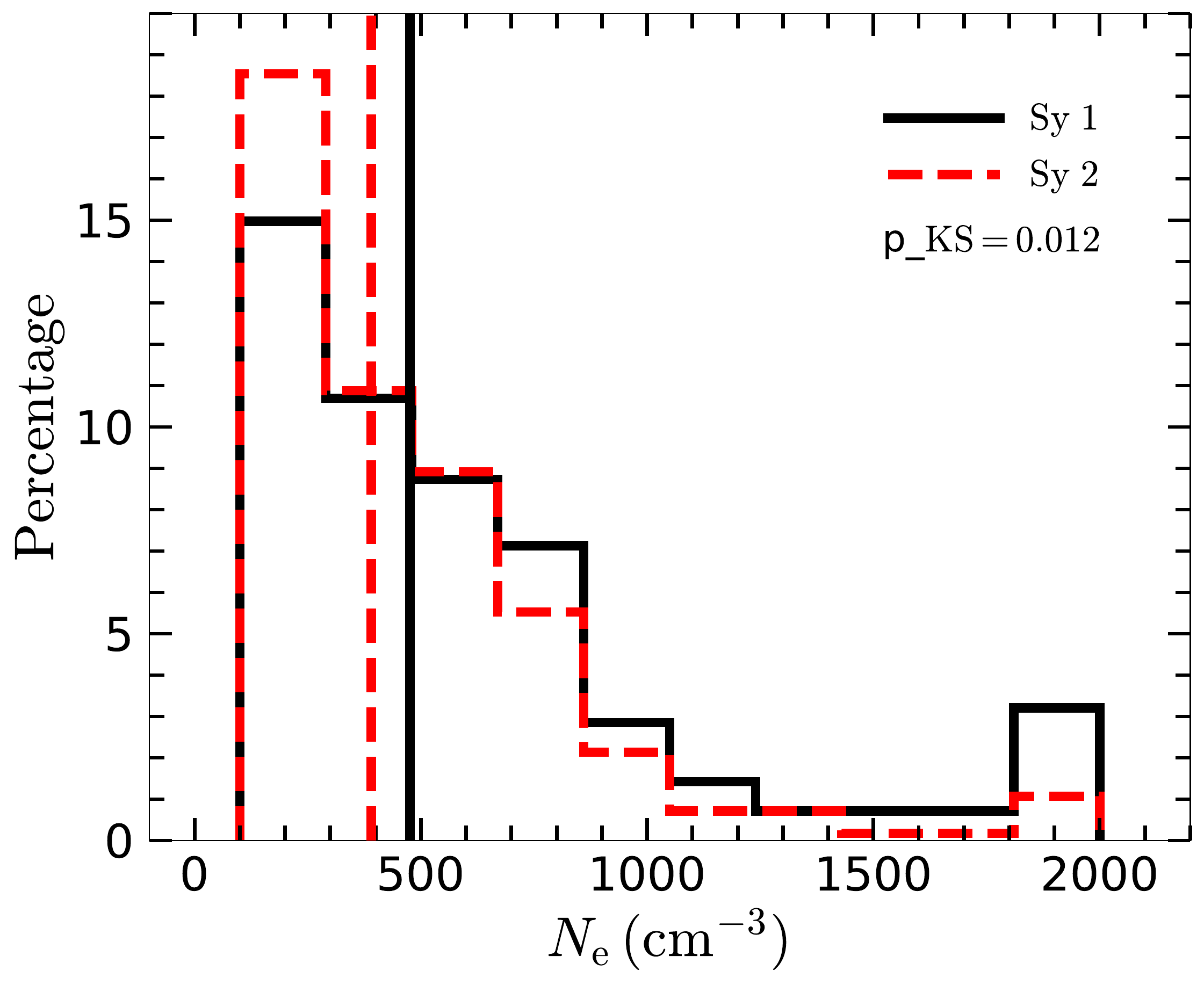}
\caption{ The distribution of electron density  estimated from   287 Sy 1s and 274 Sy~2s  of the [S\,{\sc ii}]$\lambda\lambda$6717,6731 doublet line ratio  of our local samples using {\sc PyNeb} code \citep{2015A&A...573A..42L}. The electron density is uniformly distributed between Sy~1s and Sy~2s except for few outliers.  The median values of $\rm \sim 480 \: cm^{-3}$ and $\rm \sim 390 \: cm^{-3}$ for Sy~1 and  Sy~2 are denoted by the  solid black and dashed red vertical lines, respectively. }
\label{SLDenSy12}
\end{figure}

The observed forbidden lines from the NLRs are indicative of a low density and are very useful to measure several physical parameters of the region, such as temperature and electron density. In order to examine the distributions of the electron density between Sy~1 and Sy~2, we used a two-sample KS test, as shown in Fig.~\ref{SLDenSy12}.  We find that the  significant difference between the samples with the {\it p-value},    p\_KS $\sim0.012$, indicates that Sy~1 and Sy~2 are from different distribution for a confidence level of  95\,\%, although, the same ionization mechanism is responsible for the trends between the electron density distributions of Sy~1 and Sy~2. Our derived aforementioned  electron density values (see \S~\ref{eden}) are more in agreement with the typical densities
in the NLRs of Seyferts  \citep[$N_{\rm e}\lesssim 10^4\, {\rm cm}^{-3}$; e.g.][]{ost06,2008ARA&A..46..475H}.

From the unified model scheme, the  narrow  forbidden and permitted emission lines in Sy~1 and Sy~2 sources come from the NLR, a region well outside the BLR, spanning a few tens of pc to about 1\,kpc \citep[][]{antonucci1993unified}. 
The mere presence of forbidden lines in the NLR indicates that the gas densities are lower than in the BLR. However, the electron density distribution in Fig.~\ref{SLDenSy12} shows that Sy~1 still exhibits slightly higher electron density \citep[also see][]{2013ApJ...779..109P} in comparison to Sy~2  (Sy~1 dominates the density distribution at $N_{\rm e} > 600 \: \rm cm^{-3}$, and vice versa), suggesting that the Sy~1 and Sy~2 are not only a geometrical phenomenon but sources with different physical properties \citep[see][for other examples]{2017MNRAS.464.2139A}. It is noteworthy to emphasize that the $N_{\rm e}$ values from both Sy~1s and Sy~2s are lower than the critical density ([S\,{\sc ii}]$\lambda$6717: $N_{\rm c}\sim 10^{3.2}\, {\rm cm^{-3}}$ and [S\,{\sc ii}]$\lambda$6731: $N_{\rm c}\sim 10^{3.6}\, {\rm cm^{-3}}$, see \citealt{2012MNRAS.427.1266V}) for
the emission lines involved  in the derivation of the electron density in this work, therefore, collisional de-excitation effect has no influence on our abundance estimates \citep[e.g.][]{ost06,2012MNRAS.427.1266V}.  It is worth mentioning that similar trends are found even using  higher ionization potential lines,  such as [Ar\iv]$\lambda$4711, $\lambda$4740 \citep{2012MNRAS.427.1266V,2017MNRAS.471..562C, 2021MNRAS.500.2666C}.

Additionally, the electron density determined from the $R_{\rm [S\ii]}$  is much lower than those obtained using auroral and transauroral lines, as well as ionization parameter based approach \citep[e.g.][]{2020MNRAS.498.4150D}. 
Furthermore, even considering the highest $N_{\rm e}$ value we assumed for $R_{\rm [S\ii]}\lesssim0.45$  ($N_{\rm e}=2\,000\, \rm cm^{-3}$), the O/H correction is $\sim0.1$ dex, which is in order of  the uncertainty of  abundances via direct measurements of the electron temperature \citep[e.g.][]{2003ApJ...591..801K, 2008MNRAS.383..209H} and even lower
than those ($\sim 0.2$ dex) via strong line methods
\citep[e.g.][]{2002MNRAS.330...69D, mar13}.
 Thus, the abundances derived from our sample are only marginally influenced by the electron density.

   The NLR can be explained by high-density gas clouds transfer from the nuclear region to the outer part in the host galaxy through the outflows driven by the AGN as shown by photoionization models \citep[e.g.][]{2018ApJ...867...49W}, implying that the NLR gas density increases toward small radii  \citep[e.g. ][]{2006A&A...456..953B,2006A&A...459...55B}. For instance, spatially resolved observational studies of the NLRs have revealed a profile of electron density along the AGN radius, i.e.
electron densities ranging from $\sim2\,500\, \rm cm^{-3}$ at the central parts to $\sim100\, \rm cm^{-3}$ at the outskirts 
\citep[e.g.][]{2017MNRAS.471..562C, 2018A&A...618A...6K, 2018MNRAS.476.2760F, 2018ApJ...856...46R, 2019A&A...622A.146M}.
Since our spectra were integrated in the central parts of the galaxies, the electron density values obtained in the present work must be considered as mean values for the NLRs. 
The existence of the density effect has already been taken into account in the strong-line calibration by \textcolor{blue}{C20}, while   \textcolor{blue}{SB98f1} has proposed a correction (Eq.~\ref{cal_SB2} ), which is valid for the gas density in the range $10^2\lesssim N_{\rm e} \rm (cm^{-3})\lesssim 10^4$, and would have a maximum impact of  0.1 dex \citep[e.g.][]{dors2021} on the metallicity correction.  While the use of  $R_{\rm [S\ii]}$  is still a subject of ongoing debate \citep[e.g.][]{2019MNRAS.490.5860S, 2020MNRAS.498.4150D, 2020PASP..132c3001N}, it is still better suited to probe the NLR gas density for the proposed correction by  \textcolor{blue}{SB98f1}.
Finally, even electron density values derived from spatially observed AGNs and from emission lines emitted by ions (e.g. $\rm Ar^{3+}$) with  higher ionization potentials than $\rm S^{+}$ support the phenomenon that the NLR is indicative of a low density regime ($N_{\rm e} \: \lesssim \: 10^4 \: \rm cm^{-3}$). Thus,  collisional de-excitation has a minimum effect on the formation of the emission lines we used to estimate the metallicity (e.g.
[O\iii]$\lambda5007$: $N_{\rm c}\sim10^{5.8} \: \rm cm^{-3}$ and [N\ii]$\lambda6584$: $N_{\rm c}\sim10^{4.9} \: \rm cm^{-3}$, see \citealt{2012MNRAS.427.1266V}).

\subsection{Caveats for the Seyfert~1 metallicity }

As widely accepted in the AGNs phenomena \citep[e.g.][]{antonucci1993unified}, two
distinct emission-line regions are expected to exist around the
accretion disk:
i) A BLR, which consists of relatively dense clouds ($N_{\rm e}  \gtrsim  \:10^{5} \rm cm^{-3}$) and highly perturbed gas.
    In this region forbidden emission lines are suppressed by collisional de-excitation while broad permitted lines are emitted with $\rm FWHM \: \gtrsim \: 1000 \, \rm km \: s^{-1}$, and ii) the NLR with low electron density clouds ($N_{\rm e} \: \lesssim 10^4 \: \rm cm^{-3}$)
where the narrow forbidden  and permitted emission lines ($\rm FWHM \: \lesssim \: 1000 \, \rm km \: s^{-1}$) are produced.

Despite these two distinct regions, some studies have shown that, at least in some Sy~1 nuclei, some forbidden lines
can also be emitted near the BLRs or, in other words, the  NLRs present very high electron density values. \citet{2013ApJ...779..109P} used observational data from the International AGN Watch campaign   via the Hubble Space Telescope,  found a variability of the flux
of [O\iii]$\lambda5007$ and  the continuum ($\lambda=5100$ \AA) in the 
well-studied Seyfert 1 NGC\,5548. These authors 
 \citep[see also][]{2016MNRAS.457L..64Z, 2017MNRAS.465.1898S, 2019MNRAS.489.1572L, 2021ApJ...907...76H} showed that the 
[O\iii] emission occurs preferentially 
in a compact NLR
(1–3 pc) with an electron density $\sim10^{5}\, \rm cm^{-3}$, which is close to the critical density for this line \citep[$10^{5.8} \rm \: cm^{-3}$;][]{2012MNRAS.427.1266V}. This scenario precludes any direct electron density estimation in the gas region where most of the [O\iii] is emitted, even using the [Ar\iv] line ratio.

\begin{figure}
\includegraphics[angle=-90,width=1.\columnwidth]{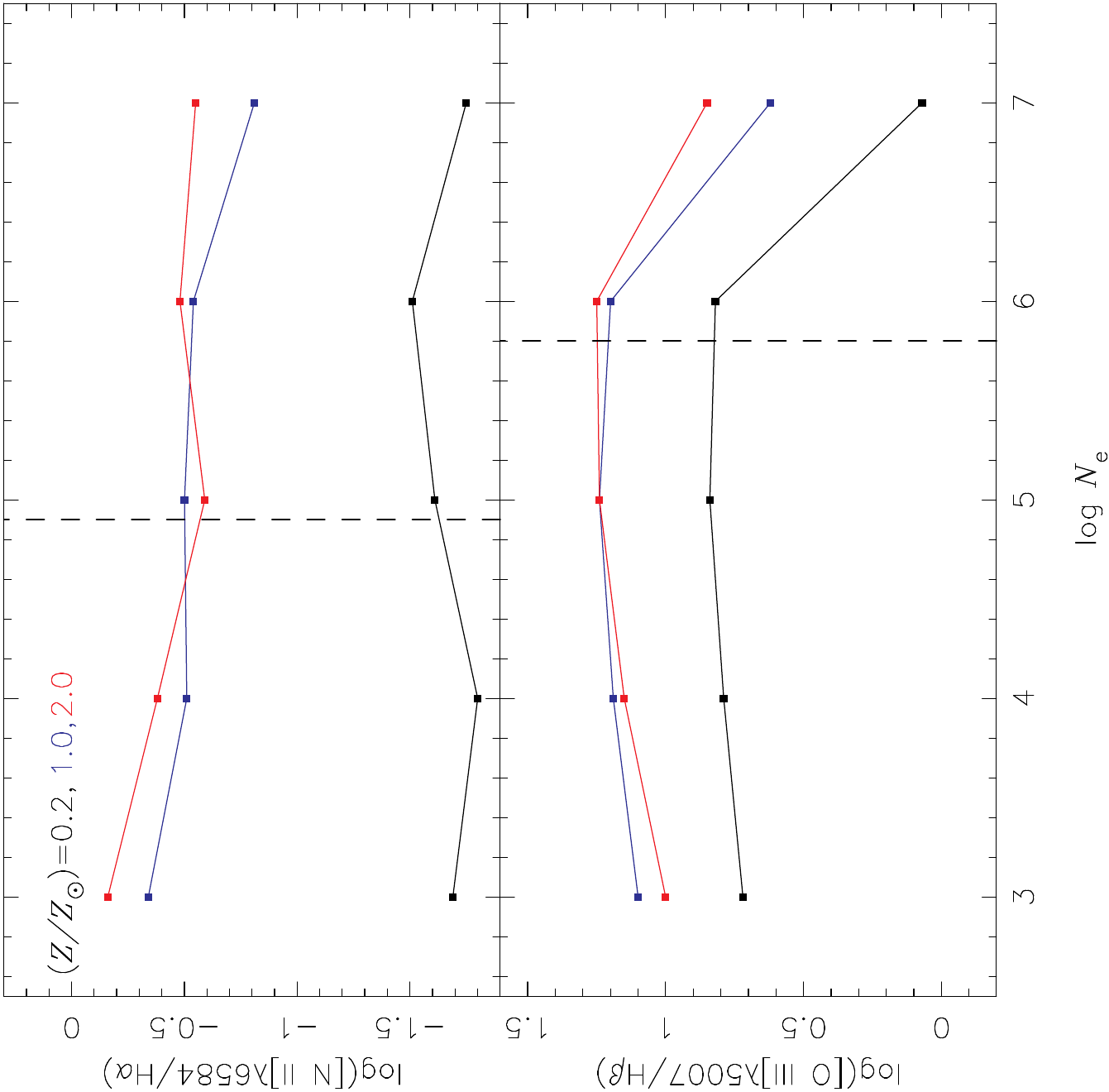}
\caption{ The intensities of 
[O\iii]$\lambda5007$/H$\beta$ and [N\ii]$\lambda6584$/H$\alpha$ emission line ratios versus the logarithm of the electron density predicted by photoionization models assuming   
radial density profiles
$N_{\rm e}\: \propto \: r^{-0.5}$, where $r$ is the distance to the centre of the AGN. Solid lines connect the points representing  model predictions with different metallicities, as indicated. Dashed lines represent the critical density values
of $10^{4.9}$ $\rm cm^{-3}$ and $10^{5.8}$ $\rm cm^{-3}$
for [\ion{N}{ii}]$\lambda6584$ and [\ion{O}{iii}]$\lambda5007$ \citep{2012MNRAS.427.1266V},
respectively.}
\label{fig122ref}
\end{figure}

Since  \textcolor{blue}{SB98f1}    and  \textcolor{blue}{C20} calibrations were obtained based on photoionization models
assuming low   
($\lesssim10^3 \, \rm cm^{-3}$) and constant electron density values along the AGN radius, we can have some biases in the Seyfert~1 oxygen abundance estimates. In order to verify the possible effect of high-density clouds on the strong emission lines used in our metallicity estimates, we performed photoionization model simulations following a similar procedure adopted by \citet{2019MNRAS.486.5853D}. We used  AGN models, which were built from version 17.00  of the spectral synthesis code  \textsc{Cloudy} \citep{2017RMxAA..53..385F}, with  radial density profiles
$N_{\rm e}\: \propto \: r^{-0.5}$, where $r$ is the distance to the center of the AGN, as derived observationally from the Sy~2 galaxy
Mrk 573 by \citet{2018ApJ...856...46R}. We assumed the same nebular parameters used by \textcolor{blue}{C20} but for
metallicity $(Z/\rm Z_{\odot})=0.2, 1.0, 2.0$, an innermost radius of 0.1 pc, the outermost radius where the gas temperature reaches 4\,000 K (standard value in
the {\sc Cloudy} code),  logarithm of the number of ionizing photons emitted by the central source is considered to be $\log[Q(\rm H)]=54$, the
slope of the spectral energy distribution, $\alpha_{ox}=-1.1$ and $\log(N_{\rm e})$ ranging from 3 to 7 with a step of 1.0 dex. In Fig.~\ref{fig122ref}, the predicted intensities of 
[O\iii]$\lambda5007$/H$\beta$, [N\ii]$\lambda6584$/H$\alpha$ emission line ratios versus the
$\log(N_{\rm e})$ are shown. Also in this figure the critical density values for [\ion{N}{ii}]$\lambda6584$ and [\ion{O}{iii}]$\lambda5007$ are indicated.
It can be seen that the variations in electron density, for density
values lower than the critical density ($N_{\rm c}$), have practically insignificant effects on the emission line ratios considered for our metallicity estimations, thus validating the employed metallicity methodology for the Sy~1 sources.

It is worth mentioning that the application of strong emission line calibrations to the derivation of metallicity must be considered in a statistical analysis considering a large object sample, as performed in this work. Obviously, estimates based on detailed photoionization
models which take into account a wide spectral wavelength range, as performed by \citet{kraemer1994spectra} for the
two Sy~2 galaxies NGC\,7674 and IZw\,92, produce more precise metallicity values, but such results are available from a few objects.

\subsection{Gas phase abundances}

\begin{figure*}
\includegraphics[width=1.\columnwidth]{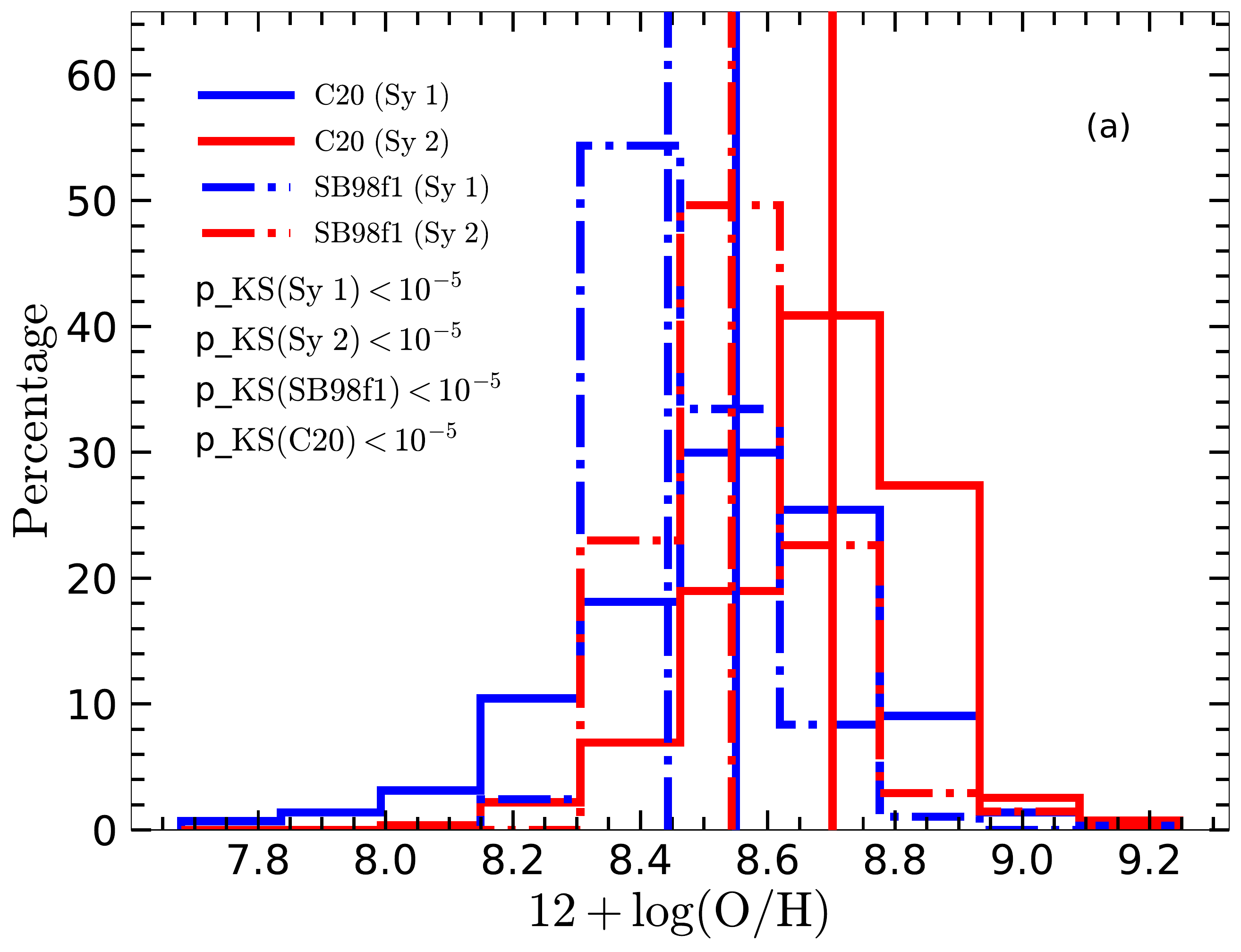}
\includegraphics[width=1.\columnwidth]{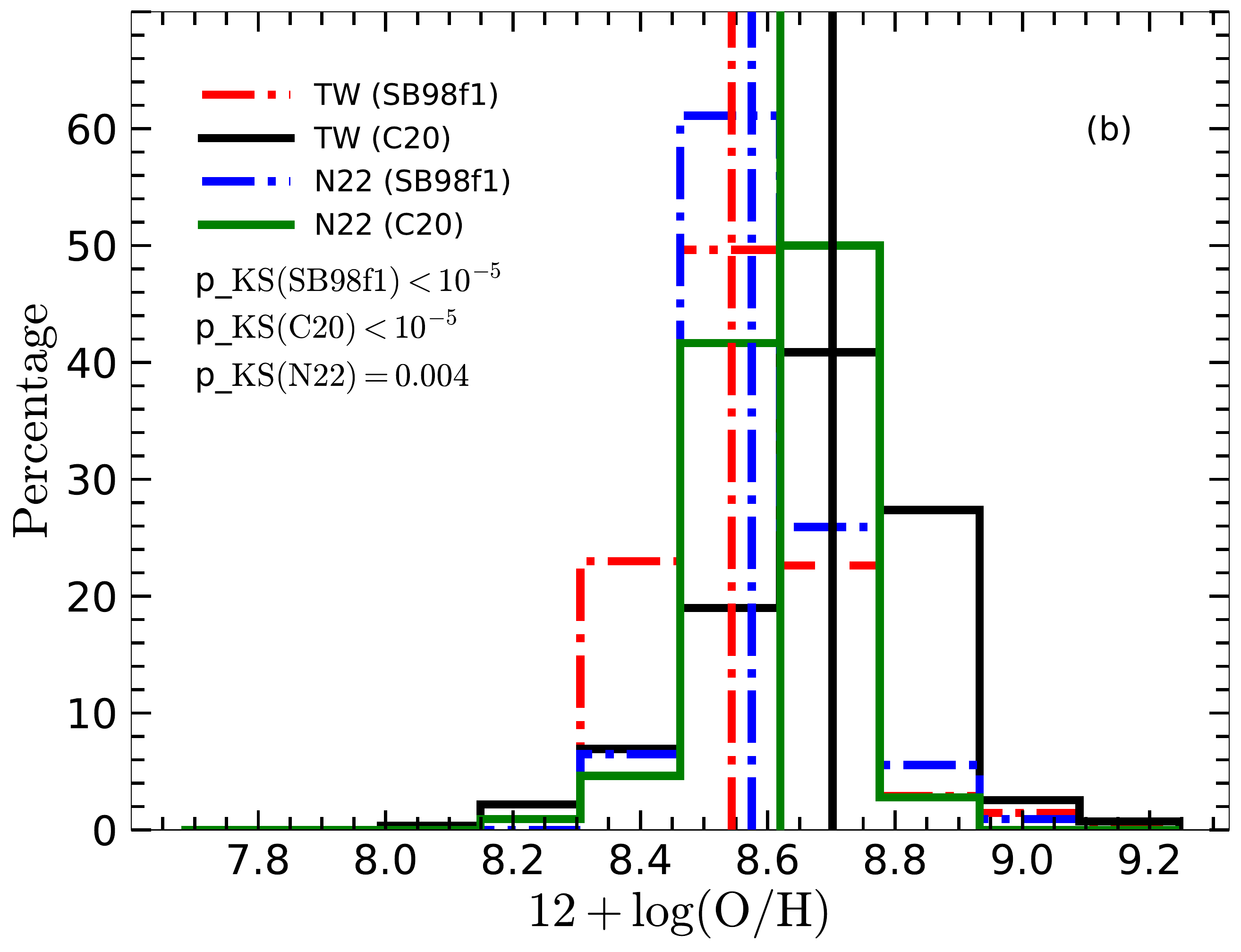}
\caption{Panel (a): distribution of oxygen abundances in this study estimated from the NLR of 287 Sy~1s and 274 Sy~2s using the calibrations by  \textcolor{blue}{SB98f1}
 and   \textcolor{blue}{C20}. The blue dashdot and red solid vertical lines represent the median values of each histogram (i.e. the same colour and line pattern are used for the histogram and median value). Panel (b):   distribution of oxygen abundances in this study for the  274 Sy~2s and  108 Sy~2s estimated by \textcolor{blue}{N22} using the  \textcolor{blue}{SB98f1} (red and blue) and   \textcolor{blue}{C20} (black and green) calibrations. A two-sample Kolmogorov--Smirnov (KS) statistical test p\_values are also shown for each calibration as well as a comparison between the SDSS-IV MaNGA survey and BASS DR2 data sets. For completeness we quote the value of 12+log(O/H)$_{\odot}=8.69$ for the solar abundance derived by  \citet{2001ApJ...556L..63A}. }
\label{Sy12}
\end{figure*}

In Fig.~\ref{Sy12}(a), we show the distribution of oxygen abundances in this study estimated from the NLRs of 287 Sy~1s and 274 Sy~2s  with respect to the calibrations by  \textcolor{blue}{SB98f1}    and  \textcolor{blue}{C20} (blue lines for Sy~1 and red for Sy~2). We also show a KS test p\_values as follows:
i) for Sy~1 sources comparing the two different methods: p\_KS(Sy~1) $<10^{-5}$;
ii) for Sy~2 sources comparing the two different methods: p\_KS(Sy~2) $<10^{-5}$;
iii) comparing Sy~1 and Sy~2 abundances  using the \textcolor{blue}{SB98f1}  method: p\_KS(SB98f1) $<10^{-5}$ and, 
iv) comparing Sy~1 and Sy~2 abundances  using the \textcolor{blue}{C20}  method: p\_KS(C20) $<10^{-5}$.

It is clear that the O/H distribution from Sy~1 covers a wide range of metallicities as compared to Sy~2, with Sy~1 sources showing median values ($8.44\pm0.03$ for SB98f1 and $8.55\pm0.04$  for \textcolor{blue}{C20}) lower than those in type~2 objects ($8.54\pm0.02$ for SB98f1 and $8.70\pm0.02$ for \textcolor{blue}{C20}). This difference is interpreted as  different chemical enrichment paths for type~1 and type~2 sources. This could be due to a more active previous star-formation in the type~2 sources than in type~1s, thus enriching the ISM with recycled material from stellar evolution. In this hypotheses, the AGN would be quenching the star-formation in the nuclear region of type~1 sources. Another possible scenario would be that type~1 sources are experiencing an inflow of low metallicity gas that is diluting the richer gas available in the centre of the galaxies, making the overall abundance lower in type~1 sources.  Indeed this hypothesis is in agreement with the findings by \textcolor{blue}{N22} when studying the abundances radial profiles of a sample  of Sy~2 sources and comparing them with the nuclear region (e.g. the disc extrapolated values are higher than those obtained for the nuclear region).

To compare our results with  the literature, in Fig.~\ref{Sy12}(b), we show the distribution of the gas phase abundance estimates for the Sy~2 galaxies in this study  in comparison with similar estimations by \citet[][hereafter \textcolor{blue}{N22}]{2022MNRAS.513..807D} using observational data from the SDSS-IV MaNGA survey. In addition we also performed KS tests to compare both samples, as follows: 
i) comparison between our results and those from \textcolor{blue}{N22} using the \textcolor{blue}{SB98f1} method: p\_KS(SB98f1)$<10^{-5}$;
ii) comparison between our results and those from \textcolor{blue}{N22} using the \textcolor{blue}{C20} method: p\_KS(C20) $<10^{-5}$ and, iii) a comparison between the results found by \textcolor{blue}{N22} from the MaNGA Sy~2 sources using the two different calibrations (\textcolor{blue}{SB98f1} and \textcolor{blue}{C20}): p\_KS(N22) = 0.004. 
We observe that there is a good agreement between our estimates and those obtained by \textcolor{blue}{N22} from the two different calibrations. However, comparing the results from Sy~2s using the two calibrations  we notice that there is a slight difference, which is obvious in the KS p\_values [p\_KS(Sy~2) $\times$ p\_KS(N22)] between our estimates  and those obtained by \textcolor{blue}{N22}, but the difference is not statistically significant.

\subsection{Oxygen abundance and AGN properties}
\subsubsection{X-ray luminosity}
\label{xlum}

\begin{figure*}
\includegraphics[width=1.\columnwidth]{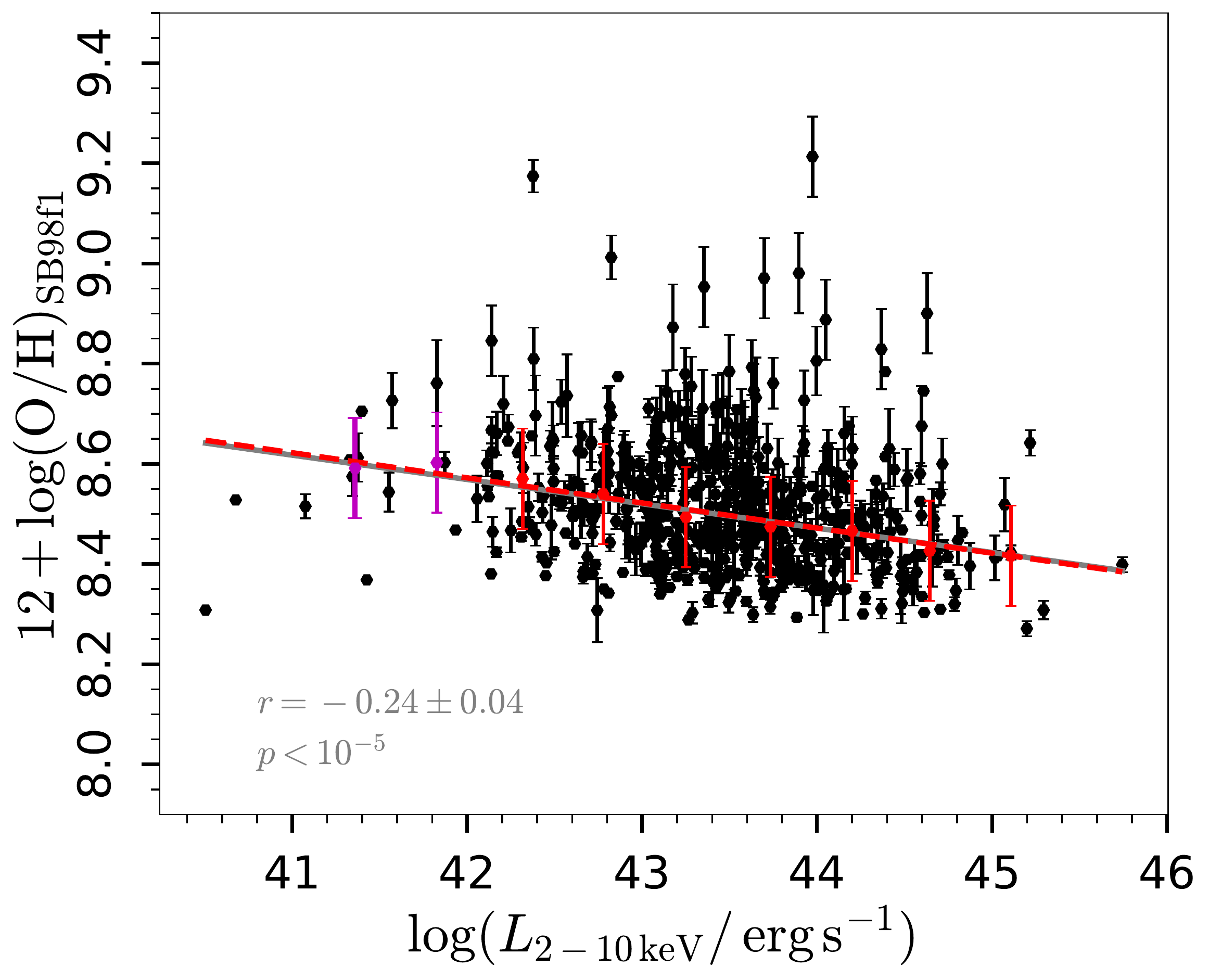}
\includegraphics[width=1.\columnwidth]{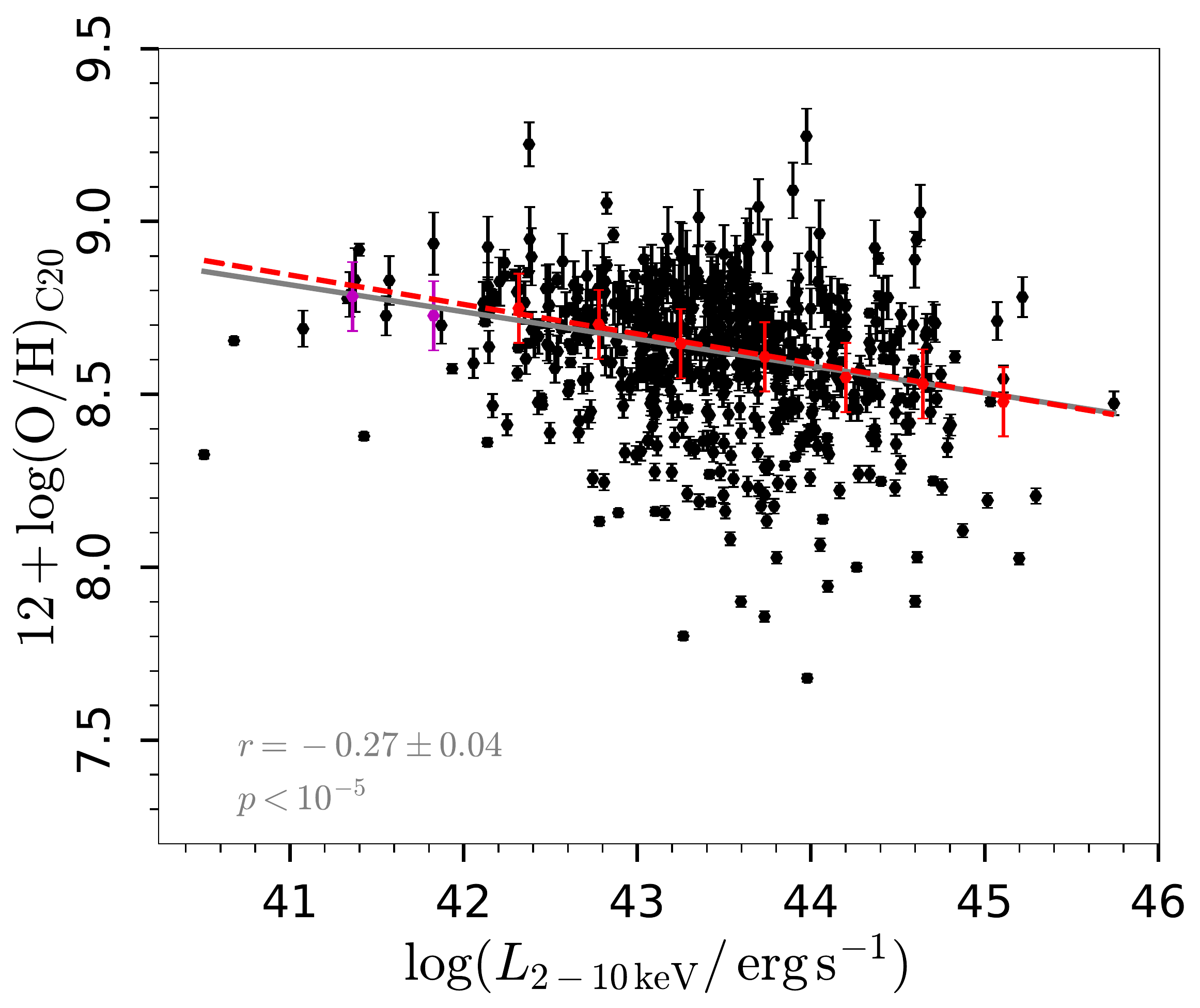}
\caption{Metallicity versus observed  ({$\log L_{2-10}$}) luminosity using the strong-line calibrations by  \textcolor{blue}{SB98f1}
 and   \textcolor{blue}{C20}.  The grey correlation parameters correspond to the solid grey lines, which represent the linear fits to the  black points. The red points and dashed line represent average  oxygen abundance values in bins of 0.5\,dex of $L_{\rm X}$ and line of best fit to the red points, respectively. }
\label{fig5}
\end{figure*}

\begin{figure*}
\includegraphics[width=1.\columnwidth]{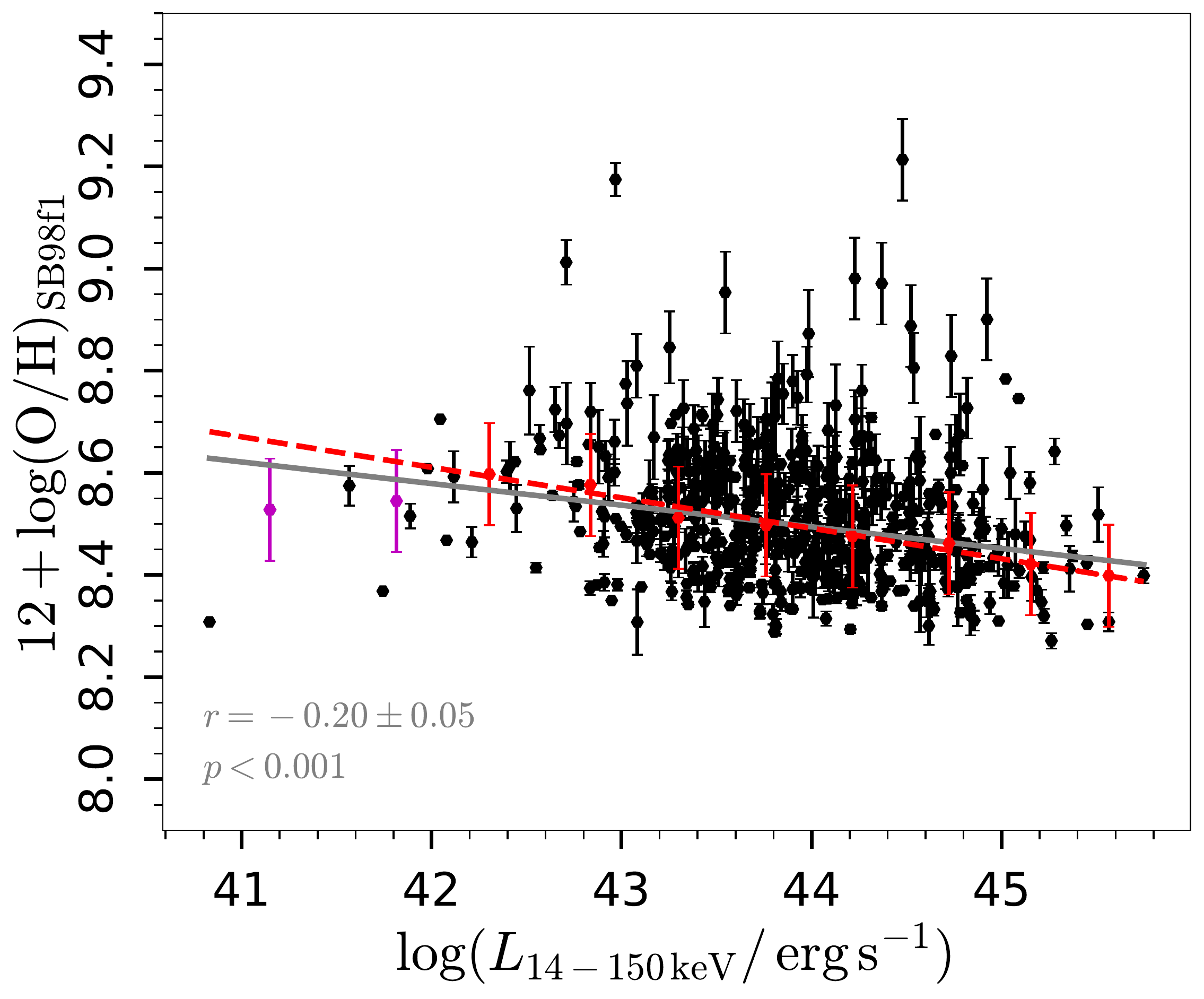}
\includegraphics[width=1.\columnwidth]{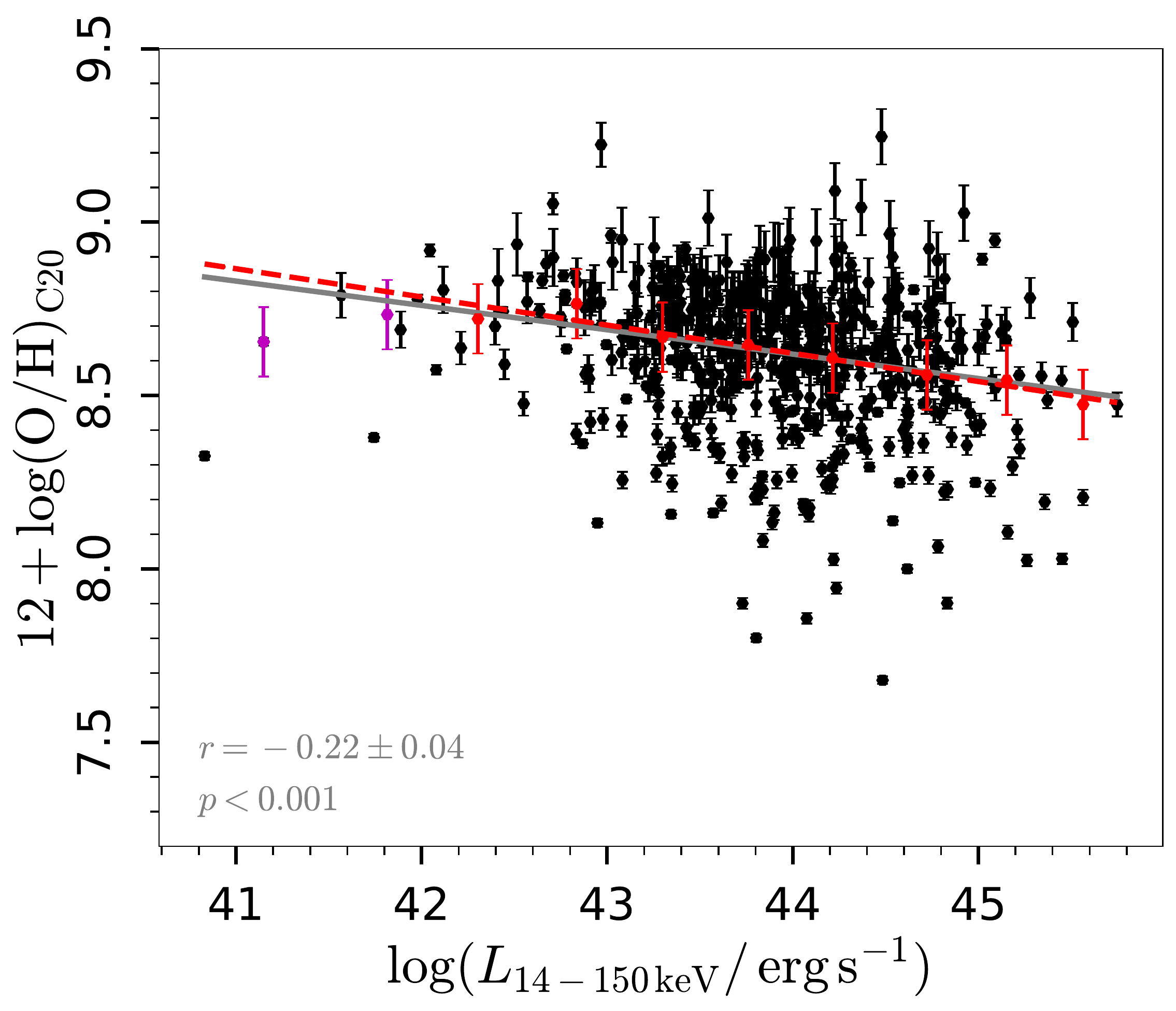}
\caption{Same as Fig.~\ref{fig5} but for the metallicity versus  intrinsic  ({$L_{14-150}$}) luminosity. }
\label{fig6}
\end{figure*}

 In this section we compare the X-ray luminosities with the oxygen abundances determined for the 561 sources using the different calibrations (see Sect.~\ref{sfinal}). In Figs.~\ref{fig5} and \ref{fig6}, we show the  the $L_{\rm X}$-$Z_{\rm NLR}$ relations which indicate   anti-correlations between the X-ray luminosities (observed and intrinsic, respectively) and metallicities from the NLR. The solid grey lines show the linear relation  between the two parameters ($L_{\rm X}$ and $Z_{\rm NLR}$). These lines were obtained following \citet{2021MNRAS.501.4064R} using 1000 bootstrap realisations \citep{davison1997bootstrap} with Huber Regressor model that is robust to outliers \citep{owen2007robust}. The Pearson correlation coefficients and {\it p-values} are quoted (they are the mean value of the bootstrap realizations). We also average the oxygen abundance values in bins of 0.5\,dex of $L_{\rm X}$, which are shown in red, and a liner regression (red dashed line) fit was performed over them (removing the points which are the average of a small number of objects -- shown in magenta). We notice that the metallicity estimates show a dependence on the various X-ray luminosities, as seen from the Pearson correlation  {\it p-values}.  In addition, there is a higher correlation from the calibration by  \textcolor{blue}{C20} ($r=-0.27\pm0.04; -0.22\pm0.04$) in comparison with  \textcolor{blue}{SB98f1}
 ($r=-0.24\pm0.04; -0.20\pm0.05$) at both the observed  ($\log L_{2-10}^{\rm\,obs}$) and intrinsic ($\log L_{14-150}^{\rm\,int}$) X-ray luminosities.  At this point, it is difficult to reconcile which physical quantity is responsible for the observed correlation difference. However, the metallicity exhibits somewhat dependence on the X-ray luminosity regardless of the method used \citep[also see][]{Oh2017}, indicating that there may be a common physical driving mechanism.  It is important to highlight that we tested for the correlations separating type~1 and type~2 sources, as a result of this exercise, we note that the metallicity estimates from Sy~1 and Sy~2  follow the same trend with respect to  $L_{\rm X}$ (e.g. the $r$ values are the same within the margin of error).

\subsubsection{The mass-metallicity relation}

\begin{figure*}
\includegraphics[width=2.\columnwidth]{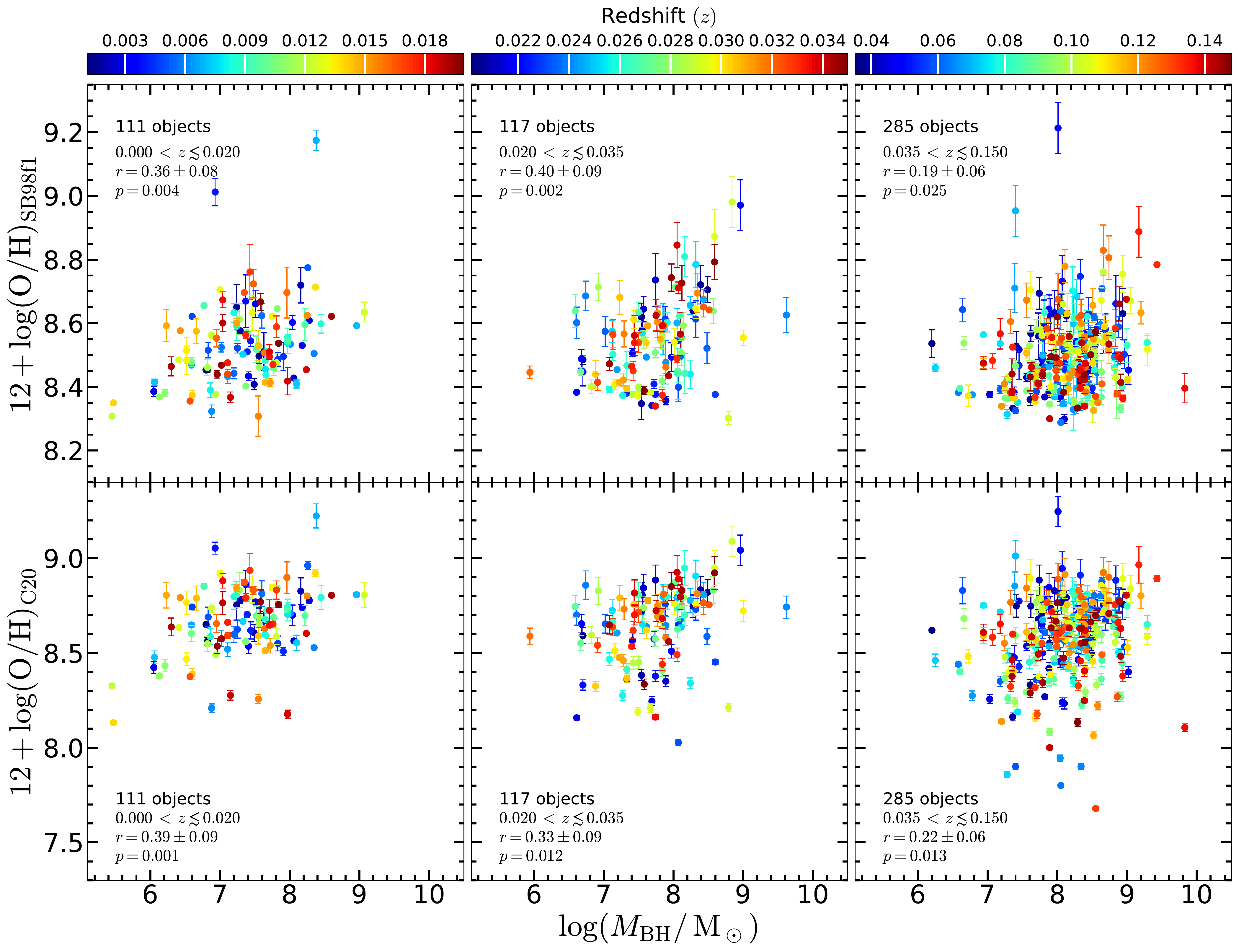}
\caption{Bottom panels: the mass-metallicity relations between the black hole mass and the metallicity from the calibration by  
 \textcolor{blue}{C20}. Top panels: same as bottom panels but for the calibration by    \textcolor{blue}{SB98f1}.  The points in all panels are colour-coded by redshift, as indicated by the colour bar. The number of sources is the same in the corresponding bottom and top panels. The results from the correlation analysis in redshift bins ($0.000<z\lesssim0.020$, $0.020<z\lesssim0.035$ and $0.035<z\lesssim0.150$) are indicated in each panel.  There is no correlation parameters at $z>0.15$.}
\label{mzr1}
\end{figure*}

\begin{figure*}
\includegraphics[width=2.\columnwidth]{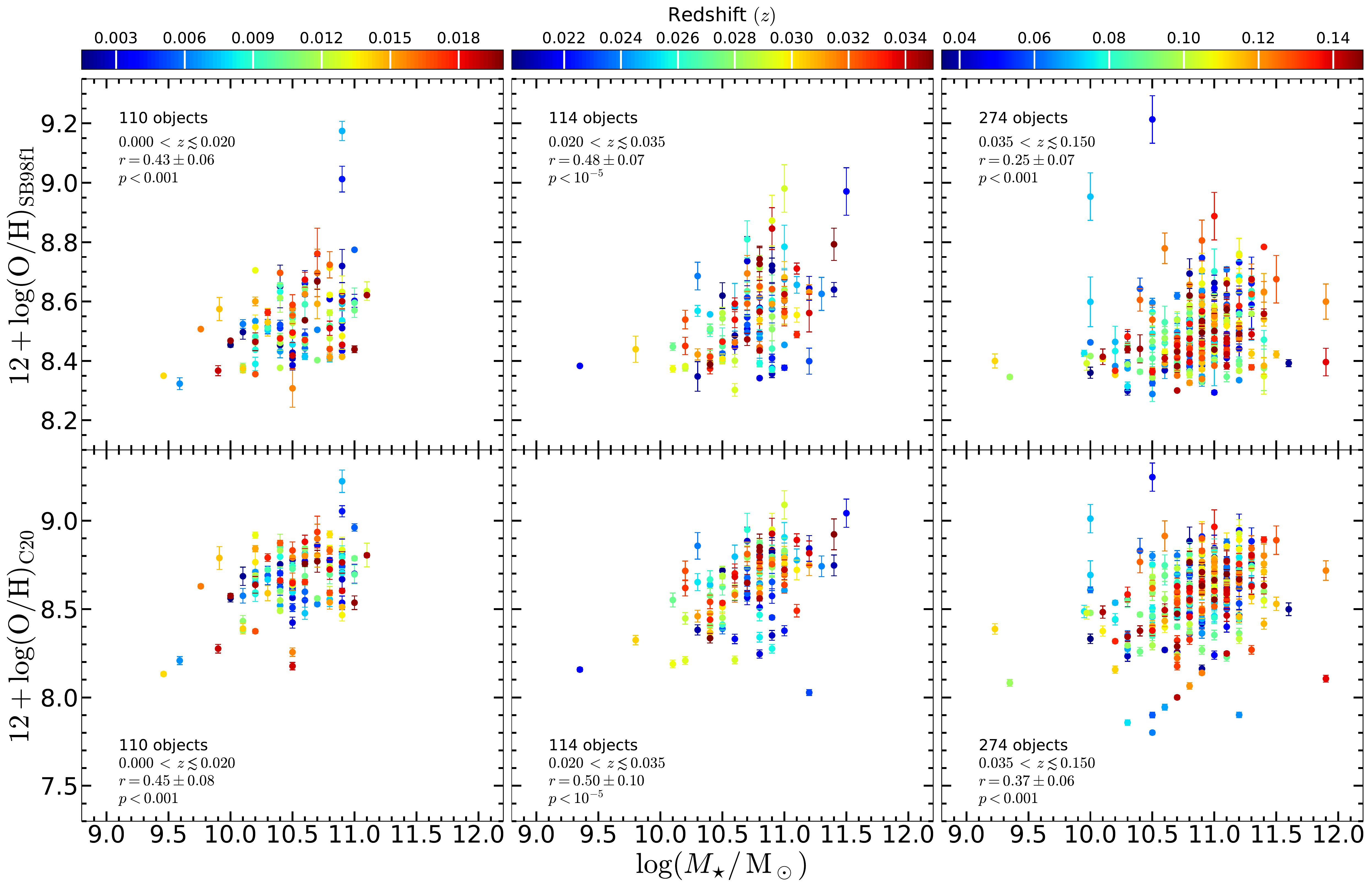}
\caption{Same as Fig.~\ref{mzr1} but for the stellar mass of the host galaxies. }
\label{mzr2}
\end{figure*}

The inflows of metal-poor gas
which activate the star formation and dilute the metallicity of the ISM or the outflows of metal-rich gas which stop the star formation are some mechanisms which depend on MZR.  In Figs.~\ref{mzr1} and \ref{mzr2}, we plot the metallicity as a function of the black hole mass and the host stellar
mass for individual galaxies in the sample, respectively.
The MZR has been shown to be dependent on redshift and various galaxy properties in the NLRs of AGNs \citep[][]{2018A&A...616L...4M} and other SFs \citep[e.g][]{kew08,2019ApJ...886...31H,2020MNRAS.491..944C}. It has generally been shown that,  at a given stellar mass, lower redshift galaxies have higher
gas-phase metallicities than their higher redshift counterparts, while the MZR always shows a positive correlation. In Fig.~\ref{mzr1}, we investigate the redshift evolution of the MZRs between the metallicities and the black hole mass. We found  weak correlations ($r=0.12$; $p = 0.05$, for \textcolor{blue}{SB98f1}  and $r=0.09$;  $p = 0.15$ for \textcolor{blue}{C20} across the full redshift range $z\lesssim0.31$), which is consistent with the findings by \citet{Oh2017}, who found a weak correlation of the [O\iii]/H$\beta$ line ratio with the SMBH mass. We show the distribution of the NLR metallicities as a function of the stellar masses of the hot galaxies in Fig.~\ref{mzr2}. We find positive correlations between NLR metallicities and the stellar masses at three different redshifts bins. This implies that the gas-phase metallicities in the NLRs of AGNs are connected to the properties of the hot galaxies.

 The MZRs and FMRs have widely been investigated in SFs but such relations have rarely been found in AGNs. However, the origin of the MZRs and FMRs remain active research in astrophysics. Therefore, we further test this hypothesis by comparing the oxygen abundances values computed here with those obtained in star forming galaxies.  A direct comparison between findings using samples from the same source might not reveal the actual properties which influence the MZRs, as we have demonstrated above. Therefore, it is important to show a comparison between our results and other findings from the literature. We briefly highlight some of the MZRs from the literature in the following.

 We considered the first MZR by \citet{2004ApJ...613..898T}, who used  SDSS spectroscopic data to demonstrate the MZR with $\sim0.1$ dex scatter at $z\sim0.1$ considering stellar masses in the range $8.5\lesssim\log \left(M_\star/{\rm M_\odot} \right)\lesssim11$. We also took into consideration the  MZR at $z\sim0.07$ derived by  \citet{kew08}  based on  the calibration by \citet{2004ApJ...617..240K}.
  Additionally, \citet{2018A&A...616L...4M} found the MZR for type-2 AGNs at  $z\sim3.0$, using high-$z$ radio galaxies (HzRGs)  and X-ray selected radio-quiet AGNs.  Furthermore, \cite{2019ApJ...886...31H} used the composite spectra of galaxies  from the extended Baryon Oscillation Spectroscopic Survey of the Sloan
Digit Sky Survey (SDSS IV/eBOSS) with a median redshift of $\sim0.83$. They found a redshift evolution of the MZR described by the relation
\begin{equation}
    12 + \log \left( \text{O}/\text{H} \right) = Z_0 + \log \left[ 1 - \exp \left( - \left[ \frac{{M}_\star}{{M}_0} \right]^\gamma \right) \right]
\end{equation}
where $Z_0 = 8.977$, $\log M_0 = 9.961$ and $\gamma=0.661$ for the redshift range $0.60$--$1.05$  with  the stellar mass covering the range $9<\log \left(M_\star/{\rm M_\odot} \right)<12$. Finally, \cite{2020MNRAS.491..944C} parametrized the median MZR  via the relation:
\begin{equation}
    12 + \log \left( \text{O}/\text{H} \right) = Z_0 - \frac{\gamma}{\beta} \log \left( 1 + \left( \frac{{M}}{{M}_0} \right)^{- \beta} \right)
\end{equation}
where $Z_0 = 8.793$ is the asymptotic metallicity at high $M_\star$, $\log \left(M_0/{\rm M_\odot} \right)= 10.02$ is the turnover mass below which the MZR reduces to a power law of index $\gamma = 0.28$ and $\beta = 1.2$ is a measure of how fast the relation reach the asymptotic value.

\begin{figure*}
\includegraphics[width=2.\columnwidth]{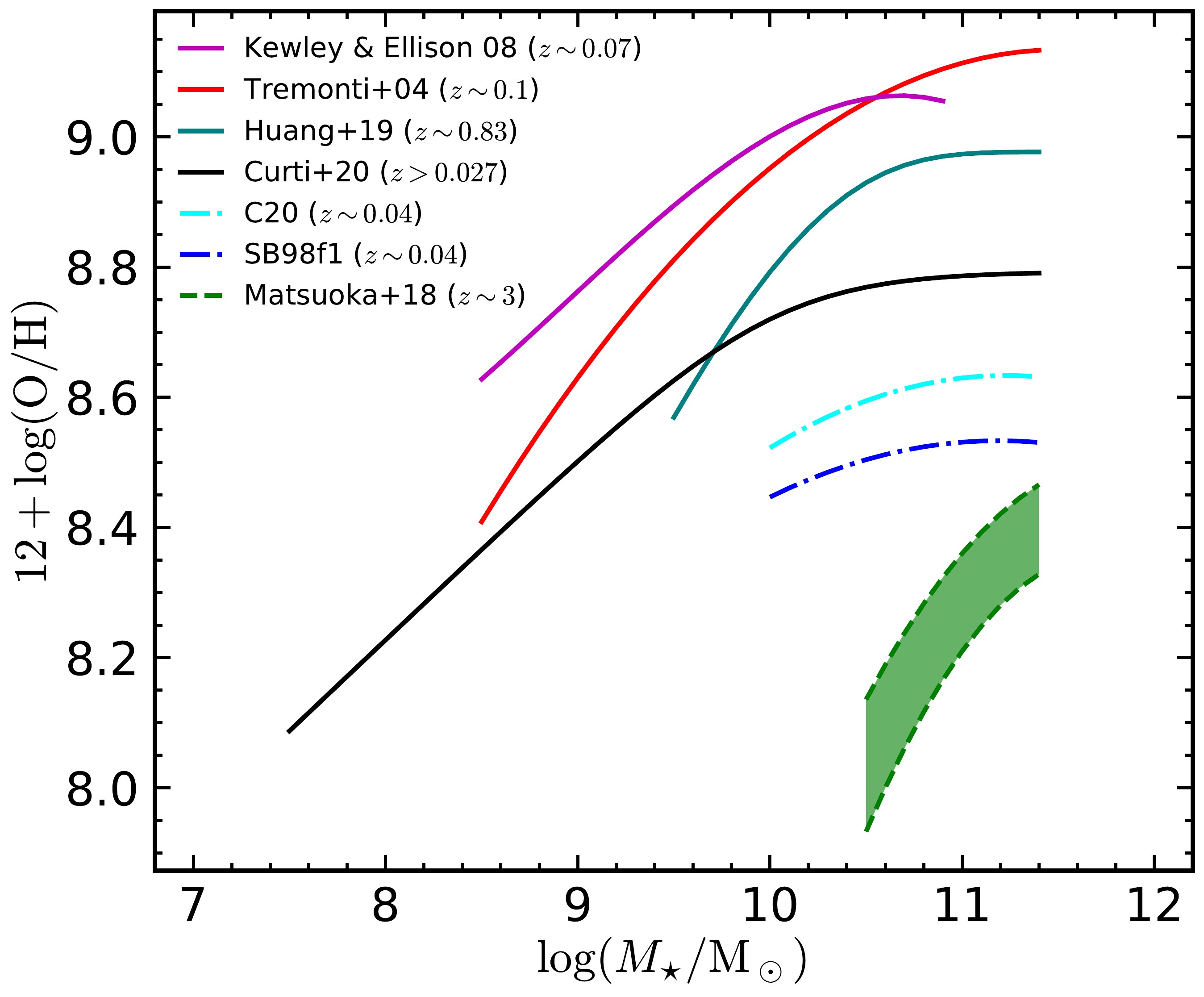}
\caption{Comparison between the mass-metallicity relations of the BASS sample (with a  median redshift at $z\sim0.04$)  and from the literature. The solid magenta, red, teal and black curves denote the MZRs for star-forming galaxies at $z \sim 0.07$, $z \sim 0.1$, $z \sim0.83$ and $z>0.027$    derived by  \citet{kew08}, \citet{2004ApJ...613..898T},  \citet{2019ApJ...886...31H} and \citet{2020MNRAS.491..944C}, respectively. The dashdot cyan and  blue  curves represent our estimates from    \textcolor{blue}{C20} and \textcolor{blue}{SB98f1} calibrations respectively, while the  green filled area is from type-2 AGNs at $z \sim 3$ by \citet{2018A&A...616L...4M}.}
\label{mzr3}
\end{figure*}

 In Fig.~\ref{mzr3}, we show the redshift evolution of the MZR in terms of the metallicities and the stellar mass of the host galaxies in this work as well as a comparison with similar MZRs from the literature.  The comparison among  previous studies of SFs in Fig.~\ref{mzr3} show a unique MZR  downward evolution trend from $z \sim 0.07$ \citep{kew08}, $z \sim 0.1$ \citet{2004ApJ...613..898T}, $z \sim0.83$ \citet{2019ApJ...886...31H}, to $z>0.027$ \citet{2020MNRAS.491..944C}, which is consistent with the fact that as redshift increases, the MZR shifts downward, indicating that more evolved galaxies tend to be more metal-rich \citep[e.g.][]{maiolino08,2019ApJ...886...31H}. Similarly, a comparison between our result at a median redshift of $z \sim 0.04$ and the MZR   derived by  \citet{2018A&A...616L...4M} for AGNs at $z \sim 3$ follow the same downward trend.
The X-ray selected AGNs have reveal  the redshift evolution of the $M_\star$-$Z_{\rm NLR}$ relations, which is consistent with the findings by \citet{2018A&A...616L...4M} and other star-forming galaxies \citep[e.g.][]{2019ApJ...886...31H,2021ApJ...914...19S}.  We note that the origin of the differences between the MZRs is outside the purview of this study. However, we highlight some possible scenarios which could be attributed to the differences in MZRs between SFs and AGNs. 
Comparing our results with those by \citet[][ for AGN with $z\sim$3]{2018A&A...616L...4M}, we find that the curves from our sample  is more flat and that the values are on average $\sim 0.3$ dex higher than those estimated from higher redshift sources. We interpret this as the redshift evolution of MZR in AGNs, in the sense that metal-rich gas are usually found in the more evolved galaxies. However, when comparing the mean values obtained from SF galaxies, reported by \citet[][red line]{2004ApJ...613..898T}, \citet[][magenta line]{kew08}, \citet[][teal line]{2019ApJ...886...31H} and \citet[][black line]{2020MNRAS.491..944C} in the stellar mass range of our sample (see Fig.~\ref{mzr3}), it can be seen that our AGN hosts (dashed cyan and blue lines) do present  values that are $0.2-0.5$~dex lower than SF galaxies (solid lines).  We note that different metallicity calibrations, even when based on the same diagnostics, are typically not consistent with one another and usually result in systematic abundance discrepancies from -0.09 up to 0.8~dex  \citep[e.g.][]{kew08,2015ApJ...798...99B, 2017ApJ...834...51B, 2020MNRAS.492..468D},  which is consistent with the correlation differences between our estimates and the aforementioned previous results obtained from SFs and AGNs.  This discrepancy can not, however, be associated with the redshift evolution of the MZR, since the sources studied here are in the lower redshift range (see Fig.~\ref{mzr3}) with different ionization mechanisms from SFs, thus, it indicates that AGNs and SF galaxies have different chemical evolution paths, suggesting that the abundances of AGN hosts are somehow affected by the central engine.

In fact, the SF around SMBH may be affected in many aspects (as a consequence of the abundances too).  For example, while some studies associate the AGN outflows with the  suppression of the SF \citep[e.g.][]{Granato+04,Fabian+12,King+15,Zubovas+17a,Trussler+20} other studies suggest that these outflows and jets can compress the galactic gas, thus acting as a catalyzer and boosting the SF \citep[e.g.][]{Rees+89,Hopkins+12,Nayakshin+12,Bieri+16,Zubovas+13,Zubovas+17a} and even form stars inside the outflow  \citep[e.g.][]{Ishibashi+12,Zubovas+13,El-Badry+16,Wang+18,2019MNRAS.485.3409G}. Using high spatial resolution observations taken with adaptive-optics assisted integral field spectroscopy \citet[][see also references therein]{2022MNRAS.512.3906R} have shown that, once the AGN is triggered, it precludes further SF, in the sense that it can be associated with the lack of new star formation in the inner few hundred of pc of AGN hosts. \textcolor{blue}{N22} studied spatial variation of oxygen abundances on a sample of AGN, and has suggested that the drop in the O/H abundance in the AGN dominated region when compared with the values obtained for the disc region is due to the inflow of less metallic gas towards the central region of the galaxy.  All these processes contribute to a {\it chaotic} chemical evolution of AGNs.

\subsubsection{Metallicity and accretion rate}

\begin{figure*}
\includegraphics[width=2.\columnwidth]{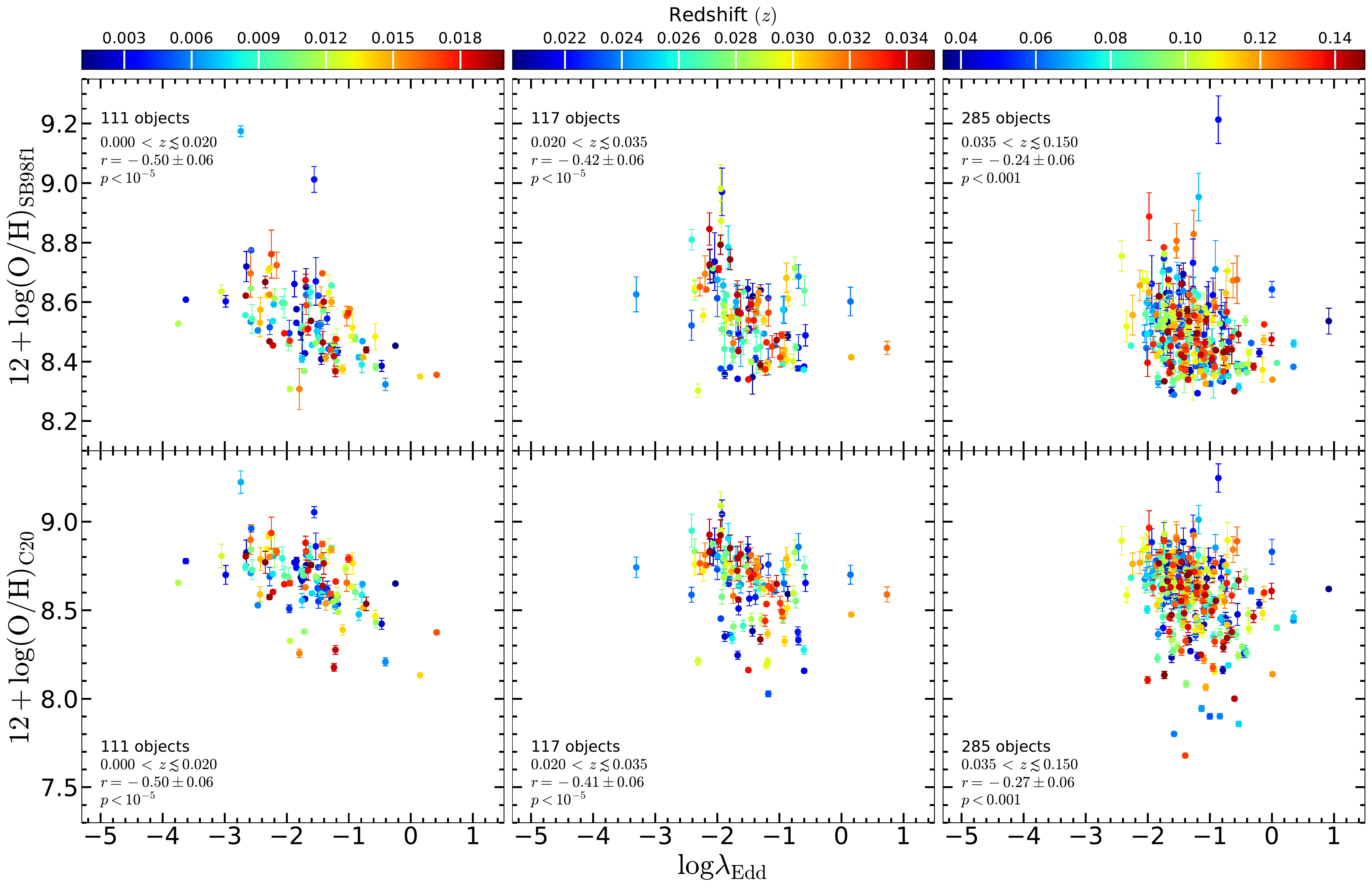}
\caption{  Bottom  panels: the relation between the metallicity from the calibration by   \textcolor{blue}{C20} and the Eddington ratios from the intrinsic  X-ray luminosity ($ L_{14-150}$). Top panels: same as bottom panels but for the calibration by  \textcolor{blue}{SB98f1}.  The points in all panels are colour-coded by redshift, as indicated by the colour bar. The number of sources is the same in the corresponding bottom and top panels. The results from the correlation analysis in redshift bins ($0.000<z\lesssim0.020$, $0.020<z\lesssim0.035$ and $0.035<z\lesssim0.150$) are indicated in each panel.  There is no correlation parameters at $z>0.15$.}
\label{edd1}
\end{figure*}

In order to test for a possible influence of the AGN on the metallicity of the host galaxies, we have compared the  metallicities with the Eddington ratio, which is a function of luminosity and $M_{\rm BH}$, shown  in Fig.~\ref{edd1} and since there is a similar distribution on the plane log$\lambda_{\rm Edd} \times \log(L_{\rm bol})$ between Sy~1 and Sy~2  we do not discriminate between them in subsequent analysis.  This assertion reiterate the fact that there are no significant variations between the SFRs of type 1 and type 2 AGNs, which is consistent with earlier findings using SFRs derived from infrared indicators \citep[e.g.][]{2019ApJ...878...11Z}.

Furthermore, from Fig.~\ref{edd1}, it can be seen that there is  anti-correlation between the accretion rate and the metallicity, for different redshift bins, with Person correlation coefficients ranging from $r\sim$-0.24 to -0.50 and {\it p-value} $< 10^{-5}$ depending on the redshift range and  the adopted calibration. These results are more in agreement with those found by \citet{Oh2017} who compared  [N\ii]$\lambda$6583/H$\alpha$, which is  sensitive to O/H abundances, with $\lambda_{\rm Edd}$ for a sub-sample of  the local hard X-ray-selected  BAT AGNs  (297 sources at $0.01 < z < 0.40$) as well as  at redshift 
 beyond the local Universe \citep[e.g. 53  sources at $0.6 < z < 1.7$;][]{Oh2019}  and found a clear anti-correlation between those two quantities. Moreover, \citet{2011A&A...527A.100M} used optical spectra of high redshift ($2.3 < z < 3.0$) quasars  and compared different metallicity sensitive emission line ratios with $\lambda_{\rm Edd}$. They obtained mostly positive correlations\footnote{It is worth mentioning that depending on the line ratio and  the black hole mass bins there is no correlation found in some cases.}  from the comparison between the metalicity sensitive line ratios and $\lambda_{\rm Edd}$, as can be observed from their Fig.~6. Furthermore, they found that the $L_{\rm AGN}$-$Z_{\rm BLR}$ trend is as a result of the positive correlations between these same line ratios and the SMBH mass.  Also, \citet[][and references therein]{2004ApJ...614..547S} found that in high redshift (2.0 < z < 3.5)  quasars there is a positive correlation between luminosity (or SMBH mass) and BLR metallicity sensitive line ratios (N\V/C\iv).

The correlation of the $Z_{\rm NLR}$  and/or metallicity sensitive emission line ratios with the $L_{\rm X}$, or $\lambda_{\rm Edd}$ in Seyferts is  in contrast with the results that were obtained based on measurements of UV broad emission lines in quasar. Additionally, the correlation of $Z_{\rm AGN}$-$M_{\rm BH}$ relation, is extremely weak (or absent) in Seyferts as compared to quasars. These discrepancies might be due to the fact that Seyferts are moderate-luminous AGNs whereas quasars are the most-luminous AGNs as well as the different methodologies used in the estimation of the $\lambda_{\rm Edd}$, which has a strong dependence on the $L_{\rm X}$-$M_{\rm BH}$ relation.  It is worthwhile to note that, the $\lambda_{\rm Edd}$ is responsible for the AGN fueling while the metal enrichment of the nuclear gas is due to nuclear/circumnuclear star-formation.

Moreover, it is challenging to fairly compare the results obtained from the BLR using high $z$ quasars with those found from the NLR of local Universe sources. Even, to compare the NLR gas from local Universe sources with quasar is very difficult, as has been shown by \citet{2004ApJ...614..558N}, the NLR properties in high-luminosity  quasars are very different from those observed in nearby AGNs,  for example, they may be in a phase of violent  star-forming events that release high quantities of metals into the gas. This would naturally explain the differences we  observe between our  results and those from high redshift objects. In fact, this is fully supported by our finding in Fig.~\ref{edd1}, since the "strength" of the anti-correlation (measured by the $r$) drops when we consider the higher redshift sources (0.035 < z < 0.150), indicating that there is a change in the chemical evolution path of AGNs with redshift.

The anti-correlation observed in Fig.~\ref{edd1}, where $\lambda_{\rm Edd}$ increases with decreasing metallicity suggest that  the chemical evolution of the host galaxy seems to be  affected by the AGN activity through  suppression of their nuclear SF, and thus stopping the enrichment, or the sources with higher Eddington ratio experiencing an inflow of a lower metallicity gas from the outskirts of the galaxy and/or from the cosmic web. Thus, this metal-poor gas would  be diluting the more metal-rich gas in the inner regions and feeding/triggering the AGN. The scatter we observe from the $Z_{\rm NLR}$  in Fig.~\ref{edd1} (e.g. for a fixed luminosity, there is a range of possible metallicity values), could be explained by the fact that the gas reaching the SMBH is probably originating from mass loss from intermediate-age stellar population  \citep[][]{2022MNRAS.512.3906R}. In this sense, the outskirts low metallicity gas reaching the AGN, will get an extra supply of gas that has already been processed by stellar nucleo-synthesis, and thus enhancing the $Z_{\rm NLR}$. Therefore, the values of oxygen abundances, for a fixed $L_{\rm X}$ would be the balance between the pristine and the processed gas phases.

From the above, we interpret the anti-correlation found in Fig.~\ref{edd1} as due to the combination of the inflow of pristine gas diluting the metal-rich gas, activating the AGNs, which in turn suppress the nuclear SF, thus stopping the gas enriching process. In this framework, the more luminous sources will have lower metallicities (no SF + pristine gas), while the less luminous would have higher  $Z_{\rm NLR}$  values since they may still have some level of nuclear SF but  no (or a small amount of) pristine gas is reaching their SMBHs.

\section{Concluding remarks}
\label{conc}

We derived metallicities via the strong-line calibrations by \citet{sb98} and \citet{2020MNRAS.492.5675C} for a sample of 561 Seyfert  nuclei in the local universe ($z \:\lesssim \:0.31$) selected from the  \textit{Swift}-BAT 70-month AGN Data Release 2 (DR2) Catalog. These metallicities and the hard X-ray ($\gtrsim10$ keV) luminosities were used to study the $L_{\rm X}$-$Z_{\rm NLR}$ relation  for the first time in Seyfert galaxies. We also studied the relation between the metallicities and the AGN properties. The physical properties comparison between Sy~1s and Sy~2s indicate that even in the narrow line regions of both AGN classes, Sy~1 still exhibit higher electron  density in comparison to Sy~2.  We  found that the AGN metallicities are related to the hosts stellar masses following a downward redshift evolution, similar to that of SF galaxies, from lower to higher redshifts, but with  lower values of O/H abundances (with a  mean difference of  0.2-0.5~dex)  in AGN hosts than in SF galaxies. We also found that the metallicities decrease with increasing X-ray luminosities and have significant correlations with $\lambda_{\rm Edd}$ at the redshift range $z\lesssim0.02$ (the overall oxygen and nitrogen dependent: {\it r}$\sim -0.50\pm 0.06 $,   $p< 10^{-5}$, $\sigma = 0.20$ dex) and at the full redshift range $z \:\lesssim \:0.31$ (oxygen dependent: {\it r}$\sim -0.42\pm 0.04 $,   $p< 10^{-5}$, $\sigma = 0.12$ dex and nitrogen dependent: {\it r}$\sim -0.45\pm 0.04$,   $p< 10^{-5}$, $\sigma = 0.18$ dex). We argue that, these anti-correlations are driven by the X-ray luminosities, and they change with redshift, indicating that the AGNs are somehow driving the  chemical enrichment of their host galaxies, as a result of the  inflow of pristine gases that are diluting the metal-rich gases, together with a recent cessations on the star formation.

\section*{Acknowledgements}

We thank the anonymous referee for insightful comments and suggestions. MA gratefully acknowledges support from Coordenação de Aperfeiçoamento de Pessoal de Nível Superior (CAPES). 
RR thanks  Conselho Nacional de Desenvolvimento Cient\'{i}fico e Tecnol\'ogico  (CNPq, Proj. 311223/2020-6,  304927/2017-1 and  400352/2016-8), Funda\c{c}\~ao de amparo \`{a} pesquisa do Rio Grande do Sul (FAPERGS, Proj. 16/2551-0000251-7 and 19/1750-2), Coordena\c{c}\~ao de Aperfei\c{c}oamento de Pessoal de N\'{i}vel Superior (CAPES, Proj. 0001).
OLD is grateful to the Fundação de Amparo à Pesquisa do Estado de São Paulo (FAPESP) and to Conselho Nacional de Desenvolvimento Científico e Tecnológico (CNPq)  for the financial support. KO acknowledges support from the Korea Astronomy and Space Science Institute under the R\&D program (Project No. 2023-1-830-01) supervised by the Ministry of Science and ICT and from the National Research Foundation of Korea (NRF- 2020R1C1C1005462). BT acknowledges support from the European Research Council (ERC) under the European Union's Horizon 2020 research and innovation program (grant agreement 950533) and from the Israel Science Foundation (grant 1849/19).
M.V. acknowledges support from grant "ESTANCIAS POSDOCTORALES POR M{\'E}XICO 2022" by CONACYT.
CR acknowledges support from the Fondecyt Iniciacion grant 11190831 and ANID BASAL project FB210003.

\section{DATA AVAILABILITY}
The data underlying this article will be shared on reasonable request
with the corresponding author.

\bibliographystyle{marc}
\bibliography{marc}

\begin{thebibliography}{}
\makeatletter
\relax
\def\mn@urlcharsother{\let\do\@makeother \do\$\do\&\do\#\do\^\do\_\do\%\do\~}
\def\mn@doi{\begingroup\mn@urlcharsother \@ifnextchar [ {\mn@doi@}
  {\mn@doi@[]}}
\def\mn@doi@[#1]#2{\def\@tempa{#1}\ifx\@tempa\@empty \href
  {http://dx.doi.org/#2} {doi:#2}\else \href {http://dx.doi.org/#2} {#1}\fi
  \endgroup}
\def\mn@eprint#1#2{\mn@eprint@#1:#2::\@nil}
\def\mn@eprint@arXiv#1{\href {http://arxiv.org/abs/#1} {{\tt arXiv:#1}}}
\def\mn@eprint@dblp#1{\href {http://dblp.uni-trier.de/rec/bibtex/#1.xml}
  {dblp:#1}}
\def\mn@eprint@#1:#2:#3:#4\@nil{\def\@tempa {#1}\def\@tempb {#2}\def\@tempc
  {#3}\ifx \@tempc \@empty \let \@tempc \@tempb \let \@tempb \@tempa \fi \ifx
  \@tempb \@empty \def\@tempb {arXiv}\fi \@ifundefined
  {mn@eprint@\@tempb}{\@tempb:\@tempc}{\expandafter \expandafter \csname
  mn@eprint@\@tempb\endcsname \expandafter{\@tempc}}}

\bibitem[\protect\citeauthoryear{{Abazajian} et~al.,}{{Abazajian}
  et~al.}{2009}]{2009ApJS..182..543A}
{Abazajian} K.~N.,  et~al., 2009, \mn@doi [\apjs]
  {10.1088/0067-0049/182/2/543}, \href
  {https://ui.adsabs.harvard.edu/abs/2009ApJS..182..543A} {182, 543}

\bibitem[\protect\citeauthoryear{{Agostino} et~al.,}{{Agostino}
  et~al.}{2021}]{2021ApJ...922..156A}
{Agostino} C.~J.,  et~al., 2021, \mn@doi [\apj] {10.3847/1538-4357/ac1e8d},
  \href {https://ui.adsabs.harvard.edu/abs/2021ApJ...922..156A} {922, 156}

\bibitem[\protect\citeauthoryear{{Aguado} et~al.,}{{Aguado}
  et~al.}{2019}]{aguado19}
{Aguado} D.~S.,  et~al., 2019, \mn@doi [\apjs] {10.3847/1538-4365/aaf651},
  \href {https://ui.adsabs.harvard.edu/abs/2019ApJS..240...23A} {240, 23}

\bibitem[\protect\citeauthoryear{{Allende Prieto}, {Lambert}  \&
  {Asplund}}{{Allende Prieto} et~al.}{2001}]{2001ApJ...556L..63A}
{Allende Prieto} C.,  {Lambert} D.~L.,   {Asplund} M.,  2001, \mn@doi [\apjl]
  {10.1086/322874}, \href
  {https://ui.adsabs.harvard.edu/abs/2001ApJ...556L..63A} {556, L63}

\bibitem[\protect\citeauthoryear{Ananna et~al.,}{Ananna
  et~al.}{2022}]{Ananna_2022}
Ananna T.~T.,  et~al., 2022, \mn@doi [ApJS] {10.3847/1538-4365/ac5b64}, 261, 9

\bibitem[\protect\citeauthoryear{{Antonucci}}{{Antonucci}}{1993}]{antonucci1993unified}
{Antonucci} R.,  1993, \mn@doi
  [\href{https://ui.adsabs.harvard.edu/abs/1993ARA&A..31..473A}{\araa}]
  {10.1146/annurev.aa.31.090193.002353}, \href
  {https://ui.adsabs.harvard.edu/abs/1993ARA&A..31..473A} {31, 473}

\bibitem[\protect\citeauthoryear{{Armah} et~al.,}{{Armah}
  et~al.}{2021}]{2021MNRAS.508..371A}
{Armah} M.,  et~al., 2021, \mn@doi [\mnras] {10.1093/mnras/stab2610}, \href
  {https://ui.adsabs.harvard.edu/abs/2021MNRAS.508..371A} {508, 371}

\bibitem[\protect\citeauthoryear{{Audibert}, {Riffel}, {Sales}, {Pastoriza}  \&
  {Ruschel-Dutra}}{{Audibert} et~al.}{2017}]{2017MNRAS.464.2139A}
{Audibert} A.,  {Riffel} R.,  {Sales} D.~A.,  {Pastoriza} M.~G.,
  {Ruschel-Dutra} D.,  2017, \mn@doi [\mnras] {10.1093/mnras/stw2477}, \href
  {https://ui.adsabs.harvard.edu/abs/2017MNRAS.464.2139A} {464, 2139}

\bibitem[\protect\citeauthoryear{{Baldwin}, {Phillips}  \&
  {Terlevich}}{{Baldwin} et~al.}{1981}]{bpt81}
{Baldwin} J.~A.,  {Phillips} M.~M.,   {Terlevich} R.,  1981, \mn@doi [\pasp]
  {10.1086/130766}, \href {http://adsabs.harvard.edu/abs/1981PASP...93....5B}
  {93, 5}

\bibitem[\protect\citeauthoryear{{Baldwin}, {Ferland}, {Korista}, {Hamann}  \&
  {Dietrich}}{{Baldwin} et~al.}{2003a}]{2003ApJ...582..590B}
{Baldwin} J.~A.,  {Ferland} G.~J.,  {Korista} K.~T.,  {Hamann} F.,   {Dietrich}
  M.,  2003a, \mn@doi [\apj] {10.1086/344788}, \href
  {https://ui.adsabs.harvard.edu/abs/2003ApJ...582..590B} {582, 590}

\bibitem[\protect\citeauthoryear{{Baldwin}, {Hamann}, {Korista}, {Ferland},
  {Dietrich}  \& {Warner}}{{Baldwin} et~al.}{2003b}]{2003ApJ...583..649B}
{Baldwin} J.~A.,  {Hamann} F.,  {Korista} K.~T.,  {Ferland} G.~J.,  {Dietrich}
  M.,   {Warner} C.,  2003b, \mn@doi [\apj] {10.1086/345449}, \href
  {https://ui.adsabs.harvard.edu/abs/2003ApJ...583..649B} {583, 649}

\bibitem[\protect\citeauthoryear{{Baumgartner}, {Tueller}, {Markwardt},
  {Skinner}, {Barthelmy}, {Mushotzky}, {Evans}  \& {Gehrels}}{{Baumgartner}
  et~al.}{2013}]{Baumgartner2013}
{Baumgartner} W.~H.,  {Tueller} J.,  {Markwardt} C.~B.,  {Skinner} G.~K.,
  {Barthelmy} S.,  {Mushotzky} R.~F.,  {Evans} P.~A.,   {Gehrels} N.,  2013,
  \mn@doi [\apjs] {10.1088/0067-0049/207/2/19}, \href
  {https://ui.adsabs.harvard.edu/abs/2013ApJS..207...19B} {207, 19}

\bibitem[\protect\citeauthoryear{{Bennert}, {Jungwiert}, {Komossa}, {Haas}  \&
  {Chini}}{{Bennert} et~al.}{2006a}]{2006A&A...456..953B}
{Bennert} N.,  {Jungwiert} B.,  {Komossa} S.,  {Haas} M.,   {Chini} R.,  2006a,
  \mn@doi [\aap] {10.1051/0004-6361:20065319}, \href
  {https://ui.adsabs.harvard.edu/abs/2006A&A...456..953B} {456, 953}

\bibitem[\protect\citeauthoryear{{Bennert}, {Jungwiert}, {Komossa}, {Haas}  \&
  {Chini}}{{Bennert} et~al.}{2006b}]{2006A&A...459...55B}
{Bennert} N.,  {Jungwiert} B.,  {Komossa} S.,  {Haas} M.,   {Chini} R.,  2006b,
  \mn@doi [\aap] {10.1051/0004-6361:20065477}, \href
  {https://ui.adsabs.harvard.edu/abs/2006A&A...459...55B} {459, 55}

\bibitem[\protect\citeauthoryear{{Bentz}, {Hall}  \& {Osmer}}{{Bentz}
  et~al.}{2004}]{2004AJ....128..561B}
{Bentz} M.~C.,  {Hall} P.~B.,   {Osmer} P.~S.,  2004, \mn@doi [\aj]
  {10.1086/422346}, \href
  {https://ui.adsabs.harvard.edu/abs/2004AJ....128..561B} {128, 561}

\bibitem[\protect\citeauthoryear{{Berg}, {Pogge}, {Skillman}, {Croxall},
  {Moustakas}, {Rogers}  \& {Sun}}{{Berg} et~al.}{2020}]{2020ApJ...893...96B}
{Berg} D.~A.,  {Pogge} R.~W.,  {Skillman} E.~D.,  {Croxall} K.~V.,  {Moustakas}
  J.,  {Rogers} N. S.~J.,   {Sun} J.,  2020, \mn@doi [\apj]
  {10.3847/1538-4357/ab7eab}, \href
  {https://ui.adsabs.harvard.edu/abs/2020ApJ...893...96B} {893, 96}

\bibitem[\protect\citeauthoryear{{Berney} et~al.,}{{Berney}
  et~al.}{2015}]{Berney2015}
{Berney} S.,  et~al., 2015, \mn@doi [\mnras] {10.1093/mnras/stv2181}, \href
  {https://ui.adsabs.harvard.edu/abs/2015MNRAS.454.3622B} {454, 3622}

\bibitem[\protect\citeauthoryear{{Bian}, {Kewley}, {Dopita}  \& {Blanc}}{{Bian}
  et~al.}{2017}]{2017ApJ...834...51B}
{Bian} F.,  {Kewley} L.~J.,  {Dopita} M.~A.,   {Blanc} G.~A.,  2017, \mn@doi
  [\apj] {10.3847/1538-4357/834/1/51}, \href
  {https://ui.adsabs.harvard.edu/abs/2017ApJ...834...51B} {834, 51}

\bibitem[\protect\citeauthoryear{Bieri, Dubois, Silk, Mamon  \& Gaibler}{Bieri
  et~al.}{2016}]{Bieri+16}
Bieri R.,  Dubois Y.,  Silk J.,  Mamon G.~A.,   Gaibler V.,  2016, \mn@doi
  [Monthly Notices of the Royal Astronomical Society] {10.1093/mnras/stv2551},
  455, 4166

\bibitem[\protect\citeauthoryear{{Blanc}, {Kewley}, {Vogt}  \&
  {Dopita}}{{Blanc} et~al.}{2015}]{2015ApJ...798...99B}
{Blanc} G.~A.,  {Kewley} L.,  {Vogt} F. P.~A.,   {Dopita} M.~A.,  2015, \mn@doi
  [\apj] {10.1088/0004-637X/798/2/99}, \href
  {https://ui.adsabs.harvard.edu/abs/2015ApJ...798...99B} {798, 99}

\bibitem[\protect\citeauthoryear{{Boardman} et~al.,}{{Boardman}
  et~al.}{2022}]{2022MNRAS.514.2298B}
{Boardman} N.,  et~al., 2022, \mn@doi [\mnras] {10.1093/mnras/stac1475}, \href
  {https://ui.adsabs.harvard.edu/abs/2022MNRAS.514.2298B} {514, 2298}

\bibitem[\protect\citeauthoryear{{Bowen}}{{Bowen}}{1960}]{1960ApJ...132....1B}
{Bowen} I.~S.,  1960, \mn@doi [\apj] {10.1086/146893}, \href
  {https://ui.adsabs.harvard.edu/abs/1960ApJ...132....1B} {132, 1}

\bibitem[\protect\citeauthoryear{{Brownson}, {Belfiore}, {Maiolino}, {Lin}  \&
  {Carniani}}{{Brownson} et~al.}{2020}]{2020MNRAS.498L..66B}
{Brownson} S.,  {Belfiore} F.,  {Maiolino} R.,  {Lin} L.,   {Carniani} S.,
  2020, \mn@doi [\mnras] {10.1093/mnrasl/slaa128}, \href
  {https://ui.adsabs.harvard.edu/abs/2020MNRAS.498L..66B} {498, L66}

\bibitem[\protect\citeauthoryear{{Cardelli}, {Clayton}  \& {Mathis}}{{Cardelli}
  et~al.}{1989}]{cardelli89}
{Cardelli} J.~A.,  {Clayton} G.~C.,   {Mathis} J.~S.,  1989, \mn@doi [\apj]
  {10.1086/167900}, \href
  {https://ui.adsabs.harvard.edu/abs/1989ApJ...345..245C} {345, 245}

\bibitem[\protect\citeauthoryear{{Carvalho} et~al.,}{{Carvalho}
  et~al.}{2020}]{2020MNRAS.492.5675C}
{Carvalho} S.~P.,  et~al., 2020, \mn@doi [\mnras] {10.1093/mnras/staa193},
  \href {https://ui.adsabs.harvard.edu/abs/2020MNRAS.492.5675C} {492, 5675}

\bibitem[\protect\citeauthoryear{{Castro}, {Dors}, {Cardaci}  \&
  {H{\"a}gele}}{{Castro} et~al.}{2017}]{castro2017}
{Castro} C.~S.,  {Dors} O.~L.,  {Cardaci} M.~V.,   {H{\"a}gele} G.~F.,  2017,
  \mn@doi [\mnras] {10.1093/mnras/stx150}, \href
  {https://ui.adsabs.harvard.edu/abs/2017MNRAS.467.1507C} {467, 1507}

\bibitem[\protect\citeauthoryear{{Cerqueira-Campos}, {Rodr{\'\i}guez-Ardila},
  {Riffel}, {Marinello}, {Prieto}  \& {Dahmer-Hahn}}{{Cerqueira-Campos}
  et~al.}{2021}]{2021MNRAS.500.2666C}
{Cerqueira-Campos} F.~C.,  {Rodr{\'\i}guez-Ardila} A.,  {Riffel} R.,
  {Marinello} M.,  {Prieto} A.,   {Dahmer-Hahn} L.~G.,  2021, \mn@doi [\mnras]
  {10.1093/mnras/staa3320}, \href
  {https://ui.adsabs.harvard.edu/abs/2021MNRAS.500.2666C} {500, 2666}

\bibitem[\protect\citeauthoryear{{Chen} et~al.,}{{Chen}
  et~al.}{2019}]{2019MNRAS.489..855C}
{Chen} J.,  et~al., 2019, \mn@doi [\mnras] {10.1093/mnras/stz2183}, \href
  {https://ui.adsabs.harvard.edu/abs/2019MNRAS.489..855C} {489, 855}

\bibitem[\protect\citeauthoryear{{Cicone} et~al.,}{{Cicone}
  et~al.}{2014}]{2014A&A...562A..21C}
{Cicone} C.,  et~al., 2014, \mn@doi [\aap] {10.1051/0004-6361/201322464}, \href
  {https://ui.adsabs.harvard.edu/abs/2014A&A...562A..21C} {562, A21}

\bibitem[\protect\citeauthoryear{{Cid Fernandes}, {Stasi{\'n}ska},
  {Schlickmann}, {Mateus}, {Vale Asari}, {Schoenell}  \& {Sodr{\'e}}}{{Cid
  Fernandes} et~al.}{2010}]{2010MNRAS.403.1036C}
{Cid Fernandes} R.,  {Stasi{\'n}ska} G.,  {Schlickmann} M.~S.,  {Mateus} A.,
  {Vale Asari} N.,  {Schoenell} W.,   {Sodr{\'e}} L.,  2010, \mn@doi [\mnras]
  {10.1111/j.1365-2966.2009.16185.x}, \href
  {https://ui.adsabs.harvard.edu/abs/2010MNRAS.403.1036C} {403, 1036}

\bibitem[\protect\citeauthoryear{{Clemens}, {Crain}  \& {Anderson}}{{Clemens}
  et~al.}{2004}]{cemens04}
{Clemens} J.~C.,  {Crain} J.~A.,   {Anderson} R.,  2004, in {Moorwood} A.
  F.~M.,  {Iye} M.,  eds,  Society of Photo-Optical Instrumentation Engineers
  (SPIE) Conference Series Vol. 5492, Ground-based Instrumentation for
  Astronomy. pp 331--340, \mn@doi{10.1117/12.550069}

\bibitem[\protect\citeauthoryear{{Congiu} et~al.,}{{Congiu}
  et~al.}{2017}]{2017MNRAS.471..562C}
{Congiu} E.,  et~al., 2017, \mn@doi [\mnras] {10.1093/mnras/stx1628}, \href
  {https://ui.adsabs.harvard.edu/abs/2017MNRAS.471..562C} {471, 562}

\bibitem[\protect\citeauthoryear{{Curti}, {Cresci}, {Mannucci}, {Marconi},
  {Maiolino}  \& {Esposito}}{{Curti} et~al.}{2017}]{cur17}
{Curti} M.,  {Cresci} G.,  {Mannucci} F.,  {Marconi} A.,  {Maiolino} R.,
  {Esposito} S.,  2017, \mn@doi [\mnras] {10.1093/mnras/stw2766}, \href
  {http://adsabs.harvard.edu/abs/2017MNRAS.465.1384C} {465, 1384}

\bibitem[\protect\citeauthoryear{{Curti}, {Mannucci}, {Cresci}  \&
  {Maiolino}}{{Curti} et~al.}{2020}]{2020MNRAS.491..944C}
{Curti} M.,  {Mannucci} F.,  {Cresci} G.,   {Maiolino} R.,  2020, \mn@doi
  [\mnras] {10.1093/mnras/stz2910}, \href
  {https://ui.adsabs.harvard.edu/abs/2020MNRAS.491..944C} {491, 944}

\bibitem[\protect\citeauthoryear{{Davies} et~al.,}{{Davies}
  et~al.}{2020}]{2020MNRAS.498.4150D}
{Davies} R.,  et~al., 2020, \mn@doi [\mnras] {10.1093/mnras/staa2413}, \href
  {https://ui.adsabs.harvard.edu/abs/2020MNRAS.498.4150D} {498, 4150}

\bibitem[\protect\citeauthoryear{Davison \& Hinkley}{Davison \&
  Hinkley}{1997}]{davison1997bootstrap}
Davison A.,  Hinkley D.,  1997, Technical report, Bootstrap Methods and Their
  Application.
Cambridge University Press

\bibitem[\protect\citeauthoryear{{Dempsey} \& {Zakamska}}{{Dempsey} \&
  {Zakamska}}{2018}]{2018MNRAS.477.4615D}
{Dempsey} R.,  {Zakamska} N.~L.,  2018, \mn@doi [\mnras]
  {10.1093/mnras/sty941}, \href
  {https://ui.adsabs.harvard.edu/abs/2018MNRAS.477.4615D} {477, 4615}

\bibitem[\protect\citeauthoryear{{Denicol{\'o}}, {Terlevich}  \&
  {Terlevich}}{{Denicol{\'o}} et~al.}{2002}]{2002MNRAS.330...69D}
{Denicol{\'o}} G.,  {Terlevich} R.,   {Terlevich} E.,  2002, \mn@doi [\mnras]
  {10.1046/j.1365-8711.2002.05041.x}, \href
  {https://ui.adsabs.harvard.edu/abs/2002MNRAS.330...69D} {330, 69}

\bibitem[\protect\citeauthoryear{{D{\'\i}az}, {Terlevich}, {Castellanos}  \&
  {H{\"a}gele}}{{D{\'\i}az} et~al.}{2007}]{2007MNRAS.382..251D}
{D{\'\i}az} {\'A}.~I.,  {Terlevich} E.,  {Castellanos} M.,   {H{\"a}gele}
  G.~F.,  2007, \mn@doi [\mnras] {10.1111/j.1365-2966.2007.12351.x}, \href
  {https://ui.adsabs.harvard.edu/abs/2007MNRAS.382..251D} {382, 251}

\bibitem[\protect\citeauthoryear{{Dors}}{{Dors}}{2021}]{dors2021}
{Dors} O.~L.,  2021, \mn@doi [\mnras] {10.1093/mnras/stab2166}, \href
  {https://ui.adsabs.harvard.edu/abs/2021MNRAS.507..466D} {507, 466}

\bibitem[\protect\citeauthoryear{{Dors}, {Storchi-Bergmann}, {Riffel}  \&
  {Schimdt}}{{Dors} et~al.}{2008}]{2008A&A...482...59D}
{Dors} O.~L. J.,  {Storchi-Bergmann} T.,  {Riffel} R.~A.,   {Schimdt} A.~A.,
  2008, \mn@doi [\aap] {10.1051/0004-6361:20078960}, \href
  {https://ui.adsabs.harvard.edu/abs/2008A&A...482...59D} {482, 59}

\bibitem[\protect\citeauthoryear{{Dors}, {Cardaci}, {H{\"a}gele}  \&
  {Krabbe}}{{Dors} et~al.}{2014}]{dors2014}
{Dors} O.~L.,  {Cardaci} M.~V.,  {H{\"a}gele} G.~F.,   {Krabbe} {\^A}.~C.,
  2014, \mn@doi [\mnras] {10.1093/mnras/stu1218}, \href
  {https://ui.adsabs.harvard.edu/abs/2014MNRAS.443.1291D} {443, 1291}

\bibitem[\protect\citeauthoryear{{Dors}, {Monteiro}, {Cardaci}, {H{\"a}gele}
  \& {Krabbe}}{{Dors} et~al.}{2019}]{2019MNRAS.486.5853D}
{Dors} O.~L.,  {Monteiro} A.~F.,  {Cardaci} M.~V.,  {H{\"a}gele} G.~F.,
  {Krabbe} A.~C.,  2019, \mn@doi [\mnras] {10.1093/mnras/stz1242}, \href
  {https://ui.adsabs.harvard.edu/abs/2019MNRAS.486.5853D} {486, 5853}

\bibitem[\protect\citeauthoryear{{Dors} et~al.,}{{Dors}
  et~al.}{2020a}]{2020MNRAS.492..468D}
{Dors} O.~L.,  et~al., 2020a, \mn@doi [\mnras] {10.1093/mnras/stz3492}, \href
  {https://ui.adsabs.harvard.edu/abs/2020MNRAS.492..468D} {492, 468}

\bibitem[\protect\citeauthoryear{{Dors}, {Maiolino}, {Cardaci}, {H{\"a}gele},
  {Krabbe}, {P{\'e}rez-Montero}  \& {Armah}}{{Dors}
  et~al.}{2020b}]{2020MNRAS.496.3209D}
{Dors} O.~L.,  {Maiolino} R.,  {Cardaci} M.~V.,  {H{\"a}gele} G.~F.,  {Krabbe}
  A.~C.,  {P{\'e}rez-Montero} E.,   {Armah} M.,  2020b, \mn@doi [\mnras]
  {10.1093/mnras/staa1781}, \href
  {https://ui.adsabs.harvard.edu/abs/2020MNRAS.496.3209D} {496, 3209}

\bibitem[\protect\citeauthoryear{{Dors} et~al.,}{{Dors}
  et~al.}{2022}]{2022MNRAS.514.5506D}
{Dors} O.~L.,  et~al., 2022, \mn@doi [\mnras] {10.1093/mnras/stac1722}, \href
  {https://ui.adsabs.harvard.edu/abs/2022MNRAS.514.5506D} {514, 5506}

\bibitem[\protect\citeauthoryear{{El-Badry}, Wetzel, Geha, Hopkins, Kere{\v s},
  Chan  \& {Faucher-Gigu{\`e}re}}{{El-Badry} et~al.}{2016}]{El-Badry+16}
{El-Badry} K.,  Wetzel A.,  Geha M.,  Hopkins P.~F.,  Kere{\v s} D.,  Chan
  T.~K.,   {Faucher-Gigu{\`e}re} C.-A.,  2016, \mn@doi [The Astrophysical
  Journal] {10.3847/0004-637X/820/2/131}, 820, 131

\bibitem[\protect\citeauthoryear{{Elbaz}, {Jahnke}, {Pantin}, {Le Borgne}  \&
  {Letawe}}{{Elbaz} et~al.}{2009}]{elbaz09}
{Elbaz} D.,  {Jahnke} K.,  {Pantin} E.,  {Le Borgne} D.,   {Letawe} G.,  2009,
  \mn@doi [\aap] {10.1051/0004-6361/200912848}, \href
  {https://ui.adsabs.harvard.edu/abs/2009A&A...507.1359E} {507, 1359}

\bibitem[\protect\citeauthoryear{Fabian}{Fabian}{2012}]{Fabian+12}
Fabian A.~C.,  2012, \mn@doi [Annual Review of Astronomy and Astrophysics]
  {10.1146/annurev-astro-081811-125521}, 50, 455

\bibitem[\protect\citeauthoryear{{Feltre}, {Charlot}  \& {Gutkin}}{{Feltre}
  et~al.}{2016}]{feltre2016nuclear}
{Feltre} A.,  {Charlot} S.,   {Gutkin} J.,  2016, \mn@doi [\mnras]
  {10.1093/mnras/stv2794}, \href
  {https://ui.adsabs.harvard.edu/abs/2016MNRAS.456.3354F} {456, 3354}

\bibitem[\protect\citeauthoryear{{Ferland} \& {Netzer}}{{Ferland} \&
  {Netzer}}{1983}]{ferland1983shock}
{Ferland} G.~J.,  {Netzer} H.,  1983, \mn@doi [\apj] {10.1086/160577}, \href
  {https://ui.adsabs.harvard.edu/abs/1983ApJ...264..105F} {264, 105}

\bibitem[\protect\citeauthoryear{{Ferland} et~al.,}{{Ferland}
  et~al.}{2017}]{2017RMxAA..53..385F}
{Ferland} G.~J.,  et~al., 2017, \rmxaa, \href
  {https://ui.adsabs.harvard.edu/abs/2017RMxAA..53..385F} {53, 385}

\bibitem[\protect\citeauthoryear{{Freitas} et~al.,}{{Freitas}
  et~al.}{2018}]{2018MNRAS.476.2760F}
{Freitas} I.~C.,  et~al., 2018, \mn@doi [\mnras] {10.1093/mnras/sty303}, \href
  {https://ui.adsabs.harvard.edu/abs/2018MNRAS.476.2760F} {476, 2760}

\bibitem[\protect\citeauthoryear{{Gallagher}, {Maiolino}, {Belfiore}, {Drory},
  {Riffel}  \& {Riffel}}{{Gallagher} et~al.}{2019}]{2019MNRAS.485.3409G}
{Gallagher} R.,  {Maiolino} R.,  {Belfiore} F.,  {Drory} N.,  {Riffel} R.,
  {Riffel} R.~A.,  2019, \mn@doi [\mnras] {10.1093/mnras/stz564}, \href
  {https://ui.adsabs.harvard.edu/abs/2019MNRAS.485.3409G} {485, 3409}

\bibitem[\protect\citeauthoryear{{Garnica}, {Negrete}, {Marziani}, {Dultzin},
  {{\'S}niegowska}  \& {Panda}}{{Garnica} et~al.}{2022}]{2022A&A...667A.105G}
{Garnica} K.,  {Negrete} C.~A.,  {Marziani} P.,  {Dultzin} D.,
  {{\'S}niegowska} M.,   {Panda} S.,  2022, \mn@doi [\aap]
  {10.1051/0004-6361/202142837}, \href
  {https://ui.adsabs.harvard.edu/abs/2022A&A...667A.105G} {667, A105}

\bibitem[\protect\citeauthoryear{{Gaskell}}{{Gaskell}}{1982}]{1982PASP...94..891G}
{Gaskell} C.~M.,  1982, \mn@doi [\pasp] {10.1086/131080}, \href
  {https://ui.adsabs.harvard.edu/abs/1982PASP...94..891G} {94, 891}

\bibitem[\protect\citeauthoryear{{Gaskell}}{{Gaskell}}{1984}]{gaskell1984red}
{Gaskell} C.~M.,  1984, \aplett, \href
  {https://ui.adsabs.harvard.edu/abs/1984ApL....24...43G} {24, 43}

\bibitem[\protect\citeauthoryear{{Gaskell} \& {Ferland}}{{Gaskell} \&
  {Ferland}}{1984}]{gaskell1984theo}
{Gaskell} C.~M.,  {Ferland} G.~J.,  1984, \mn@doi [\pasp] {10.1086/131352},
  \href {https://ui.adsabs.harvard.edu/abs/1984PASP...96..393G} {96, 393}

\bibitem[\protect\citeauthoryear{{Gehrels} et~al.,}{{Gehrels}
  et~al.}{2004}]{Gehrels2004}
{Gehrels} N.,  et~al., 2004, \mn@doi [\apj] {10.1086/422091}, \href
  {https://ui.adsabs.harvard.edu/abs/2004ApJ...611.1005G} {611, 1005}

\bibitem[\protect\citeauthoryear{{Graham}}{{Graham}}{2016}]{2016ASSL..418..263G}
{Graham} A.~W.,  2016, in {Laurikainen} E.,  {Peletier} R.,   {Gadotti} D.,
  eds,  Astrophysics and Space Science Library Vol. 418, Galactic Bulges.
  p.~263 (\mn@eprint {arXiv} {1501.02937}),
  \mn@doi{10.1007/978-3-319-19378-6_11}

\bibitem[\protect\citeauthoryear{Granato, De~Zotti, Silva, Bressan  \&
  Danese}{Granato et~al.}{2004}]{Granato+04}
Granato G.~L.,  De~Zotti G.,  Silva L.,  Bressan A.,   Danese L.,  2004,
  \mn@doi [The Astrophysical Journal] {10.1086/379875}, 600, 580

\bibitem[\protect\citeauthoryear{{H{\"a}gele}, {D{\'\i}az}, {Terlevich},
  {Terlevich}, {P{\'e}rez-Montero}  \& {Cardaci}}{{H{\"a}gele}
  et~al.}{2008}]{2008MNRAS.383..209H}
{H{\"a}gele} G.~F.,  {D{\'\i}az} {\'A}.~I.,  {Terlevich} E.,  {Terlevich} R.,
  {P{\'e}rez-Montero} E.,   {Cardaci} M.~V.,  2008, \mn@doi [\mnras]
  {10.1111/j.1365-2966.2007.12527.x}, \href
  {https://ui.adsabs.harvard.edu/abs/2008MNRAS.383..209H} {383, 209}

\bibitem[\protect\citeauthoryear{{Halpern}}{{Halpern}}{1982}]{halpern1982xray}
{Halpern} J.~P.,  1982, PhD thesis, Harvard University, Cambridge, MA.

\bibitem[\protect\citeauthoryear{{Halpern} \& {Steiner}}{{Halpern} \&
  {Steiner}}{1983}]{halpern1983ionization}
{Halpern} J.~P.,  {Steiner} J.~E.,  1983, \mn@doi [\apjl] {10.1086/184051},
  \href {https://ui.adsabs.harvard.edu/abs/1983ApJ...269L..37H} {269, L37}

\bibitem[\protect\citeauthoryear{{Hamann} \& {Ferland}}{{Hamann} \&
  {Ferland}}{1993}]{1993ApJ...418...11H}
{Hamann} F.,  {Ferland} G.,  1993, \mn@doi [\apj] {10.1086/173366}, \href
  {https://ui.adsabs.harvard.edu/abs/1993ApJ...418...11H} {418, 11}

\bibitem[\protect\citeauthoryear{{Heckman}, {Miley}, {van Breugel}  \&
  {Butcher}}{{Heckman} et~al.}{1981}]{1981ApJ...247..403H}
{Heckman} T.~M.,  {Miley} G.~K.,  {van Breugel} W.~J.~M.,   {Butcher} H.~R.,
  1981, \mn@doi [\apj] {10.1086/159050}, \href
  {https://ui.adsabs.harvard.edu/abs/1981ApJ...247..403H} {247, 403}

\bibitem[\protect\citeauthoryear{{Ho}}{{Ho}}{2008}]{2008ARA&A..46..475H}
{Ho} L.~C.,  2008, \mn@doi [\araa] {10.1146/annurev.astro.45.051806.110546},
  \href {https://ui.adsabs.harvard.edu/abs/2008ARA&A..46..475H} {46, 475}

\bibitem[\protect\citeauthoryear{Hopkins}{Hopkins}{2012}]{Hopkins+12}
Hopkins P.~F.,  2012, \mn@doi [Monthly Notices of the RAS]
  {10.1111/j.1745-3933.2011.01179.x}, \href
  {http://adsabs.harvard.edu/abs/2012MNRAS.420L...8H} {420, L8}

\bibitem[\protect\citeauthoryear{{Hopkins} \& {Elvis}}{{Hopkins} \&
  {Elvis}}{2010}]{hopkins10}
{Hopkins} P.~F.,  {Elvis} M.,  2010, \mn@doi [\mnras]
  {10.1111/j.1365-2966.2009.15643.x}, \href
  {https://ui.adsabs.harvard.edu/abs/2010MNRAS.401....7H} {401, 7}

\bibitem[\protect\citeauthoryear{{Horne} et~al.,}{{Horne}
  et~al.}{2021}]{2021ApJ...907...76H}
{Horne} K.,  et~al., 2021, \mn@doi [\apj] {10.3847/1538-4357/abce60}, \href
  {https://ui.adsabs.harvard.edu/abs/2021ApJ...907...76H} {907, 76}

\bibitem[\protect\citeauthoryear{{Huang} et~al.,}{{Huang}
  et~al.}{2019}]{2019ApJ...886...31H}
{Huang} C.,  et~al., 2019, \mn@doi [\apj] {10.3847/1538-4357/ab4902}, \href
  {https://ui.adsabs.harvard.edu/abs/2019ApJ...886...31H} {886, 31}

\bibitem[\protect\citeauthoryear{Ishibashi \& Fabian}{Ishibashi \&
  Fabian}{2012}]{Ishibashi+12}
Ishibashi W.,  Fabian A.~C.,  2012, \mn@doi [Monthly Notices of the Royal
  Astronomical Society] {10.1111/j.1365-2966.2012.22074.x}, 427, 2998

\bibitem[\protect\citeauthoryear{{Jensen}, {Strom}  \& {Strom}}{{Jensen}
  et~al.}{1976}]{1976ApJ...209..748J}
{Jensen} E.~B.,  {Strom} K.~M.,   {Strom} S.~E.,  1976, \mn@doi [\apj]
  {10.1086/154773}, \href
  {https://ui.adsabs.harvard.edu/abs/1976ApJ...209..748J} {209, 748}

\bibitem[\protect\citeauthoryear{{Ji}, {Yan}, {Riffel}, {Drory}  \&
  {Zhang}}{{Ji} et~al.}{2020}]{2020MNRAS.496.1262J}
{Ji} X.,  {Yan} R.,  {Riffel} R.,  {Drory} N.,   {Zhang} K.,  2020, \mn@doi
  [\mnras] {10.1093/mnras/staa1521}, \href
  {https://ui.adsabs.harvard.edu/abs/2020MNRAS.496.1262J} {496, 1262}

\bibitem[\protect\citeauthoryear{{Jiang}, {Malhotra}, {Rhoads}  \&
  {Yang}}{{Jiang} et~al.}{2019}]{2019ApJ...872..145J}
{Jiang} T.,  {Malhotra} S.,  {Rhoads} J.~E.,   {Yang} H.,  2019, \mn@doi [\apj]
  {10.3847/1538-4357/aaee8a}, \href
  {https://ui.adsabs.harvard.edu/abs/2019ApJ...872..145J} {872, 145}

\bibitem[\protect\citeauthoryear{{Kakkad} et~al.,}{{Kakkad}
  et~al.}{2018}]{2018A&A...618A...6K}
{Kakkad} D.,  et~al., 2018, \mn@doi [\aap] {10.1051/0004-6361/201832790}, \href
  {https://ui.adsabs.harvard.edu/abs/2018A&A...618A...6K} {618, A6}

\bibitem[\protect\citeauthoryear{{Kakkad} et~al.,}{{Kakkad}
  et~al.}{2022}]{kakkad2022}
{Kakkad} D.,  et~al., 2022, \mn@doi [\mnras] {10.1093/mnras/stac103}, \href
  {https://ui.adsabs.harvard.edu/abs/2022MNRAS.511.2105K} {511, 2105}

\bibitem[\protect\citeauthoryear{{Kaspi}, {Smith}, {Netzer}, {Maoz}, {Jannuzi}
  \& {Giveon}}{{Kaspi} et~al.}{2000}]{2000ApJ...533..631K}
{Kaspi} S.,  {Smith} P.~S.,  {Netzer} H.,  {Maoz} D.,  {Jannuzi} B.~T.,
  {Giveon} U.,  2000, \mn@doi [\apj] {10.1086/308704}, \href
  {https://ui.adsabs.harvard.edu/abs/2000ApJ...533..631K} {533, 631}

\bibitem[\protect\citeauthoryear{{Kawamuro}, {Ricci}, {Izumi}, {Imanishi},
  {Baba}, {Nguyen}  \& {Onishi}}{{Kawamuro} et~al.}{2021}]{2021ApJS..257...64K}
{Kawamuro} T.,  {Ricci} C.,  {Izumi} T.,  {Imanishi} M.,  {Baba} S.,  {Nguyen}
  D.~D.,   {Onishi} K.,  2021, \mn@doi [\apjs] {10.3847/1538-4365/ac2891},
  \href {https://ui.adsabs.harvard.edu/abs/2021ApJS..257...64K} {257, 64}

\bibitem[\protect\citeauthoryear{{Kennicutt}, {Bresolin}  \&
  {Garnett}}{{Kennicutt} et~al.}{2003}]{2003ApJ...591..801K}
{Kennicutt} Robert~C. J.,  {Bresolin} F.,   {Garnett} D.~R.,  2003, \mn@doi
  [\apj] {10.1086/375398}, \href
  {https://ui.adsabs.harvard.edu/abs/2003ApJ...591..801K} {591, 801}

\bibitem[\protect\citeauthoryear{{Kewley} \& {Ellison}}{{Kewley} \&
  {Ellison}}{2008}]{kew08}
{Kewley} L.~J.,  {Ellison} S.~L.,  2008, \mn@doi [\apj] {10.1086/587500}, \href
  {http://adsabs.harvard.edu/abs/2008ApJ...681.1183K} {681, 1183}

\bibitem[\protect\citeauthoryear{{Kewley}, {Dopita}, {Sutherland}, {Heisler}
  \& {Trevena}}{{Kewley} et~al.}{2001}]{kewley2001}
{Kewley} L.~J.,  {Dopita} M.~A.,  {Sutherland} R.~S.,  {Heisler} C.~A.,
  {Trevena} J.,  2001, \mn@doi [\apj] {10.1086/321545}, \href
  {https://ui.adsabs.harvard.edu/abs/2001ApJ...556..121K} {556, 121}

\bibitem[\protect\citeauthoryear{{Kewley}, {Jansen}  \& {Geller}}{{Kewley}
  et~al.}{2005}]{2005PASP..117..227K}
{Kewley} L.~J.,  {Jansen} R.~A.,   {Geller} M.~J.,  2005, \mn@doi [\pasp]
  {10.1086/428303}, \href
  {https://ui.adsabs.harvard.edu/abs/2005PASP..117..227K} {117, 227}

\bibitem[\protect\citeauthoryear{{Kewley}, {Groves}, {Kauffmann}  \&
  {Heckman}}{{Kewley} et~al.}{2006}]{kewley2006}
{Kewley} L.~J.,  {Groves} B.,  {Kauffmann} G.,   {Heckman} T.,  2006, \mn@doi
  [\mnras] {10.1111/j.1365-2966.2006.10859.x}, \href
  {https://ui.adsabs.harvard.edu/abs/2006MNRAS.372..961K} {372, 961}

\bibitem[\protect\citeauthoryear{King \& Pounds}{King \&
  Pounds}{2015}]{King+15}
King A.,  Pounds K.,  2015, \mn@doi [Annual Review of Astronomy and
  Astrophysics] {10.1146/annurev-astro-082214-122316}, 53, 115

\bibitem[\protect\citeauthoryear{{Kobulnicky} \& {Kewley}}{{Kobulnicky} \&
  {Kewley}}{2004}]{2004ApJ...617..240K}
{Kobulnicky} H.~A.,  {Kewley} L.~J.,  2004, \mn@doi [\apj] {10.1086/425299},
  \href {https://ui.adsabs.harvard.edu/abs/2004ApJ...617..240K} {617, 240}

\bibitem[\protect\citeauthoryear{{Kormendy} \& {Ho}}{{Kormendy} \&
  {Ho}}{2013}]{2013ARA&A..51..511K}
{Kormendy} J.,  {Ho} L.~C.,  2013, \mn@doi [\araa]
  {10.1146/annurev-astro-082708-101811}, \href
  {https://ui.adsabs.harvard.edu/abs/2013ARA&A..51..511K} {51, 511}

\bibitem[\protect\citeauthoryear{{Koss} et~al.,}{{Koss}
  et~al.}{2016}]{2016ApJ...825...85K}
{Koss} M.~J.,  et~al., 2016, \mn@doi [\apj] {10.3847/0004-637X/825/2/85}, \href
  {https://ui.adsabs.harvard.edu/abs/2016ApJ...825...85K} {825, 85}

\bibitem[\protect\citeauthoryear{{Koss} et~al.,}{{Koss} et~al.}{2017}]{koss17}
{Koss} M.,  et~al., 2017, \mn@doi [\apj] {10.3847/1538-4357/aa8ec9}, \href
  {https://ui.adsabs.harvard.edu/abs/2017ApJ...850...74K} {850, 74}

\bibitem[\protect\citeauthoryear{{Koss} et~al.,}{{Koss}
  et~al.}{2021}]{koss2021}
{Koss} M.~J.,  et~al., 2021, \mn@doi [\apjs] {10.3847/1538-4365/abcbfe}, \href
  {https://ui.adsabs.harvard.edu/abs/2021ApJS..252...29K} {252, 29}

\bibitem[\protect\citeauthoryear{{Koss} et~al.,}{{Koss}
  et~al.}{2022a}]{2022ApJS..261....2K}
{Koss} M.~J.,  et~al., 2022a, \mn@doi [\apjs] {10.3847/1538-4365/ac6c05}, \href
  {https://ui.adsabs.harvard.edu/abs/2022ApJS..261....2K} {261, 2}

\bibitem[\protect\citeauthoryear{{Koss} et~al.,}{{Koss}
  et~al.}{2022b}]{2022ApJS..261....6K}
{Koss} M.~J.,  et~al., 2022b, \mn@doi [\apjs] {10.3847/1538-4365/ac650b}, \href
  {https://ui.adsabs.harvard.edu/abs/2022ApJS..261....6K} {261, 6}

\bibitem[\protect\citeauthoryear{{Kraemer}, {Wu}, {Crenshaw}  \&
  {Harrington}}{{Kraemer} et~al.}{1994}]{kraemer1994spectra}
{Kraemer} S.~B.,  {Wu} C.-C.,  {Crenshaw} D.~M.,   {Harrington} J.~P.,  1994,
  \mn@doi [\apj] {10.1086/174803}, \href
  {https://ui.adsabs.harvard.edu/abs/1994ApJ...435..171K} {435, 171}

\bibitem[\protect\citeauthoryear{{Lamperti} et~al.,}{{Lamperti}
  et~al.}{2017}]{Lamperti2017}
{Lamperti} I.,  et~al., 2017, \mn@doi [\mnras] {10.1093/mnras/stx055}, \href
  {https://ui.adsabs.harvard.edu/abs/2017MNRAS.467..540L} {467, 540}

\bibitem[\protect\citeauthoryear{{Landt} et~al.,}{{Landt}
  et~al.}{2019}]{2019MNRAS.489.1572L}
{Landt} H.,  et~al., 2019, \mn@doi [\mnras] {10.1093/mnras/stz2212}, \href
  {https://ui.adsabs.harvard.edu/abs/2019MNRAS.489.1572L} {489, 1572}

\bibitem[\protect\citeauthoryear{{Lilly}, {Carollo}, {Pipino}, {Renzini}  \&
  {Peng}}{{Lilly} et~al.}{2013}]{2013ApJ...772..119L}
{Lilly} S.~J.,  {Carollo} C.~M.,  {Pipino} A.,  {Renzini} A.,   {Peng} Y.,
  2013, \mn@doi [\apj] {10.1088/0004-637X/772/2/119}, \href
  {https://ui.adsabs.harvard.edu/abs/2013ApJ...772..119L} {772, 119}

\bibitem[\protect\citeauthoryear{{Liu} et~al.,}{{Liu} et~al.}{2020}]{liu2020}
{Liu} T.,  et~al., 2020, \mn@doi [\apj] {10.3847/1538-4357/ab952d}, \href
  {https://ui.adsabs.harvard.edu/abs/2020ApJ...896..122L} {896, 122}

\bibitem[\protect\citeauthoryear{{Lopez-Sanchez} \& {Esteban}}{{Lopez-Sanchez}
  \& {Esteban}}{2010}]{2010arXiv1004.5251L}
{Lopez-Sanchez} A.~R.,  {Esteban} C.,  2010, arXiv e-prints, \href
  {https://ui.adsabs.harvard.edu/abs/2010arXiv1004.5251L} {p. arXiv:1004.5251}

\bibitem[\protect\citeauthoryear{{Lu}, {Zhao}, {Bai}  \& {Fan}}{{Lu}
  et~al.}{2019}]{2019MNRAS.483.1722L}
{Lu} K.-X.,  {Zhao} Y.,  {Bai} J.-M.,   {Fan} X.-L.,  2019, \mn@doi [\mnras]
  {10.1093/mnras/sty3229}, \href
  {https://ui.adsabs.harvard.edu/abs/2019MNRAS.483.1722L} {483, 1722}

\bibitem[\protect\citeauthoryear{{Luridiana}, {Morisset}  \&
  {Shaw}}{{Luridiana} et~al.}{2015}]{2015A&A...573A..42L}
{Luridiana} V.,  {Morisset} C.,   {Shaw} R.~A.,  2015, \mn@doi [\aap]
  {10.1051/0004-6361/201323152}, \href
  {https://ui.adsabs.harvard.edu/abs/2015A&A...573A..42L} {573, A42}

\bibitem[\protect\citeauthoryear{{Maiolino} et~al.,}{{Maiolino}
  et~al.}{2008}]{maiolino08}
{Maiolino} R.,  et~al., 2008, \mn@doi [\aap] {10.1051/0004-6361:200809678},
  \href {https://ui.adsabs.harvard.edu/abs/2008A&A...488..463M} {488, 463}

\bibitem[\protect\citeauthoryear{{Maiolino} et~al.,}{{Maiolino}
  et~al.}{2017}]{2017Natur.544..202M}
{Maiolino} R.,  et~al., 2017, \mn@doi [\nat] {10.1038/nature21677}, \href
  {https://ui.adsabs.harvard.edu/abs/2017Natur.544..202M} {544, 202}

\bibitem[\protect\citeauthoryear{{Marino} et~al.,}{{Marino}
  et~al.}{2013}]{mar13}
{Marino} R.~A.,  et~al., 2013, \mn@doi [\aap] {10.1051/0004-6361/201321956},
  \href {https://ui.adsabs.harvard.edu/abs/2013A\%26A...559A.114M} {559, A114}

\bibitem[\protect\citeauthoryear{{Matsuoka}, {Nagao}, {Maiolino}, {Marconi}  \&
  {Taniguchi}}{{Matsuoka} et~al.}{2009}]{2009A&A...503..721M}
{Matsuoka} K.,  {Nagao} T.,  {Maiolino} R.,  {Marconi} A.,   {Taniguchi} Y.,
  2009, \mn@doi [\aap] {10.1051/0004-6361/200811478}, \href
  {https://ui.adsabs.harvard.edu/abs/2009A&A...503..721M} {503, 721}

\bibitem[\protect\citeauthoryear{{Matsuoka}, {Nagao}, {Marconi}, {Maiolino}  \&
  {Taniguchi}}{{Matsuoka} et~al.}{2011}]{2011A&A...527A.100M}
{Matsuoka} K.,  {Nagao} T.,  {Marconi} A.,  {Maiolino} R.,   {Taniguchi} Y.,
  2011, \mn@doi [\aap] {10.1051/0004-6361/201015584}, \href
  {https://ui.adsabs.harvard.edu/abs/2011A&A...527A.100M} {527, A100}

\bibitem[\protect\citeauthoryear{{Matsuoka}, {Nagao}, {Marconi}, {Maiolino},
  {Mannucci}, {Cresci}, {Terao}  \& {Ikeda}}{{Matsuoka}
  et~al.}{2018}]{2018A&A...616L...4M}
{Matsuoka} K.,  {Nagao} T.,  {Marconi} A.,  {Maiolino} R.,  {Mannucci} F.,
  {Cresci} G.,  {Terao} K.,   {Ikeda} H.,  2018, \mn@doi [\aap]
  {10.1051/0004-6361/201833418}, \href
  {https://ui.adsabs.harvard.edu/abs/2018A&A...616L...4M} {616, L4}

\bibitem[\protect\citeauthoryear{{Mej{\'\i}a-Restrepo}, {Trakhtenbrot}, {Lira},
  {Netzer}  \& {Capellupo}}{{Mej{\'\i}a-Restrepo}
  et~al.}{2016}]{2016MNRAS.460..187M}
{Mej{\'\i}a-Restrepo} J.~E.,  {Trakhtenbrot} B.,  {Lira} P.,  {Netzer} H.,
  {Capellupo} D.~M.,  2016, \mn@doi [\mnras] {10.1093/mnras/stw568}, \href
  {https://ui.adsabs.harvard.edu/abs/2016MNRAS.460..187M} {460, 187}

\bibitem[\protect\citeauthoryear{{Mingozzi} et~al.,}{{Mingozzi}
  et~al.}{2019}]{2019A&A...622A.146M}
{Mingozzi} M.,  et~al., 2019, \mn@doi [\aap] {10.1051/0004-6361/201834372},
  \href {https://ui.adsabs.harvard.edu/abs/2019A&A...622A.146M} {622, A146}

\bibitem[\protect\citeauthoryear{{Monteiro} \& {Dors}}{{Monteiro} \&
  {Dors}}{2021}]{2021MNRAS.508.3023M}
{Monteiro} A.~F.,  {Dors} O.~L.,  2021, \mn@doi [\mnras]
  {10.1093/mnras/stab2750}, \href
  {https://ui.adsabs.harvard.edu/abs/2021MNRAS.508.3023M} {508, 3023}

\bibitem[\protect\citeauthoryear{{Moore}}{{Moore}}{1945}]{1945CoPri..20....1M}
{Moore} C.~E.,  1945, Contributions from the Princeton University Observatory,
  \href {https://ui.adsabs.harvard.edu/abs/1945CoPri..20....1M} {20, 1}

\bibitem[\protect\citeauthoryear{{Nagao}, {Maiolino}  \& {Marconi}}{{Nagao}
  et~al.}{2006}]{2006A&A...447..863N}
{Nagao} T.,  {Maiolino} R.,   {Marconi} A.,  2006, \mn@doi [\aap]
  {10.1051/0004-6361:20054127}, \href
  {https://ui.adsabs.harvard.edu/abs/2006A&A...447..863N} {447, 863}

\bibitem[\protect\citeauthoryear{Nayakshin \& Zubovas}{Nayakshin \&
  Zubovas}{2012}]{Nayakshin+12}
Nayakshin S.,  Zubovas K.,  2012, \mn@doi [Monthly Notices of the Royal
  Astronomical Society] {10.1111/j.1365-2966.2012.21950.x}, 427, 372

\bibitem[\protect\citeauthoryear{{Netzer}, {Shemmer}, {Maiolino}, {Oliva},
  {Croom}, {Corbett}  \& {di Fabrizio}}{{Netzer}
  et~al.}{2004}]{2004ApJ...614..558N}
{Netzer} H.,  {Shemmer} O.,  {Maiolino} R.,  {Oliva} E.,  {Croom} S.,
  {Corbett} E.,   {di Fabrizio} L.,  2004, \mn@doi [\apj] {10.1086/423608},
  \href {https://ui.adsabs.harvard.edu/abs/2004ApJ...614..558N} {614, 558}

\bibitem[\protect\citeauthoryear{{Nicholls}, {Kewley}  \&
  {Sutherland}}{{Nicholls} et~al.}{2020}]{2020PASP..132c3001N}
{Nicholls} D.~C.,  {Kewley} L.~J.,   {Sutherland} R.~S.,  2020, \mn@doi [\pasp]
  {10.1088/1538-3873/ab6818}, \href
  {https://ui.adsabs.harvard.edu/abs/2020PASP..132c3001N} {132, 033001}

\bibitem[\protect\citeauthoryear{{Oh} et~al.,}{{Oh} et~al.}{2017}]{Oh2017}
{Oh} K.,  et~al., 2017, \mn@doi [\mnras] {10.1093/mnras/stw2467}, \href
  {https://ui.adsabs.harvard.edu/abs/2017MNRAS.464.1466O} {464, 1466}

\bibitem[\protect\citeauthoryear{{Oh}, {Ueda}, {Akiyama}, {Suh}, {Koss},
  {Kashino}  \& {Hasinger}}{{Oh} et~al.}{2019}]{Oh2019}
{Oh} K.,  {Ueda} Y.,  {Akiyama} M.,  {Suh} H.,  {Koss} M.~J.,  {Kashino} D.,
  {Hasinger} G.,  2019, \mn@doi [\apj] {10.3847/1538-4357/ab288b}, \href
  {https://ui.adsabs.harvard.edu/abs/2019ApJ...880..112O} {880, 112}

\bibitem[\protect\citeauthoryear{{Oh} et~al.,}{{Oh}
  et~al.}{2022}]{2022ApJS..261....4O}
{Oh} K.,  et~al., 2022, \mn@doi [\apjs] {10.3847/1538-4365/ac5b68}, \href
  {https://ui.adsabs.harvard.edu/abs/2022ApJS..261....4O} {261, 4}

\bibitem[\protect\citeauthoryear{{Oke} et~al.,}{{Oke} et~al.}{1995}]{oke95}
{Oke} J.~B.,  et~al., 1995, \mn@doi [\pasp] {10.1086/133562}, \href
  {https://ui.adsabs.harvard.edu/abs/1995PASP..107..375O} {107, 375}

\bibitem[\protect\citeauthoryear{{Osterbrock}}{{Osterbrock}}{1981}]{1981ApJ...249..462O}
{Osterbrock} D.~E.,  1981, \mn@doi [\apj] {10.1086/159306}, \href
  {https://ui.adsabs.harvard.edu/abs/1981ApJ...249..462O} {249, 462}

\bibitem[\protect\citeauthoryear{{Osterbrock}}{{Osterbrock}}{1989}]{ost1989}
{Osterbrock} D.~E.,  1989, {Astrophysics of gaseous nebulae and active galactic
  nuclei}.
UNIVERSITY SCIENCE BOOKS

\bibitem[\protect\citeauthoryear{{Osterbrock} \& {Ferland}}{{Osterbrock} \&
  {Ferland}}{2006}]{ost06}
{Osterbrock} D.~E.,  {Ferland} G.~J.,  2006, {Astrophysics of gaseous nebulae
  and active galactic nuclei}.
UNIVERSITY SCIENCE BOOKS

\bibitem[\protect\citeauthoryear{Owen}{Owen}{2007}]{owen2007robust}
Owen A.~B.,  2007, Contemporary Mathematics, 443, 59

\bibitem[\protect\citeauthoryear{{Pagel}, {Edmunds}, {Blackwell}, {Chun}  \&
  {Smith}}{{Pagel} et~al.}{1979}]{pagel1979composition}
{Pagel} B.~E.~J.,  {Edmunds} M.~G.,  {Blackwell} D.~E.,  {Chun} M.~S.,
  {Smith} G.,  1979, \mn@doi [\mnras] {10.1093/mnras/189.1.95}, \href
  {https://ui.adsabs.harvard.edu/abs/1979MNRAS.189...95P} {189, 95}

\bibitem[\protect\citeauthoryear{{Peimbert}, {Peimbert}  \&
  {Delgado-Inglada}}{{Peimbert} et~al.}{2017}]{2017PASP..129h2001P}
{Peimbert} M.,  {Peimbert} A.,   {Delgado-Inglada} G.,  2017, \mn@doi [\pasp]
  {10.1088/1538-3873/aa72c3}, \href
  {https://ui.adsabs.harvard.edu/abs/2017PASP..129h2001P} {129, 082001}

\bibitem[\protect\citeauthoryear{{P{\'e}rez-Montero}}{{P{\'e}rez-Montero}}{2017}]{2017PASP..129d3001P}
{P{\'e}rez-Montero} E.,  2017, \mn@doi [\pasp] {10.1088/1538-3873/aa5abb},
  \href {https://ui.adsabs.harvard.edu/abs/2017PASP..129d3001P} {129, 043001}

\bibitem[\protect\citeauthoryear{{Peterson} et~al.,}{{Peterson}
  et~al.}{2013}]{2013ApJ...779..109P}
{Peterson} B.~M.,  et~al., 2013, \mn@doi [\apj] {10.1088/0004-637X/779/2/109},
  \href {https://ui.adsabs.harvard.edu/abs/2013ApJ...779..109P} {779, 109}

\bibitem[\protect\citeauthoryear{{Pilyugin}}{{Pilyugin}}{2003}]{2003A&A...399.1003P}
{Pilyugin} L.~S.,  2003, \mn@doi [\aap] {10.1051/0004-6361:20021669}, \href
  {https://ui.adsabs.harvard.edu/abs/2003A&A...399.1003P} {399, 1003}

\bibitem[\protect\citeauthoryear{{Pilyugin} \& {Grebel}}{{Pilyugin} \&
  {Grebel}}{2016}]{pil16}
{Pilyugin} L.~S.,  {Grebel} E.~K.,  2016, \mn@doi [\mnras]
  {10.1093/mnras/stw238}, \href
  {http://adsabs.harvard.edu/abs/2016MNRAS.457.3678P} {457, 3678}

\bibitem[\protect\citeauthoryear{{Pistis} et~al.,}{{Pistis}
  et~al.}{2022}]{2022A&A...663A.162P}
{Pistis} F.,  et~al., 2022, \mn@doi [\aap] {10.1051/0004-6361/202142430}, \href
  {https://ui.adsabs.harvard.edu/abs/2022A&A...663A.162P} {663, A162}

\bibitem[\protect\citeauthoryear{Rees}{Rees}{1989}]{Rees+89}
Rees M.~J.,  1989, \mn@doi [Monthly Notices of the Royal Astronomical Society]
  {10.1093/mnras/239.1.1P}, 239, 1P

\bibitem[\protect\citeauthoryear{{Revalski}, {Crenshaw}  \&
  {Kraemer}}{{Revalski} et~al.}{2018}]{2018ApJ...856...46R}
{Revalski} M.,  {Crenshaw} D.~M.,   {Kraemer} 2018, \mn@doi [\apj]
  {10.3847/1538-4357/aab107}, \href
  {https://ui.adsabs.harvard.edu/abs/2018ApJ...856...46R} {856, 46}

\bibitem[\protect\citeauthoryear{{Ricci}, {Ueda}, {Koss}, {Trakhtenbrot},
  {Bauer}  \& {Gandhi}}{{Ricci} et~al.}{2015}]{2015ApJ...815L..13R}
{Ricci} C.,  {Ueda} Y.,  {Koss} M.~J.,  {Trakhtenbrot} B.,  {Bauer} F.~E.,
  {Gandhi} P.,  2015, \mn@doi [\apjl] {10.1088/2041-8205/815/1/L13}, \href
  {https://ui.adsabs.harvard.edu/abs/2015ApJ...815L..13R} {815, L13}

\bibitem[\protect\citeauthoryear{{Ricci} et~al.,}{{Ricci}
  et~al.}{2017a}]{Ricci2017b}
{Ricci} C.,  et~al., 2017a, \mn@doi [\apjs] {10.3847/1538-4365/aa96ad}, \href
  {https://ui.adsabs.harvard.edu/abs/2017ApJS..233...17R} {233, 17}

\bibitem[\protect\citeauthoryear{{Ricci} et~al.,}{{Ricci}
  et~al.}{2017b}]{Ricci2017a}
{Ricci} C.,  et~al., 2017b, \mn@doi [\nat] {10.1038/nature23906}, \href
  {https://ui.adsabs.harvard.edu/abs/2017Natur.549..488R} {549, 488}

\bibitem[\protect\citeauthoryear{{Ricci} et~al.,}{{Ricci}
  et~al.}{2018}]{2018MNRAS.480.1819R}
{Ricci} C.,  et~al., 2018, \mn@doi [\mnras] {10.1093/mnras/sty1879}, \href
  {https://ui.adsabs.harvard.edu/abs/2018MNRAS.480.1819R} {480, 1819}

\bibitem[\protect\citeauthoryear{{Ricci} et~al.,}{{Ricci}
  et~al.}{2022}]{2022ApJ...938...67R}
{Ricci} C.,  et~al., 2022, \mn@doi [\apj] {10.3847/1538-4357/ac8e67}, \href
  {https://ui.adsabs.harvard.edu/abs/2022ApJ...938...67R} {938, 67}

\bibitem[\protect\citeauthoryear{{Riffel} et~al.,}{{Riffel}
  et~al.}{2021}]{2021MNRAS.501.4064R}
{Riffel} R.,  et~al., 2021, \mn@doi [\mnras] {10.1093/mnras/staa3907}, \href
  {https://ui.adsabs.harvard.edu/abs/2021MNRAS.501.4064R} {501, 4064}

\bibitem[\protect\citeauthoryear{{Riffel} et~al.,}{{Riffel}
  et~al.}{2022}]{2022MNRAS.512.3906R}
{Riffel} R.,  et~al., 2022, \mn@doi [\mnras] {10.1093/mnras/stac740}, \href
  {https://ui.adsabs.harvard.edu/abs/2022MNRAS.512.3906R} {512, 3906}

\bibitem[\protect\citeauthoryear{{Rojas} et~al.,}{{Rojas}
  et~al.}{2020}]{rojas2020}
{Rojas} A.~F.,  et~al., 2020, \mn@doi [\mnras] {10.1093/mnras/stz3386}, \href
  {https://ui.adsabs.harvard.edu/abs/2020MNRAS.491.5867R} {491, 5867}

\bibitem[\protect\citeauthoryear{{Salom{\'e}}, {Salom{\'e}},
  {Miville-Desch{\^e}nes}, {Combes}  \& {Hamer}}{{Salom{\'e}}
  et~al.}{2017}]{2017A&A...608A..98S}
{Salom{\'e}} Q.,  {Salom{\'e}} P.,  {Miville-Desch{\^e}nes} M.~A.,  {Combes}
  F.,   {Hamer} S.,  2017, \mn@doi [\aap] {10.1051/0004-6361/201731429}, \href
  {https://ui.adsabs.harvard.edu/abs/2017A&A...608A..98S} {608, A98}

\bibitem[\protect\citeauthoryear{{Sanders} et~al.,}{{Sanders}
  et~al.}{2016}]{san16a}
{Sanders} R.~L.,  et~al., 2016, \mn@doi [\apj] {10.3847/0004-637X/816/1/23},
  \href {http://adsabs.harvard.edu/abs/2016ApJ...816...23S} {816, 23}

\bibitem[\protect\citeauthoryear{{Sanders} et~al.,}{{Sanders}
  et~al.}{2021}]{2021ApJ...914...19S}
{Sanders} R.~L.,  et~al., 2021, \mn@doi [\apj] {10.3847/1538-4357/abf4c1},
  \href {https://ui.adsabs.harvard.edu/abs/2021ApJ...914...19S} {914, 19}

\bibitem[\protect\citeauthoryear{{Schawinski}, {Thomas}, {Sarzi}, {Maraston},
  {Kaviraj}, {Joo}, {Yi}  \& {Silk}}{{Schawinski} et~al.}{2007}]{schawinski07}
{Schawinski} K.,  {Thomas} D.,  {Sarzi} M.,  {Maraston} C.,  {Kaviraj} S.,
  {Joo} S.-J.,  {Yi} S.~K.,   {Silk} J.,  2007, \mn@doi [\mnras]
  {10.1111/j.1365-2966.2007.12487.x}, \href
  {https://ui.adsabs.harvard.edu/abs/2007MNRAS.382.1415S} {382, 1415}

\bibitem[\protect\citeauthoryear{{Schnorr-M{\"u}ller}
  et~al.,}{{Schnorr-M{\"u}ller} et~al.}{2016}]{2016MNRAS.462.3570S}
{Schnorr-M{\"u}ller} A.,  et~al., 2016, \mn@doi [\mnras]
  {10.1093/mnras/stw1865}, \href
  {https://ui.adsabs.harvard.edu/abs/2016MNRAS.462.3570S} {462, 3570}

\bibitem[\protect\citeauthoryear{{Sergeev}, {Nazarov}  \& {Borman}}{{Sergeev}
  et~al.}{2017}]{2017MNRAS.465.1898S}
{Sergeev} S.~G.,  {Nazarov} S.~V.,   {Borman} G.~A.,  2017, \mn@doi [\mnras]
  {10.1093/mnras/stw2857}, \href
  {https://ui.adsabs.harvard.edu/abs/2017MNRAS.465.1898S} {465, 1898}

\bibitem[\protect\citeauthoryear{{Shemmer}, {Netzer}, {Maiolino}, {Oliva},
  {Croom}, {Corbett}  \& {di Fabrizio}}{{Shemmer}
  et~al.}{2004}]{2004ApJ...614..547S}
{Shemmer} O.,  {Netzer} H.,  {Maiolino} R.,  {Oliva} E.,  {Croom} S.,
  {Corbett} E.,   {di Fabrizio} L.,  2004, \mn@doi [\apj] {10.1086/423607},
  \href {https://ui.adsabs.harvard.edu/abs/2004ApJ...614..547S} {614, 547}

\bibitem[\protect\citeauthoryear{{Shimizu} et~al.,}{{Shimizu}
  et~al.}{2019}]{2019MNRAS.490.5860S}
{Shimizu} T.~T.,  et~al., 2019, \mn@doi [\mnras] {10.1093/mnras/stz2802}, \href
  {https://ui.adsabs.harvard.edu/abs/2019MNRAS.490.5860S} {490, 5860}

\bibitem[\protect\citeauthoryear{{Silk}}{{Silk}}{2013}]{2013ApJ...772..112S}
{Silk} J.,  2013, \mn@doi [\apj] {10.1088/0004-637X/772/2/112}, \href
  {https://ui.adsabs.harvard.edu/abs/2013ApJ...772..112S} {772, 112}

\bibitem[\protect\citeauthoryear{{Smith}, {Koss}, {Mushotzky}, {Wong},
  {Shimizu}, {Ricci}  \& {Ricci}}{{Smith} et~al.}{2020}]{2020ApJ...904...83S}
{Smith} K.~L.,  {Koss} M.,  {Mushotzky} R.,  {Wong} O.~I.,  {Shimizu} T.~T.,
  {Ricci} C.,   {Ricci} F.,  2020, \mn@doi [\apj] {10.3847/1538-4357/abc3c4},
  \href {https://ui.adsabs.harvard.edu/abs/2020ApJ...904...83S} {904, 83}

\bibitem[\protect\citeauthoryear{{Somerville} \& {Dav{\'e}}}{{Somerville} \&
  {Dav{\'e}}}{2015}]{2015ARA&A..53...51S}
{Somerville} R.~S.,  {Dav{\'e}} R.,  2015, \mn@doi [\araa]
  {10.1146/annurev-astro-082812-140951}, \href
  {https://ui.adsabs.harvard.edu/abs/2015ARA&A..53...51S} {53, 51}

\bibitem[\protect\citeauthoryear{{Springel}, {Di Matteo}  \&
  {Hernquist}}{{Springel} et~al.}{2005}]{springel05b}
{Springel} V.,  {Di Matteo} T.,   {Hernquist} L.,  2005, \mn@doi [\mnras]
  {10.1111/j.1365-2966.2005.09238.x}, \href
  {https://ui.adsabs.harvard.edu/abs/2005MNRAS.361..776S} {361, 776}

\bibitem[\protect\citeauthoryear{{Stanley}, {Harrison}, {Alexander},
  {Swinbank}, {Aird}, {Del Moro}, {Hickox}  \& {Mullaney}}{{Stanley}
  et~al.}{2015}]{2015MNRAS.453..591S}
{Stanley} F.,  {Harrison} C.~M.,  {Alexander} D.~M.,  {Swinbank} A.~M.,  {Aird}
  J.~A.,  {Del Moro} A.,  {Hickox} R.~C.,   {Mullaney} J.~R.,  2015, \mn@doi
  [\mnras] {10.1093/mnras/stv1678}, \href
  {https://ui.adsabs.harvard.edu/abs/2015MNRAS.453..591S} {453, 591}

\bibitem[\protect\citeauthoryear{{Storchi-Bergmann}, {Calzetti}  \&
  {Kinney}}{{Storchi-Bergmann} et~al.}{1994}]{1994ApJ...429..572S}
{Storchi-Bergmann} T.,  {Calzetti} D.,   {Kinney} A.~L.,  1994, \mn@doi [\apj]
  {10.1086/174345}, \href
  {https://ui.adsabs.harvard.edu/abs/1994ApJ...429..572S} {429, 572}

\bibitem[\protect\citeauthoryear{{Storchi-Bergmann}, {Schmitt}, {Calzetti}  \&
  {Kinney}}{{Storchi-Bergmann} et~al.}{1998}]{sb98}
{Storchi-Bergmann} T.,  {Schmitt} H.~R.,  {Calzetti} D.,   {Kinney} A.~L.,
  1998, \mn@doi [\aj] {10.1086/300242}, \href
  {https://ui.adsabs.harvard.edu/abs/1998AJ....115..909S} {115, 909}

\bibitem[\protect\citeauthoryear{{Suganuma} et~al.,}{{Suganuma}
  et~al.}{2006}]{2006ApJ...639...46S}
{Suganuma} M.,  et~al., 2006, \mn@doi [\apj] {10.1086/499326}, \href
  {https://ui.adsabs.harvard.edu/abs/2006ApJ...639...46S} {639, 46}

\bibitem[\protect\citeauthoryear{{Suh} et~al.,}{{Suh}
  et~al.}{2019}]{2019ApJ...872..168S}
{Suh} H.,  et~al., 2019, \mn@doi [\apj] {10.3847/1538-4357/ab01fb}, \href
  {https://ui.adsabs.harvard.edu/abs/2019ApJ...872..168S} {872, 168}

\bibitem[\protect\citeauthoryear{{Sun} et~al.,}{{Sun}
  et~al.}{2018}]{2018MNRAS.480.2302S}
{Sun} A.-L.,  et~al., 2018, \mn@doi [\mnras] {10.1093/mnras/sty1394}, \href
  {https://ui.adsabs.harvard.edu/abs/2018MNRAS.480.2302S} {480, 2302}

\bibitem[\protect\citeauthoryear{{Tang} et~al.,}{{Tang}
  et~al.}{2019}]{2019MNRAS.484.2575T}
{Tang} J.-J.,  et~al., 2019, \mn@doi [\mnras] {10.1093/mnras/stz134}, \href
  {https://ui.adsabs.harvard.edu/abs/2019MNRAS.484.2575T} {484, 2575}

\bibitem[\protect\citeauthoryear{{Temple}, {Ferland}, {Rankine}, {Chatzikos}
  \& {Hewett}}{{Temple} et~al.}{2021}]{2021MNRAS.505.3247T}
{Temple} M.~J.,  {Ferland} G.~J.,  {Rankine} A.~L.,  {Chatzikos} M.,   {Hewett}
  P.~C.,  2021, \mn@doi [\mnras] {10.1093/mnras/stab1610}, \href
  {https://ui.adsabs.harvard.edu/abs/2021MNRAS.505.3247T} {505, 3247}

\bibitem[\protect\citeauthoryear{{Thomas}, {Kewley}, {Dopita}, {Groves},
  {Hopkins}  \& {Sutherland}}{{Thomas} et~al.}{2019}]{2019ApJ...874..100T}
{Thomas} A.~D.,  {Kewley} L.~J.,  {Dopita} M.~A.,  {Groves} B.~A.,  {Hopkins}
  A.~M.,   {Sutherland} R.~S.,  2019, \mn@doi [\apj]
  {10.3847/1538-4357/ab08a1}, \href
  {https://ui.adsabs.harvard.edu/abs/2019ApJ...874..100T} {874, 100}

\bibitem[\protect\citeauthoryear{{Toribio San Cipriano},
  {Dom{\'\i}nguez-Guzm{\'a}n}, {Esteban}, {Garc{\'\i}a-Rojas}, {Mesa-Delgado},
  {Bresolin}, {Rodr{\'\i}guez}  \& {Sim{\'o}n-D{\'\i}az}}{{Toribio San
  Cipriano} et~al.}{2017}]{2017MNRAS.467.3759T}
{Toribio San Cipriano} L.,  {Dom{\'\i}nguez-Guzm{\'a}n} G.,  {Esteban} C.,
  {Garc{\'\i}a-Rojas} J.,  {Mesa-Delgado} A.,  {Bresolin} F.,  {Rodr{\'\i}guez}
  M.,   {Sim{\'o}n-D{\'\i}az} S.,  2017, \mn@doi [\mnras]
  {10.1093/mnras/stx328}, \href
  {https://ui.adsabs.harvard.edu/abs/2017MNRAS.467.3759T} {467, 3759}

\bibitem[\protect\citeauthoryear{{Trakhtenbrot} et~al.,}{{Trakhtenbrot}
  et~al.}{2017}]{Trakhtenbrot2017}
{Trakhtenbrot} B.,  et~al., 2017, \mn@doi [\mnras] {10.1093/mnras/stx1117},
  \href {https://ui.adsabs.harvard.edu/abs/2017MNRAS.470..800T} {470, 800}

\bibitem[\protect\citeauthoryear{{Tremonti} et~al.,}{{Tremonti}
  et~al.}{2004}]{2004ApJ...613..898T}
{Tremonti} C.~A.,  et~al., 2004, \mn@doi [\apj] {10.1086/423264}, \href
  {https://ui.adsabs.harvard.edu/abs/2004ApJ...613..898T} {613, 898}

\bibitem[\protect\citeauthoryear{Trussler, Maiolino, Maraston, Peng, Thomas,
  Goddard  \& Lian}{Trussler et~al.}{2020}]{Trussler+20}
Trussler J.,  Maiolino R.,  Maraston C.,  Peng Y.,  Thomas D.,  Goddard D.,
  Lian J.,  2020, \mn@doi [Monthly Notices of the Royal Astronomical Society]
  {10.1093/mnras/stz3286}, 491, 5406

\bibitem[\protect\citeauthoryear{{Vaona}, {Ciroi}, {Di Mille}, {Cracco}, {La
  Mura}  \& {Rafanelli}}{{Vaona} et~al.}{2012}]{2012MNRAS.427.1266V}
{Vaona} L.,  {Ciroi} S.,  {Di Mille} F.,  {Cracco} V.,  {La Mura} G.,
  {Rafanelli} P.,  2012, \mn@doi [\mnras] {10.1111/j.1365-2966.2012.22060.x},
  \href {https://ui.adsabs.harvard.edu/abs/2012MNRAS.427.1266V} {427, 1266}

\bibitem[\protect\citeauthoryear{{Veilleux} \& {Osterbrock}}{{Veilleux} \&
  {Osterbrock}}{1987}]{1987ApJS...63..295V}
{Veilleux} S.,  {Osterbrock} D.~E.,  1987, \mn@doi [\apjs] {10.1086/191166},
  \href {https://ui.adsabs.harvard.edu/abs/1987ApJS...63..295V} {63, 295}

\bibitem[\protect\citeauthoryear{{Vernet} et~al.,}{{Vernet}
  et~al.}{2011}]{vernet11}
{Vernet} J.,  et~al., 2011, \mn@doi [\aap] {10.1051/0004-6361/201117752}, \href
  {https://ui.adsabs.harvard.edu/abs/2011A&A...536A.105V} {536, A105}

\bibitem[\protect\citeauthoryear{{Wada}, {Yonekura}  \& {Nagao}}{{Wada}
  et~al.}{2018}]{2018ApJ...867...49W}
{Wada} K.,  {Yonekura} K.,   {Nagao} T.,  2018, \mn@doi [\apj]
  {10.3847/1538-4357/aae204}, \href
  {https://ui.adsabs.harvard.edu/abs/2018ApJ...867...49W} {867, 49}

\bibitem[\protect\citeauthoryear{Wang \& Loeb}{Wang \& Loeb}{2018}]{Wang+18}
Wang X.,  Loeb A.,  2018, \mn@doi [New Astronomy]
  {10.1016/j.newast.2017.12.004}, 61, 95

\bibitem[\protect\citeauthoryear{{Wang}, {Yan}, {Gao}, {Hu}, {Li}  \&
  {Zhang}}{{Wang} et~al.}{2010}]{2010ApJ...719L.148W}
{Wang} J.-M.,  {Yan} C.-S.,  {Gao} H.-Q.,  {Hu} C.,  {Li} Y.-R.,   {Zhang} S.,
  2010, \mn@doi [\apjl] {10.1088/2041-8205/719/2/L148}, \href
  {https://ui.adsabs.harvard.edu/abs/2010ApJ...719L.148W} {719, L148}

\bibitem[\protect\citeauthoryear{{Wysota} \& {Gaskell}}{{Wysota} \&
  {Gaskell}}{1988}]{wysota88}
{Wysota} A.,  {Gaskell} C.~M.,  1988, in {Miller} H.~R.,  {Wiita} P.~J.,  eds,
  , Vol.~307, Active Galactic Nuclei.
pp 79--82, \mn@doi{10.1007/3-540-19492-4_171}

\bibitem[\protect\citeauthoryear{{Xu}, {Bian}, {Shen}, {Zuo}, {Fan}  \&
  {Zhu}}{{Xu} et~al.}{2018}]{2018MNRAS.480..345X}
{Xu} F.,  {Bian} F.,  {Shen} Y.,  {Zuo} W.,  {Fan} X.,   {Zhu} Z.,  2018,
  \mn@doi [\mnras] {10.1093/mnras/sty1763}, \href
  {https://ui.adsabs.harvard.edu/abs/2018MNRAS.480..345X} {480, 345}

\bibitem[\protect\citeauthoryear{{York} et~al.,}{{York} et~al.}{2000}]{york00}
{York} D.~G.,  et~al., 2000, \mn@doi [\aj] {10.1086/301513}, \href
  {https://ui.adsabs.harvard.edu/abs/2000AJ....120.1579Y} {120, 1579}

\bibitem[\protect\citeauthoryear{{Zhang} \& {Feng}}{{Zhang} \&
  {Feng}}{2016}]{2016MNRAS.457L..64Z}
{Zhang} X.-G.,  {Feng} L.-L.,  2016, \mn@doi [\mnras] {10.1093/mnrasl/slv204},
  \href {https://ui.adsabs.harvard.edu/abs/2016MNRAS.457L..64Z} {457, L64}

\bibitem[\protect\citeauthoryear{{Zou}, {Yang}, {Brandt}  \& {Xue}}{{Zou}
  et~al.}{2019}]{2019ApJ...878...11Z}
{Zou} F.,  {Yang} G.,  {Brandt} W.~N.,   {Xue} Y.,  2019, \mn@doi [\apj]
  {10.3847/1538-4357/ab1eb1}, \href
  {https://ui.adsabs.harvard.edu/abs/2019ApJ...878...11Z} {878, 11}

\bibitem[\protect\citeauthoryear{Zubovas \& Bourne}{Zubovas \&
  Bourne}{2017}]{Zubovas+17a}
Zubovas K.,  Bourne M.~A.,  2017, \mn@doi [Monthly Notices of the RAS]
  {10.1093/mnras/stx787}, \href
  {http://adsabs.harvard.edu/abs/2017arXiv170310782Z} {468, 4956}

\bibitem[\protect\citeauthoryear{Zubovas, Nayakshin, King  \&
  Wilkinson}{Zubovas et~al.}{2013}]{Zubovas+13}
Zubovas K.,  Nayakshin S.,  King A.,   Wilkinson M.,  2013, \mn@doi [Monthly
  Notices of the Royal Astronomical Society] {10.1093/mnras/stt952}, 433, 3079

\bibitem[\protect\citeauthoryear{{do Nascimento} et~al.,}{{do Nascimento}
  et~al.}{2022}]{2022MNRAS.513..807D}
{do Nascimento} J.~C.,  et~al., 2022, \mn@doi [\mnras] {10.1093/mnras/stac771},
  \href {https://ui.adsabs.harvard.edu/abs/2022MNRAS.513..807D} {513, 807}

\bibitem[\protect\citeauthoryear{{van Zee}, {Salzer}, {Haynes}, {O'Donoghue}
  \& {Balonek}}{{van Zee} et~al.}{1998}]{van1998spectroscopy}
{van Zee} L.,  {Salzer} J.~J.,  {Haynes} M.~P.,  {O'Donoghue} A.~A.,
  {Balonek} T.~J.,  1998, \mn@doi [\aj] {10.1086/300647}, \href
  {https://ui.adsabs.harvard.edu/abs/1998AJ....116.2805V} {116, 2805}

\makeatother
\end{thebibliography}

%\appendix
%\section{Photoionization models }
%\label{app}

% Don't change these lines
\bsp	% typesetting comment
\label{lastpage}
\end{document}